\documentclass[useAMS,usenatbib]{mnras}
\usepackage{amsmath}
\usepackage{amssymb}
\usepackage{graphicx}
\graphicspath{{./figures/}}
\usepackage[authoryear]{natbib}
\usepackage{enumitem}
\usepackage{lscape}

\usepackage{color}

\newcommand{\eq}[1]{\begin{equation}\begin{split}#1\end{split}\end{equation}}
\newcommand{\eal}[1]{\begin{align}#1\end{align}}

\hypersetup{draft}
\begin{document}

\title[BHL accretion in SgXBs]{Bondi-Hoyle-Lyttleton accretion in Supergiant X-ray binaries: stability and disk formation}
\author[Xu \& Stone]{Wenrui Xu and James M. Stone\\
Department of Astrophysical Sciences, Princeton University, Princeton, NJ 08544, USA}

\maketitle
\begin{abstract}
We use 2D (axisymmetric) and 3D hydrodynamic simulations to study Bondi-Hoyle-Lyttleton (BHL) accretion with and without transverse upstream gradients. We mainly focus on the regime of high (upstream) Mach number, weak upstream gradients and small accretor size, which is relevant to neutron star (NS) accretion in wind-fed Supergiant X-ray binaries (SgXBs).
We present a systematic exploration of the flow in this regime.
When there are no upstream gradients, the flow is always stable regardless of accretor size or Mach number.
For finite upstream gradients, there are three main types of behavior: stable flow (small upstream gradient), turbulent unstable flow without a disk (intermediate upstream gradient), and turbulent flow with a disk-like structure (relatively large upstream gradient). When the accretion flow is turbulent, the accretion rate decreases non-convergently as the accretor size decreases. The flow is more prone to instability and the disk is less likely to form than previously expected; the parameters of most observed SgXBs place them in the regime of a turbulent, disk-less accretion flow.
Among the SgXBs with relatively well-determined parameters, we find OAO 1657-415 to be the only one that is likely to host a persistent disk (or disk-like structure); this finding is consistent with observations.
\end{abstract}

\section{Introduction}
A supergiant X-ray binary (SgXB) consists of an accreting compact object, often a neutron star (NS), and a supergiant O/B star. They are among the brightest X-ray sources in our Galaxy.
For a small fraction of SgXBs, the supergiant companion fills its Roche lobe and the overflowing stellar material forms an accretion disk around the compact object \citep{Soberman1997}.
For other SgXBs, the compact object accretes from the fast stellar wind of the companion. The supergiant companion can lose up to several $10^{-6} M_\odot/$yr in stellar wind, and a small fraction of this ends up being accreted.
The two scenarios above can be called disk-fed and wind-fed\footnote{It is still possible for a disk to form around the compact object in a wind-fed system.} respectively.
In this paper we focus on the more common wind-fed SgXBs, and assume that the compact object is a NS.

Understanding the morphology of the accretion flow (especially, whether a disk can form around the NS) is crucial.
It affects the mass and angular momentum accretion rate, which then determines the X-ray luminosity, variability, and spin evolution.
It also determines what kind of model for small-scale (near and inside the magnetosphere) accretion dynamics is appropriate. Currently, most models assume accretion from a thin disk \citep{ShakuraSunyaev1973} or a quasi-spherical (and laminar) inflow \citep{Shakura2013}; but as we will show later, these assumptions may not be appropriate.

For one particular system, OAO 1657-415, observations favor the existence of a disk. Its spin evolution suggests sporadic accretion from a disk \citep{Jenke2012}, and cyclotron line observation suggests a magnetic field strength inconsistent with a disk-less wind-fed scenario \citep{Taani2018}.
However, for most observed systems there is no conclusive observational clue regarding the existence of a permanent disk (or disk-like structure) around the NS (\citealt{Shakura2012}; some evidences of transient disk: \citealt{Romano2015,Hu2017}).
Thus, theoretical modeling of the accretion flow is necessary in order to understand the accretion dynamics and better interpret observations.

The wind accretion process is often studied using the Bondi-Hoyle-Lyttleton (BHL) model (see a review in \citealt{Edgar2004}).
The standard BHL model considers a point mass accreting from a supersonic flow that is homogeneous with constant velocity at infinity. Gravitationally deflected, the flow develops a bow shock in front of the accretor and an overdense tail.
This simple model is a relatively good approximation at low NS-companion mass ratio and fast upstream wind speed ($\gg$ the orbital velocity).
More generally, the upstream flow may be asymmetric, and such asymmetry can be modeled by
imposing a transverse gradient (e.g. density or velocity gradient) on the upstream flow. This breaks the axisymmetry of standard BHL accretion, and gives the accretion flow a finite mean specific angular momentum.
In wind-fed SgXBs, such transverse gradients are mainly due to misalignment between the wind velocity (in the frame rotating with the binary) and the direction of the companion, and is usually small (see \S2.3).

Analytic and numerical studies of BHL accretion (with and without upstream gradient) have a long history (see review of earlier studies in \citealt{Edgar2004} and \citealt{Foglizzo2005}; some more recent studies are \citealt{BlondinPope2009,BR12,Blondin2013,MR15}).
Simulations of 3D BHL accretion overall find that the accretion flow is more prone to instability at higher Mach number, smaller accretor size, and larger upstream gradient.
Some important questions, however, remain unanswered:
\begin{itemize}[label=$-$]
\item The mechanism of instability remains uncertain. Several possible instability mechanisms have been proposed, but none is confirmed to be the main reason of instability (see discussion in \citealt{Foglizzo2005}).
\item The boundary (in parameter space) between stable and unstable accretion flow is unclear. No individual study has covered both sides of this boundary with sufficient number of simulations to produce a tight constraint. Knowing this boundary will also help determining the validity of proposed instability mechanisms.
\item The observed systems tend to have accretor size (for NS, this would approximately be the size of the magnetosphere) smaller than what is achievable in simulations, and it is important to know whether the system's behavior converges as the accretor size decreases. This has not been answered by previous simulations, which usually include only one or two accretor sizes.
\item The criterion for disk-formation is also uncertain. In analytic studies (e.g. \citealt{Ho1988}), it is often assumed that disk forms when the mean specific angular momentum supplied by the upstream gradients approximately exceeds the Keplerian value at the magnetosphere, but the validity of this argument is not clear when the flow near the accretor is turbulent due to instability.\footnote{In 2D (planar) BHL accretion, spontaneous disk formation has been observed when there are no upstream gradients \citep{BlondinPope2009,Blondin2013}, but similar behavior is not observed in 3D \citep{MR15}.}
\end{itemize}
In this paper, we attempt to address the above questions by numerically constraining the boundary of instability, testing convergence using multiple accretor sizes, and discussing disk formation based on understanding of flow morphology and angular momentum transport.
We focus on the regime with small accretor size and small upstream gradient, which is relevant to most SgXBs (see \S2.3) but has not been systematically explored with simulations before.

Although BHL accretion with upstream gradients captures some of the main physics of wind-fed SgXB accretion, other effects are likely non-negligible. These include orbital effects, line-driven acceleration of the upstream wind (and the inhibition of such acceleration by NS radiation feedback), and radiative cooling near the NS.
Some previous studies address these effects:
\citet{Blondin1990} and \citet{ManousakisWalter2015} perform (planar) 2D simulations including orbital effects and realistic line-driven acceleration (with NS feedback), with the latter successfully reproducing the off-states of Vela X-1;
\citet{ElMellah2018b} perform 3D simulations including orbital effects and cooling, using an upstream boundary condition derived from a realistic wind model, and demonstrate the importance of orbital effect and cooling for disk formation.
In this paper, we include orbital effects and use a parametric model of wind acceleration in some 3D simulations to compare with BHL results.
We use an adiabatic equation of state for all simulations; the neglect of cooling could affect the results, but we want to study this simple model first in order to better identify mechanisms that affect the accretion dynamics. The effect of radiative cooling should be addressed by future work.

Our paper is organized as follows: In Section 2, we parameterize the stellar wind and estimate parameters relevant to accretion (especially the strength of upstream gradient) for a number of observed SgXBs. Section 3 introduces our numerical method. Section 4 covers 2D axisymmetric simulations of BHL accretion; we discuss flow stability and convergence with respect to accretor size and resolution. Section 5 presents 3D simulations of BHL accretion with upstream gradient; we cover a wide range of accretor size and upstream gradient, and discuss flow morphology and mechanism of instability. In Section 6, we develop a model including orbital effects and parameterized wind acceleration, and apply it to observed systems. In Section 7, we summarize regimes of different behaviors in parameter space and discuss the effect of factors not included in our simulations.
We present our conclusions in Section 8.

\section{System parameters}
In this section, we start by introducing the normalization and the parametric stellar wind model we adopt, then apply the wind model to observed systems to obtain a series of parameters relevant for the accretion process.
\subsection{Normalization}
Consider a NS with mass $M_{\rm NS}$ accreting from a wind with density $\rho_\infty$ and velocity $v_\infty$ at infinity. A natural length unit of the problem is the accretion radius $R_a$ defined as
\eq{
R_a = \frac{2GM_{\rm NS}}{v_{\infty}^2}.
}
$R_a$ is also the length scale within which the flow is significantly affected by the gravity of the accretor.
In this paper, we use the normalization $GM_{\rm NS}=R_a=\rho_\infty=1$. Under this normalization, $v_\infty=\sqrt{2}$ and the unit time is the accretion time $t_a$ defined as
\eq{
t_a = \sqrt{\frac{R_a^3}{GM_{\rm NS}}} = \frac{\sqrt{2}R_a}{v_\infty}.
}
$t_a$ is roughly the time it takes for the flow to cross $R_a$. The accretion rate will be expressed in terms of the Hoyle-Lyttleton accretion rate,
\eq{
\dot M_{\rm HL} = \pi R_a^2 \rho_\infty v_\infty.
}
$\dot M_{\rm HL} = \sqrt{2}\pi$ under our normalization.

\begin{landscape}

\begin{table}
\begin{tabular*}{\linewidth}{@{\extracolsep{\fill}}lcccccccccc}
System & $M_{\rm NS}~[M_\odot]$ & $M_c~[M_\odot]$ & $R_c~[R_\odot]$ & $P_b~$[d] & $P_c~[d]$ & $P_{\rm NS}~[{\rm s}]$ & $e$ & $\dot M_{\rm wind}~[M_\odot/{\rm yr}]$ & $v_t$ [km/s] & $T_{\rm eff}$ [$10^4$K]\\
\\\hline\\
4U 1538-522          & 1.02      & 16        & 13        & 3.728 & -       & 526.8 & 0.18            & $8.3\times 10^{-7}$        & (1000) & (2.8)   \\
4U 1700-377          & 1.96      & 46        & 22        & 3.412 & -       & -     & -               & $>2.1\times 10^{-6}$       & 1700   & 4.2$^{[a]}$ \\
4U 1907+097$^{[b]}$      & -         & 27        & 26        & 8.375 & -       & 440   & 0.28            & $7\times 10^{-6}$          & 1700   & 3.05    \\
EXO 1722-363         & 1.91      & 18        & 26        & 9.740 & -       & 413.9 & \textless{}0.19 & $9.0\times 10^{-7}$        & (650)  & (2.5)   \\
GX 301-2$^{[c]}$         & 2.0       & 43        & 62        & 41.37 & 63      & 690   & 0.46            & $1\times 10^{-5}$          & 305    & 1.81    \\
OAO 1657-415         & 1.74      & 17.5      & 25        & 10.45 & 12$^d$  & 37.3  & 0.10            & $(1.1-5.6)\times 10^{-7}$  & 250    & 2.0$^{[d]}$ \\
SAX J1802.7-2017     & 1.57      & 22        & 18        & 4.570 & -       & 139.6 & -               & $6.3\times 10^{-7}$        & (680)  & (2.0)   \\
Vela X-1 (slow)      & 2.12      & 26        & 29        & 8.964 & 6.0$^{[e]}$ & 283.2 & 0.09            & $<(1.0-5.3)\times 10^{-6}$ & 600    & 2.5     \\
Vela X-1 (fast)$^{[f]}$ &           &           &           &       &         &       &                 & $4\times 10^{-6}$          & 1700   & 2.5     \\
XTE J1855-026        & 1.41      & 21        & 22        & 6.072 & -       & 360.7 & 0.04            & $<(0.2-1.1)\times 10^{-5}$ & (620)  & (3.0) 
\end{tabular*}
\caption{Physical parameters of several wind-fed SgXBs with relatively well-determined stellar properties. Subscript NS and $c$ denote the NS and the companion respectively. $P_b$ is the binary orbital period, and $P_{\rm NS},P_c$ the spin period of the NS and the companion. $v_t$ is the terminal wind velocity. Bracketed data are estimated. There are two rows for Vela X-1 since its companion's temperature lies close to the bifurcation between slow and fast winds, and earlier (fast) and more recent (slow) modeling produce very different $v_t$. References: [a] \citealt{HeapCorcoran1992}; [b] \citealt{Cox2005}; [c] \citealt{Kaper2006}; [d] \citealt{Mason2012}; [e] \citealt{Quaintrell2003}; [f] \citealt{Nagase1986}. All other data (including estimated $v_t$) come from \citet{Falanga2015} and references therein.}
\label{tab:parameters1}
\end{table}

\begin{table}
\begin{tabular*}{\linewidth}{@{\extracolsep{\fill}}lcccccccccc}
System & $R_a$ [cm] & $t_a$ [s] & $a_b/R_c$ & $v_{\Phi}/v_R$ & $R_a/D$ & $\mathcal M$ & $\epsilon_\rho$ & $\epsilon_v$ & $R_a/R_H$ & $\Omega_b t_a$\\
\\\hline\\
4U 1538-522&$6.2\times 10^{10 }$&$1.3\times 10^{3 }$&
1.96&-0.61&0.071&24&0.051&-0.016&0.19&0.026\\
4U 1700-377&$6.4\times 10^{10 }$&$1.0\times 10^{3 }$&
1.55&-0.68&0.076&26&0.052&-0.020&0.17&0.022\\
4U 1907+097&$3.5\times 10^{10 }$&$4.9\times 10^{2 }$&
2.0&-0.32&0.019&35&0.0083&-0.0024&0.058&0.0042\\
EXO 1722-363&$2.5\times 10^{11 }$&$8.0\times 10^{3 }$&
1.93&-0.72&0.15&22&0.12&-0.038&0.34&0.060\\
GX 301-2&$6.2\times 10^{11 }$&$3.0\times 10^{4 }$&
2.8&-0.92&0.078&13&0.083&-0.026&0.31&0.053\\
OAO 1657-415&$7.3\times 10^{11 }$&$4.1\times 10^{4 }$&
2.1&-1.36&0.39&11&0.44&-0.20&0.97&0.29\\
SAX J1802.7-20&$1.6\times 10^{11 }$&$4.5\times 10^{3 }$&
1.80&-1.01&0.162&22&0.153&-0.051&0.38&0.072\\
Vela X-1 (slow)&$4.2\times 10^{11 }$&$1.6\times 10^{4 }$&
1.86&-0.53&0.24&14&0.155&-0.068&0.58&0.132\\
Vela X-1 (fast)&$6.5\times 10^{10 }$&$9.9\times 10^{2 }$&
1.86&-0.188&0.038&35&0.0094&-0.0032&0.089&0.0080\\
XTE J1855-026&$1.8\times 10^{11 }$&$5.8\times 10^{3 }$&
1.76&-1.02&0.159&16&0.149&-0.050&0.37&0.069\\
\end{tabular*}
\caption{Calculated accretion parameters for SgXBs in Table \ref{tab:parameters1}.
When there is no observational data, we assume $M_{\rm NS}=1.4M_\odot$ and a non-rotating companion. (Assuming that the companion is rotating with $P_c=P_b$ gives qualitative similar parameters.)
The velocity ($v_R,v_\Phi$) is evaluated in the rotating frame in which the binary is fixed. $D = a_b-R_c$ is the separation between the NS and the surface of the companion (assuming circular orbit). $\epsilon_{\rho, v}$ are the transverse gradients of density and velocity for single-star wind profile at the NS, and physically correspond to the fractional change of $\rho,v$ per $R_a$ (see text for detailed definition).
$R_a/R_H$ characterizes the importance of the companion's gravity (with $R_H$ being the Hill sphere radius), and $\Omega_b t_a$ characterizes the importance of Coriolis force.
The last five parameters show how significantly the system differs from axisymmetric BHL accretion.}
\label{tab:parameters2}
\end{table}
\end{landscape}

\subsection{Stellar wind model}\label{subsec:wind_model}
In reality, the NS in a SgXB accretes from the wind of the companion star. Here we propose a simple parametrization of the
single-star (i.e. with NS gravity ignored) wind profile of the companion. This single-star wind model can help quantify how wind accretion in real systems differs from the ideal BHL accretion.

For simplicity, we assume that the binary orbit is circular with semi-major axis $a_b$ and the spin of the companion is aligned with the orbital angular momentum. The steady state radial velocity profile is approximately given by
\eq{
v_R = v_t\left(1-\frac{R_c}{R}\right)^\beta,\label{eq:vR}
}
with $v_t$ being the terminal velocity of the wind, $R$ the distance to the companion, $R_c$ the radius of the companion, and $\beta\approx 0.8$ \citep{FriendAbbott1986}.
The azimuthal velocity of the wind ($v_\Phi$) is determined by requiring velocity continuity at the stellar surface and constant angular momentum along streamlines.
We also assume that the wind is isothermal in steady-state, with temperature equal to the effective temperature at the stellar surface $T_{\rm eff}$.

\subsection{Parameters of observed systems}\label{subsec:parameters}
Relevant orbital and wind parameters of several observed systems are summarized in Table \ref{tab:parameters1}.
From these parameters, we can derive a set of parameters more directly related to accretion, which are given in Table \ref{tab:parameters2}. Below, we discuss the significance of some of these parameters.

The accretion radius $R_a$ is $10^{\sim 11}$ cm for all systems, with a spread of approximately one order of magnitude. This is much greater than the size of the NS ($\sim 10^6$ cm), and resolving the NS in any 3D simulation covering a few $t_a$ is highly unfeasible due to the extremely short timestep.
Meanwhile, the magnetosphere can be more than a factor of 100 larger, with \citep{DavidsonOstriker1973}
\eq{
R_{\rm mag} \approx& ~2.6\times 10^8{\rm cm}~\left(\frac{B_0}{10^{12}{\rm G}}\right)^{4/7}\left(\frac{R_{\rm NS}}{10{\rm km}}\right)^{10/7}\\
&\left(\frac{M_{\rm NS}}{M_\odot}\right)^{1/7}\left(\frac{L_x}{10^{37}{\rm erg/s}}\right)^{-2/7}.
}
Here $B_0$ is the surface magnetic field, and $L_x$ the luminosity of the NS.
Depending on the strength of magnetic field and the accretion rate (which determines $L_x$), $R_{\rm mag}$ can sometimes be $\gtrsim 10^9$ cm.
To some extend, $R_{\rm mag}$ can be considered as an effective size of the accretor,
since within $R_{\rm mag}$ the magnetic field can help remove excessive angular momentum from the flow, allowing efficient accretion.
Resolving $R_{\rm mag}\sim 10^{-2}$ - $10^{-3}R_a$ is feasible for many systems.

The azimuthal velocity $v_{\Phi}$, which is due to orbital motion and companion rotation, is often comparable to the radial velocity $v_R$. Therefore, the direction of the upstream wind (as well as the shock and the overdense region behind which can affect spectral features) can be significantly misaligned with respect to the direction of the companion, and ignoring the orbital motion is usually not a good approximation.

The Mach number $\mathcal M$ of the upstream wind all lie in the regime of high Mach number ($\mathcal M\gtrsim 10$). 
As we will show later, in this regime, the dynamics of the accretion flow is not sensitive to the Mach number, since the internal energy in the upstream flow is already negligible.

The transverse gradient parameters, $\epsilon_\rho$ and $\epsilon_v$, are defined by
\eq{
\epsilon_\rho = R_a\frac{\partial \ln\rho}{\partial y}, ~~\epsilon_v = R_a\frac{\partial \ln |v_x|}{\partial y}.
}
Here $(x,y,z)$ is the cartesian coordinate with the $xy$ plane being the orbital plane, $+z$ aligned with the orbital angular momentum, and $-x$ aligned with the (rotating frame) wind velocity $\mathbf{v}$ at NS.
$\epsilon_{\rho,v}$ approximately correspond to the fractional change of $\rho,v$ per $R_a$, and directly characterize the strength of transverse gradients in the upstream flow.
For all systems, $\epsilon_\rho$ and $\epsilon_v$ have opposite sign (i.e. the side with higher density has lower velocity). The density gradient is more important, with $\epsilon_\rho$ larger than $|\epsilon_v|$ by a factor of a few.
In general, $\rho,v$ should also have gradients along the direction of the flow, but such gradients are less important since it does not directly break axisymmetry.

The strength of transverse upstream gradients show a large scatter among systems, with $\epsilon_\rho$ ranging form $<0.01$ (e.g. 4U 1907+097) to $\sim 0.5$ (OAO 1657-415). As a result, these systems may exhibit qualitatively different behaviors (e.g. with or without the formation of a disk-like structure).
Most previous studies (e.g. \citealt{MR15,ElMellah2018b}) cover the regime of $\epsilon_\rho\gtrsim 0.1$; in this paper, we will mainly explore the regime of smaller upstream gradient, $\epsilon_\rho\lesssim 0.1$, in order to provide better coverage for the parameter space relevant for wind accretion in SgXB systems.

Finally, the ratio $R_a/R_H$ (with $R_H$ being the size of the Hill sphere) and the parameter $\Omega_b t_a$ characterize the importance of accelerations related to orbital effects.
$R_a/R_H$ characterizes the importance of the companion's gravity.
Most systems have $R_a<R_H$, i.e. the companion's gravity is unimportant. The only exception is OAO 1657-415, which has $R_a\approx R_H$.
$\Omega_b t_a$ characterizes the strength of Coriolis force in the rotating frame; it
is approximately the ratio between Coriolis force and NS gravity at $\sim 1R_a$.
Meanwhile, the ratio between centrifugal force in the rotating frame and NS gravity at $\sim 1R_a$ is $\sim(\Omega_bt_a)^2$.
All systems have $\Omega_b t_a<1$, so centrifugal force is weaker than Coriolis force, and Coriolis force is weaker than NS gravity.
Still, $\Omega_b t_a$ is often comparable to $\epsilon_\rho$, meaning that the Coriolis force may be as important as the transverse upstream gradients.

Among the many physical parameters that affect the strength of transverse upstream gradients ($\epsilon_{\rho,v}$) and the importance of orbital effects ($R_a/R_H$, $\Omega_bt_a$), 
the wind speed is the most important.\footnote{This is mainly because it shows more variation compared to other parameters. There are only three relevant dimensionless parameters (if we ignore the companion's rotation), which are $M_{\rm NS}/M_c$, $a_b/R_c$ and $v_t/v_{\rm orb}$ (where $v_{\rm orb}$ is the NS orbital velocity). Among them, $v_t/v_{\rm orb}$ shows the largest variation across systems, since the terminal velocity $v_t$ is sensitive to the temperature of the companion.}
For high wind speed, $R_a$ and $t_a$ are small, making orbital effects unimportant. Larger $v_R$ and smaller $R_a$ also give smaller transverse upstream gradient, since $\epsilon_{\rho,v}\propto R_av_\Phi/v$.
Therefore, at high wind speed, accretion can be modeled using standard BHL accretion.
On the other hand, at low wind speed (e.g. OAO 1657-415), upstream gradients and orbital effects are crucial and simple BHL model may not apply.

\section{Method}
\subsection{Equations solved}\label{subsec:eqs}
We solve the Euler equations for an inviscid compressible ideal gas using Athena++, an extension of the grid-based Godunov code package Athena \citep{Stone2008}.
The equations being solved are
\eal{
\partial_t\rho + \nabla\cdot(\rho\boldsymbol{v})&=0,\label{eq:cont}\\
\partial_t(\rho \boldsymbol{v}) + \nabla\cdot(\rho\boldsymbol{v}\boldsymbol{v} + P\boldsymbol{I})&=-\rho\nabla\Phi,\label{eq:mom}\\
\partial_t E + \nabla\cdot\left[(E+P)\boldsymbol{v}\right] &= -\rho\nabla\Phi\cdot \boldsymbol{v},\label{eq:eng}
}
with the total energy density $E$ given by
\eq{
E = \frac{P}{\gamma-1}+\frac 12 \rho v^2.
}
Throughout this paper, we adopt an ideal gas equation of state with $\gamma=5/3$ and ignore the self-gravity of the wind.
For most simulations, we also ignore the gravity of the companion, and the gravitational potential is simply $\Phi = -GM_{\rm NS}/r$.
We also perform a few simulations including the gravity of the companion in a non-inertial frame rotating at the orbital frequency. In this case, $\Phi = -GM_{\rm NS}/r-GM_c/R$ and $\nabla\Phi$ in \eqref{eq:mom} and \eqref{eq:eng} are replaced by $(\nabla\Phi - \boldsymbol a_r)$ with
\eq{
\boldsymbol a_r = -2\boldsymbol{\Omega}_b\times\boldsymbol{v}-\boldsymbol{\Omega}_b\times(\boldsymbol{\Omega}_b\times\boldsymbol{R}).
}
Since the mass ratio $M_{\rm NS}/M_c$ is small, we approximate the displacement from the center of mass with the distance to the companion $\boldsymbol R$.

\subsection{Boundary conditions}
In this paper, we use two types of mesh, spherical-polar (for axisymmetric 2D simulations) and cartesian (for 3D simulations).
Below we give a general introduction of the boundary conditions adopted,
and the implementation in each case will be discussed later in \S\ref{subsec:2D_setup} and \S\ref{subsec:3D_setup}.

The outer boundary of the domain is divided into upstream and downstream regions.
For boundary in the upstream region, we impose a pre-defined wind profile depending on the problem.
For boundary in the downstream region, we use a free boundary condition that allows both inflow and outflow.

The inner boundary, located at $r_{\rm in}$, 
physically corresponds to a surface across which all flow can be accreted (this requires an effective reduction of angular momentum for $r<r_{\rm in}$ due to, for instance, magnetic interaction; in this case $r_{\rm in}$ should be comparable to the magnetosphere size).
We use two types of inner boundary conditions, absorbing and outflow. For absorbing boundary condition, we set the velocity to zero and density and pressure to very small values in cells with $r<r_{\rm in}$. The outflow boundary condition is identical to the free flow boundary condition, except that $v_r$ in ghost cell is set to zero if the cell next to the boundary has $v_r>0$ (i.e. inflow into the domain). The outflow boundary condition is not very well-defined for cartesian grid when resolution is relatively low, thus we use it only for spherical polar grid. We show in Section \ref{subsec:compare_bc} that switching between the two inner boundary conditions does not affect the result.

\subsection{Initial condition}
By default, we specify the initial condition using the pre-defined wind profile imposed on the upstream boundary. This sometimes produces transients, which can be naturally removed after evolving the system for a few flow-crossing time. When there are multiple simulations with all parameters (e.g. $\mathcal M$, upstream gradients) being identical except $r_{\rm in}$ or resolution, we initialize a simulation with smaller $r_{\rm in}$ or higher resolution using the final state of a previous simulation with larger $r_{\rm in}$ or lower resolution, provided that the previous simulation attains a (laminar or turbulent) steady-state. This is typically more efficient, since the flow of the new simulation can usually relax into a steady-state within a few $t_a$.

\subsection{The H-correction}
It is known that numerical artifacts can develop near a strong, grid-parallel shock (the ``carbuncle" instability, see \citealt{Quirk1994}), producing large fluctuation on the shock front which can significantly perturb the flow behind the shock.
Our simulations are prone to this instability, since the shock immediately in front of the accretor is strong and approximately parallel to the grid, especially for a spherical-polar grid.
We suppress this instability via a technique called the H-correction, which adds dissipations to the transverse fluxes when the shock is grid-aligned. This technique and its implementation has been discussed in \citet{Stone2008}.
We illustrate the necessity of the H-correction through an example in Section \ref{subsec:H_necessity}.

\begin{figure*}
\centering
\includegraphics[width=.65\textwidth]{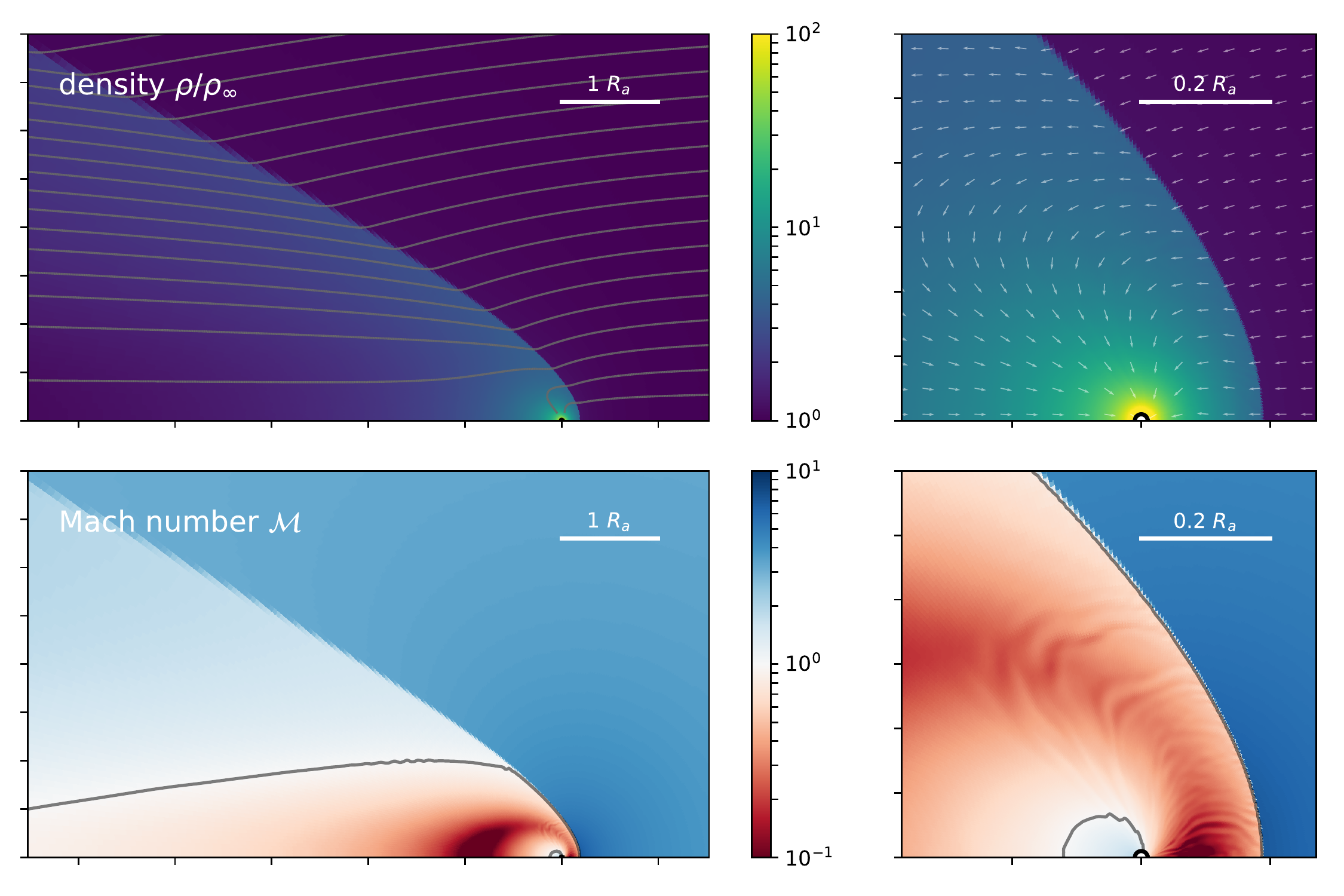}
\caption{A snapshot of the accretion flow for axisymmetric accretion with $\mathcal M=3, ~~r_{\rm in}=0.01R_a$ and $4\times$ resolution at 11$t_a$.
Top left: density and velocity streamlines on large scale. Top right: density and velocity vectors (showing only direction but not amplitude) near the accretor.
Bottom left: Mach number on large scale, with the sonic surfaces ($\mathcal M=1$) marked in grey. Bottom right: same as bottom left, but near the accretor. The inner sonic surface is attached to the inner boundary.
The flow is stable, with some small perturbation due to grid effect.}
\label{fig:2D_snapshot_3}
\end{figure*}

\begin{figure*}
\centering
\includegraphics[width=.65\textwidth]{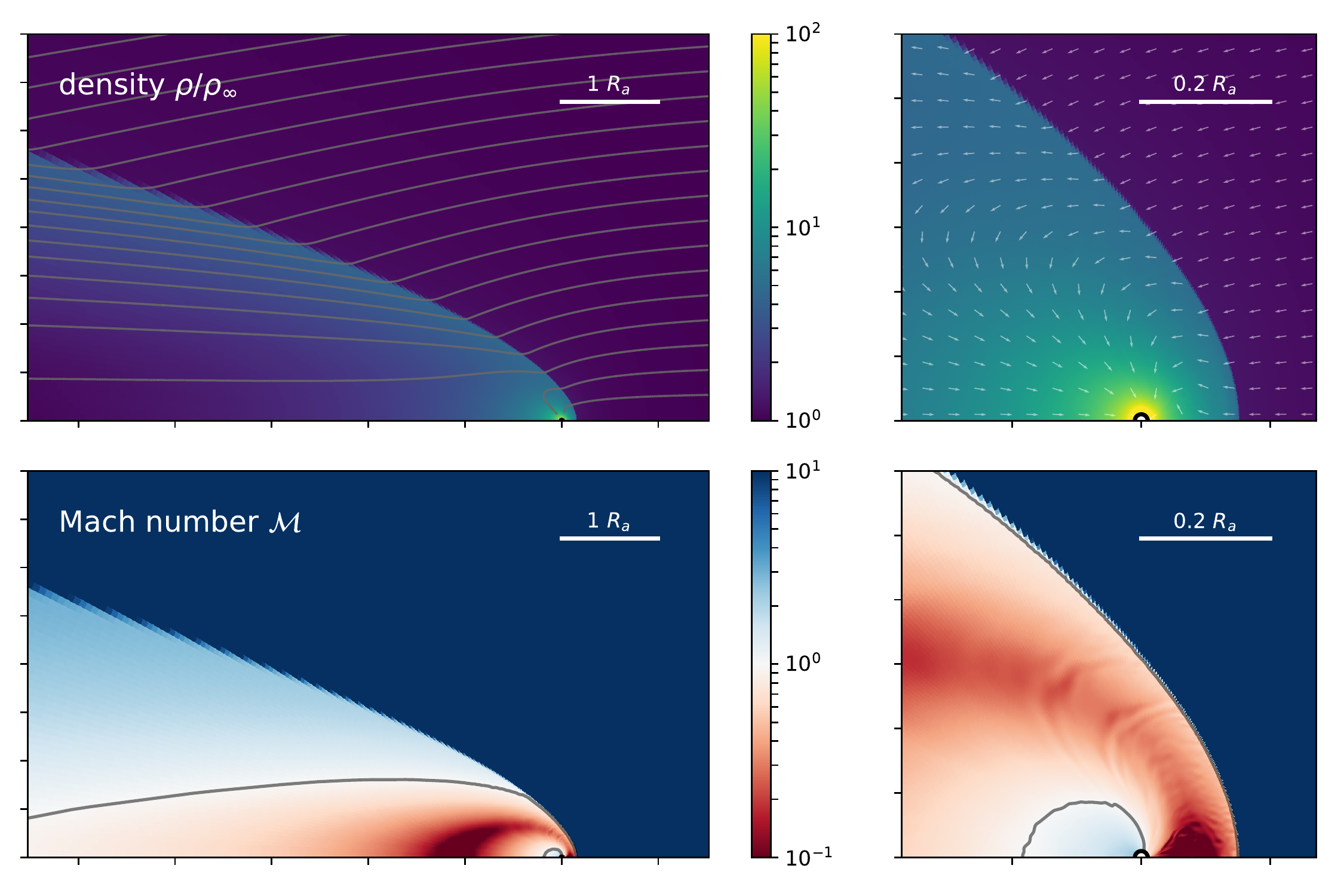}
\caption{Same as Figure \ref{fig:2D_snapshot_3} but for $\mathcal M=10$ at 15$t_a$.
}
\label{fig:2D_snapshot_10}
\end{figure*}

\section{Axisymmetric BHL accretion}\label{sec:axisym}
First we consider the problem of standard BHL accretion, where the upstream wind is axisymmetric with no transverse gradient.
In this section, we assume that there is no perturbation breaking the axisymmetry, allowing us to investigate the problem using 2D axisymmetric simulations. This case has been studied in many previous works with 2D axisymmetric simulations, starting from \citet{Hunt1971}. A review of previous works on this topic is given in \citet{Foglizzo2005}. All previous 2D axisymmetric simulations show a stable flow for $\gamma=5/3$,\footnote{Here we exclude simulations with a non-absorbing or not fully absorbing accretor, such as \citet{Fryxell1987} and \citet{Matsuda1989}.} with the only exception being \citet{Koide1991} which reports a dome-like, indented shock in front of the accretor due to the formation of a vortex when $\mathcal M\geq 5$; this leads to oscillation of the shock front and a small fluctuation of the accretion rate. The result of Koide et al., when compared to other studies, also seems to suggest that the inner boundary size may affect stability, as they use the smallest inner boundary ($r_{\rm in}=0.015$ and 0.005) among all previous 2D axisymmetric simulations.
The dependence of stability on inner boundary size is also observed by \citet{BR12}, whose 3D simulations at $\mathcal M=3$ show a stable flow for $r_{\rm in}=0.05$ and an unstable flow with a near-axisymmetric ``breathing mode" for $r_{\rm in}=0.01$, leading to an accretion rate fluctuation with $\sim 10\%$ amplitude.

In this section, we report 2D simulations with higher resolution than any previous 2D study at $\mathcal M=3$ and 10 and $r_{\rm in}=0.01$ - 0.04. We also discuss the convergence with respect to increasing resolution (which is also decreasing numerical dissipation) and the effect of changing the inner boundary condition. This allows a detailed comparison with previous works (\S \ref{subsec:axisym_comparison}).

\subsection{Setup}\label{subsec:2D_setup}
We use a 2D ($r,\theta$) spherical-polar grid, with the accretor (NS) located at the origin. The wind comes from $\theta=0$ direction, and has uniform density $\rho_\infty$ and velocity $v_\infty$ at $r\to \infty$. Given our normalization, $\rho_\infty=1$ and $v_\infty=\sqrt{2}$. The sound speed at infinity $c_\infty$ is determined by specifying the Mach number $\mathcal M\equiv v_\infty/c_\infty$.

\subsubsection{Grid and resolution}
The grids are evenly spaced in $\theta$ and $\log r$, with a lowest resolution of 96 cells in $\theta$ and 14 cells per factor of 2 in $r$. This choice gives $\delta_r/r\approx0.05$ and $\delta_\theta\approx 0.03$ where $\delta_r,\delta_\theta$ are the size of the cell.
We vary resolution from $1\times$ up to $4\times$ this base resolution to investigate numerical convergence of the result. As a reference, the 2D axisymmetric simulation by \citet{Pogorelov2000} 
is $\sim 1\times$ our base resolution (they use a special radial grid, which compared to our choice has higher resolution at $r\ll R_a$) and the 3D simulation by \citet{BR12}, with a grid evenly spaced in $\log(r)$, is $\sim2.5\times$ our base resolution; these are studies with the highest resolutions so far.

The outer boundary of the domain $r_{\rm out}$ is fixed at $10.24R_a$. For the parameters we use, this is large enough to ensure that the flow is supersonic everywhere on the outer boundary, so that the boundary cannot introduce unphysical feedback.
We vary the inner boundary $r_{\rm in}$ between $0.01R_a$ and $0.04R_a$ across different simulations, to investigate how different inner boundary size affects the result.

\subsubsection{Boundary conditions}
We define the upstream (downstream) boundary on the region with $\theta<\pi/2$ ($\theta>\pi/2$). The wind profile we impose on the upstream boundary assumes that the wind follows ballistic trajectories from infinity (see \citealt{Bisnovatyi-Kogan1979}). i.e. the gravity of the accretor is considered, but the effect of pressure gradient outside the domain is ignored. This allows analytic calculation of the flow properties on the outer boundary. It has been shown that for $\mathcal M$ as low as 1.4, the supersonic upstream flow can still be well approximated by this ballistic wind profile \citep{Koide1991}.
This ballistic wind is also used as the initial condition of the simulation. For the downstream ($\theta>\pi/2$) outer boundary, we apply free flow boundary condition.

For the inner boundary, we apply outflow boundary condition for most of the simulations. We also perform two simulations with an absorbing inner boundary condition for comparison in Section \ref{subsec:compare_bc}.

\begin{figure}
\centering
\includegraphics[width=.5\textwidth]{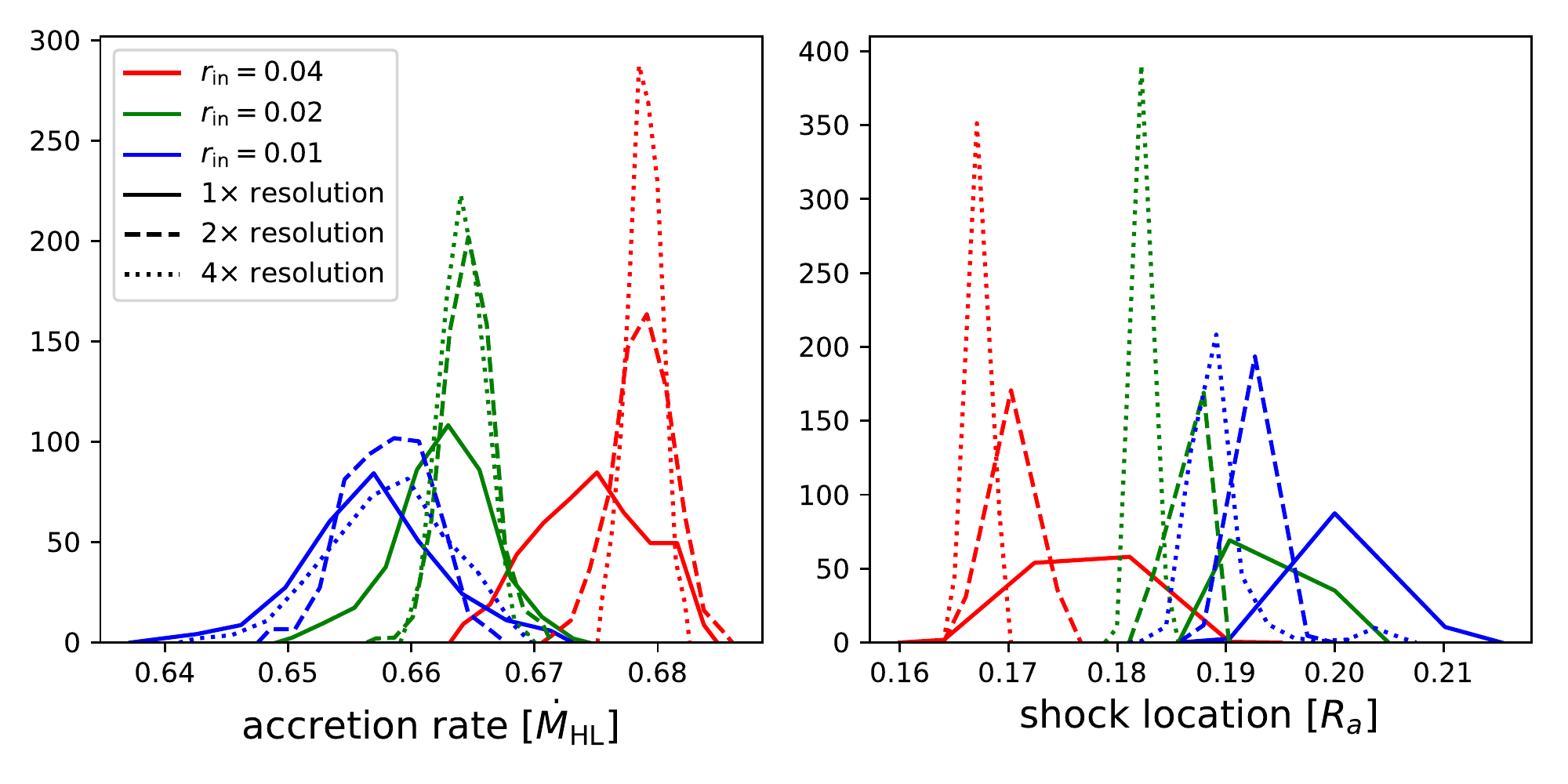}
\caption{The distribution of mass accretion rate (left panel) and shock standoff distance (right panel) for 2D axisymmetric simulations at $\mathcal M=3$ 
with different inner boundary sizes and resolutions. 
The fluctuation of accretion rate and oscillation of shock front remain similar or decrease as resolution increases, showing good numerical convergence. The result for $\mathcal M=10$ is very similar.}
\label{fig:2D_acc}
\end{figure}

\subsection{Results}
We study whether the stability of the system is affected by the Mach number, inner boundary size and resolution.
We use two Mach numbers, $\mathcal M=3$ and 10;\footnote{The real systems we are interested in all have $\mathcal M\gtrsim 10$. Here we perform $\mathcal M=3$ simulations mainly for comparison with previous studies.}
for each Mach number we use three different inner boundary sizes, $r_{\rm in}=0.04,$ 0.02 and 0.01$R_a$;
then for each pair of $(\mathcal M, r_{\rm in})$ we use three different resolutions, which are $1\times,2\times$ and $4\times$ base resolution.
This gives a set of 18 simulations in total.

\subsubsection{Properties of the flow}
For all simulations, the flow is mostly (but not exactly; see \S\ref{subsec:vortex}) laminar and quickly settles into a steady-state. The flow converges to this steady-state in $\lesssim 10 t_a$ if the simulation is initialized with the ballistic wind profile, and in $\lesssim 2t_a$ if the simulation is initialized with the steady-state of another simulation with the same $\mathcal M$. We run each simulation for 20$t_a$, thus for more than half of the time the system is in steady-state.

The density, velocity and Mach number of the steady-state accretion flow for $\mathcal M=3$ and 10, $r_{\rm in}=0.01 R_a$ and $4\times$ resolution (our smallest $r_{\rm in}$ and highest resolution) are shown in Figure \ref{fig:2D_snapshot_3} and \ref{fig:2D_snapshot_10} respectively.
The flow geometry for the two different Mach numbers are similar, with the shock front slightly closer to the accretor and the shock cone narrower for higher Mach number.
Perturbations with wavelength comparable to the cell size are visible in the Mach number profile. Such perturbations are unphysical and are due to misalignment between the grid and the shock (``stair-stepping"); its effect tends to weaken as resolution increases.

The distribution of accretion rate and shock standoff distance for $\mathcal M=3$ simulations are summarized in Figure \ref{fig:2D_acc}. The results for $\mathcal M=10$ are nearly identical and are not shown. To avoid the result being affected by the initial condition, the first 10$t_a$ of each simulation is excluded from this figure. Accretion is overall stable, with the amplitude of the accretion rate fluctuation and shock front oscillation less than a few percent for all simulations.
Figure \ref{fig:2D_acc} shows good numerical convergence with respect to increasing resolution. In general, the amplitude of accretion rate fluctuation and shock front oscillation remain similar or decrease as resolution increases, showing that there is no unresolved instability even at our lowest resolution,\footnote{Unless the instability can only be resolved for even higher ($>4\times$) resolution.} and that the unphysical perturbation due to finite grid size is indeed reduced as resolution increases.

\begin{figure}
\centering
\includegraphics[width=.35\textwidth]{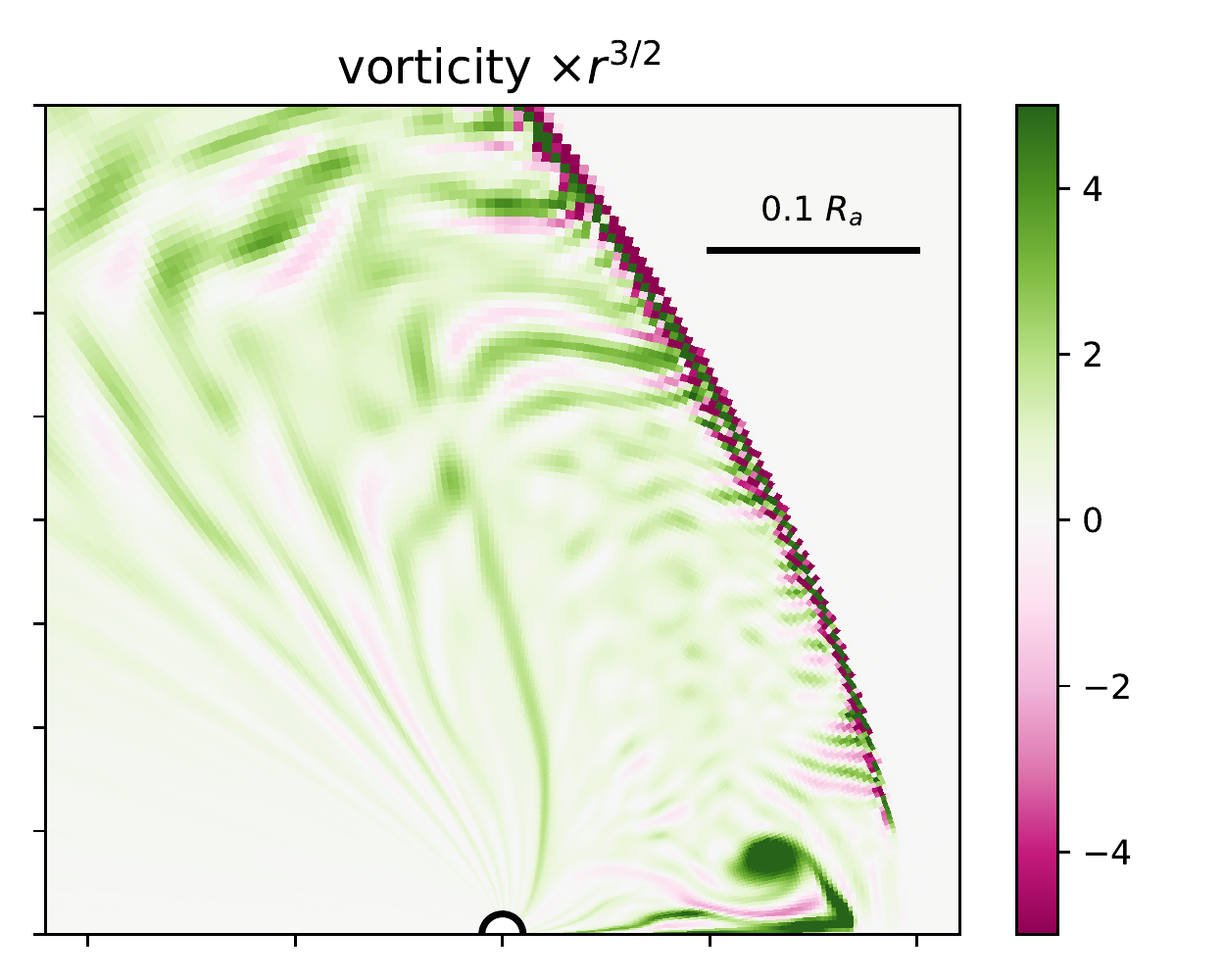}
\caption{Snapshot of vorticity profile for axisymmetric accretion with $\mathcal M=3, ~~r_{\rm in}=0.01R_a$ and $4\times$ resolution, taken at 15.5$t_a$.
Vorticity is scaled by $r^{3/2}$ to account for the larger velocity and smaller length scale when the flow is closer to the accretor.
The vortex on the upstream direction at $\sim 1.5R_a$ is the largest vortex we see in this simulation.
Vortices appear only intermittently, and for most of the time, the accretion flow does not host any vortex and is similar to the snapshot shown in Figure \ref{fig:2D_snapshot_3}.
Despite perturbation from vortices, the accretion rate and shock standoff distance remain nearly constant.
(Unphysical) vorticity perturbation behind the shock due to stair-stepping is also visible; note that the large vorticity on the shock is not due to stair-stepping but corresponds to the divergence of vorticity when there is a discontinuity.
See text for more discussion.}
\label{fig:vortex}
\end{figure}

\begin{figure}
\centering
\includegraphics[width=.35\textwidth]{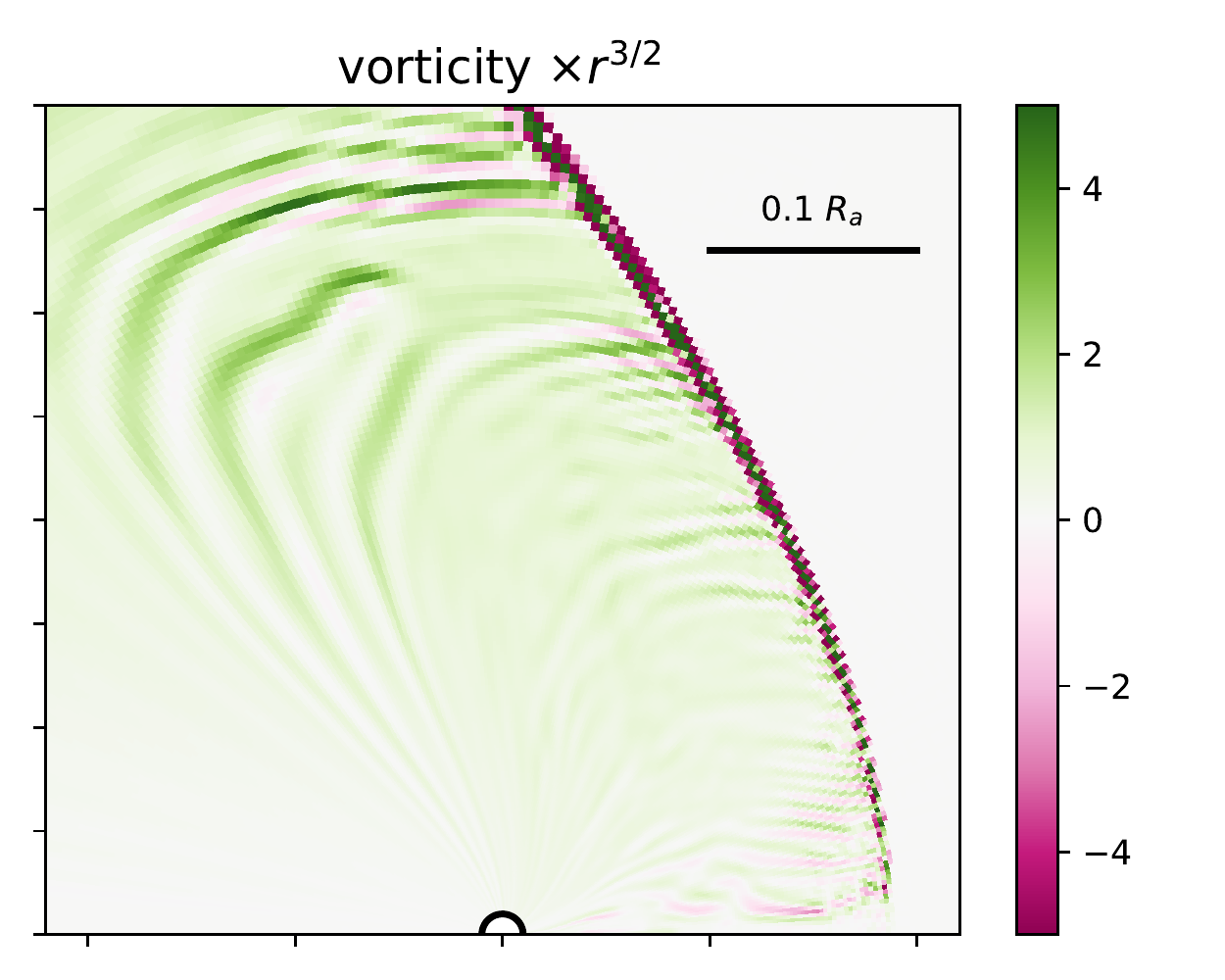}
\caption{Same as Figure \ref{fig:vortex}, but uses mesh refinement to double the resolution near the shock. (Data have been casted to unrefined resolution when making this plot.) Vortex is no longer generated in front of the accretor, and the vorticity perturbation generated by stair-stepping is weaker compared to Figure \ref{fig:vortex}.}
\label{fig:vortex_amr}
\end{figure}

\subsubsection{Vortex generation from unphysical perturbations}\label{subsec:vortex}
Although the accretion rate is nearly constant for all simulations, the accretion flow is not exactly laminar when the accretor size is small and resolution high. Instead, we observe vortices intermittently generated in front of the accretor for $\mathcal M=3,~r_{\rm in}=0.01$ and $\mathcal M=10,~r_{\rm in}=0.02$ at $4\times$ resolution, and for $\mathcal M=10,~r_{\rm in}=0.01$ at $2\times$ and $4\times$ resolution.
An example of the vortex generated is shown in Figure \ref{fig:vortex}.
Such vortex originates from the front side of the accretor, moves forward (i.e. towards upstream direction) while growing larger, then moves back downstream and is eventually absorbed by the accretor. The typical lifetime of each vortex is $\lesssim 1t_a$.
Although the vortices seem to perturb the flow significantly, it barely affects the accretion rate. This is because most of accretion happens behind the accretor (near $\theta=\pi$), and the flow there remains unperturbed as vortices only appear in front of the accretor. In addition, since vortices appear in a region of low flow velocity (this is visible in the Mach number profile in Figure \ref{fig:2D_snapshot_3} and \ref{fig:2D_snapshot_10}), it has little effect on the location of the shock front. Although the shock front moves back and forth in response to the vortex, the amplitude of such oscillation is at most a few percent of the shock standoff distance (see right panels of Figure \ref{fig:2D_acc}).

Even though they barely affect accretion rate, the origin of the vortices in axisymmetric BHL accretion is an interesting problem.
One possibility is that vortex generation in front of the accretor is related to stair-stepping at the shock front.
When the shock front is not aligned with the grid, stair-stepping generates perturbation behind the shock front, which is then advected towards the accretor. Figure \ref{fig:vortex} clearly shows the generation and advection of vorticity perturbation behind the shock.
Such perturbation can sometime make a small portion of the flow miss the accretor; this small portion of overshot flow creates a weak outflow in front of the accretor, which becomes a vortex upon encountering the incoming flow from upstream.
This mechanism can also be interpreted using vorticity conservation: Vorticity perturbation that is generated by stair-stepping and advected towards the accretor but fails to be accreted can be accumulated in front of the accretor, since in this asymmetric flow $\omega/(r\sin\theta)$ is approximately conserved.\footnote{Gravity, as a conservative force, does not change vorticity. However, pressure gradient that is misaligned with density gradient (since entropy gradient in the flow makes it non-barotropic) can break vorticity conservation. This makes the vorticity conservation argument less rigorous.}
This explanation is consistent with the observation that vortex generation happens only for flow with small $r_{\rm in}$, which makes the flow (and the advecting vorticity perturbation) easier to miss accretor, and high resolution, which corresponds to lower numerical viscosity (vorticity generation by stair-stepping, on the other hand, is not significantly reduced when resolution increases, since the characteristic wavenumber also increases).

To test our explanation, we perform a simulation with $\mathcal M=3,~ r_{\rm in}=0.01$ and $4\times$ (base) resolution with a mesh refinement that doubles the resolution near the shock. Although the increased resolution may not directly reduce vorticity production by stair-stepping, the perturbation generated by stair-stepping should be damped as it passes the refinement boundary due to the sudden drop of resolution. Therefore, vortex generation should be suppressed compared to the simulation without mesh refinement. This is indeed the case; in the simulation with mesh refinement, we never observe vortex generation, and the vorticity perturbation (shown in Figure \ref{fig:vortex_amr}) due to stair-stepping is smaller near the accretor compared to Figure \ref{fig:vortex}.
This provides strong evidence that the vortices we observe originate from numerical artifacts.

Although the perturbations that generate the vortices are unphysical, this vortex generation mechanism demonstrates that when $r_{\rm in}$ is small, even small perturbation can significantly affect the flow structure in front of the accretor. This is related to why the flow is prone to instability in 3D at small accretor size and finite upstream gradients (see \S5).

\begin{figure}
\centering
\includegraphics[width=.35\textwidth]{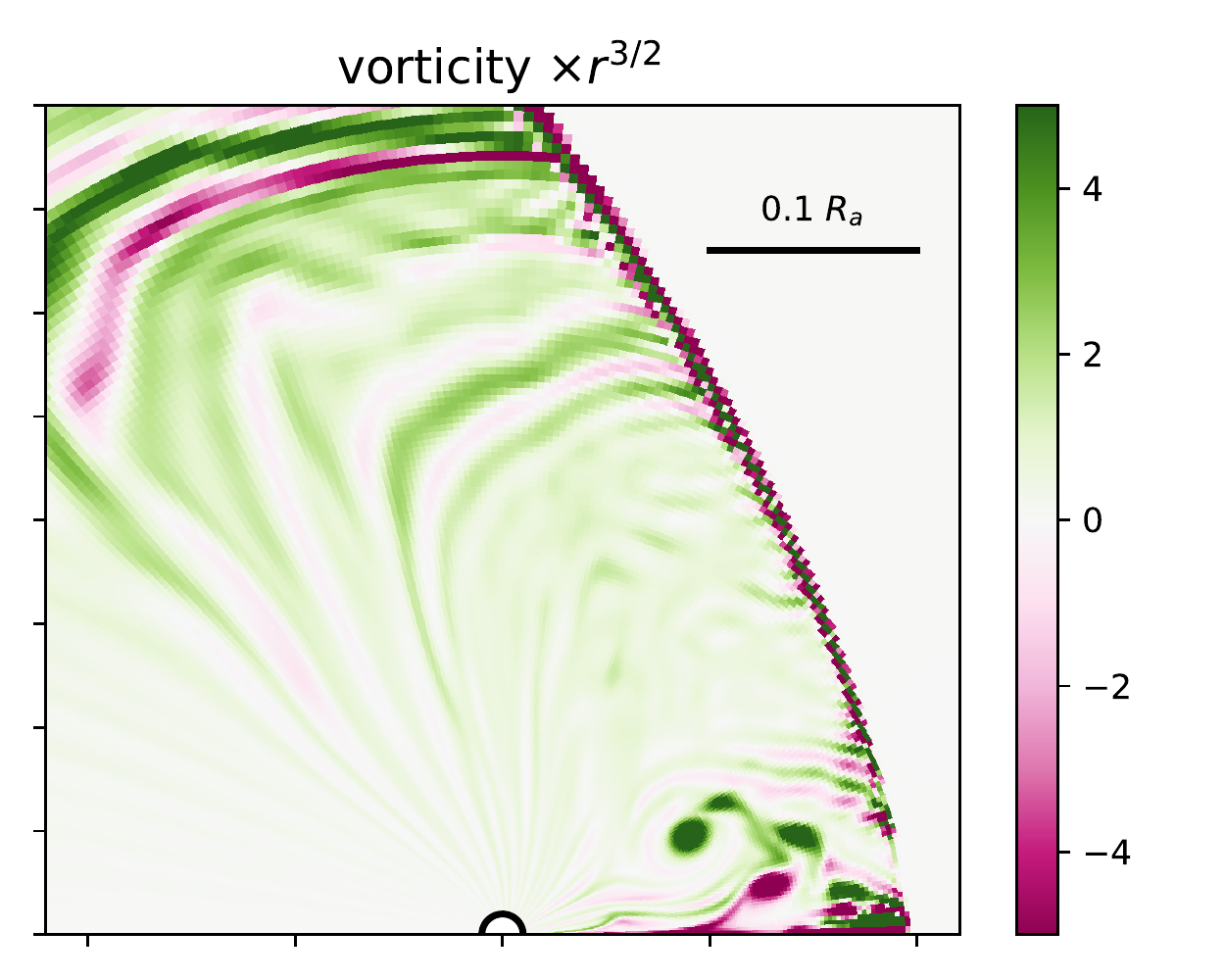}
\caption{Same as Figure \ref{fig:vortex}, but the simulation is performed without the H-correction. Unphysical vortices are generated at the shock front, in contrast to the vortices originating near the accretor observed in other simulations.}
\label{fig:vortex_no_H}
\end{figure}

\subsubsection{Necessity of the H-correction}\label{subsec:H_necessity}
Near $\theta=0$, the shock is approximately aligned with the grid, making the flow prone to numerical artifacts \citep{Quirk1994}. We suppress such artifacts by introducing extra dissipation near the shock though the H-correction \citep{Stone2008}. To illustrate the necessity of this H-correction, we perform a simulation with $\mathcal M=3,~r_{\rm in}=0.01$ and $4\times$ resolution without the H-correction. 
Figure \ref{fig:vortex_no_H} shows that removing the H-correction leads to unphysical vortex generation at the shock front; a series of vortices are generated at the shock near $\theta=0$, and advect with the flow, eventually reaching the accretor. (This is very different from the vortex in Figure \ref{fig:vortex}, which originates near the accretor.) The vorticity perturbation due to stair-stepping at the shock is also stronger compared to Figure \ref{fig:vortex} (most visible in the upper left part of the figure). The H-correction is therefore necessary, since otherwise numerical artifacts can significantly affect flow properties.

\begin{figure}
\centering
\includegraphics[width=.5\textwidth]{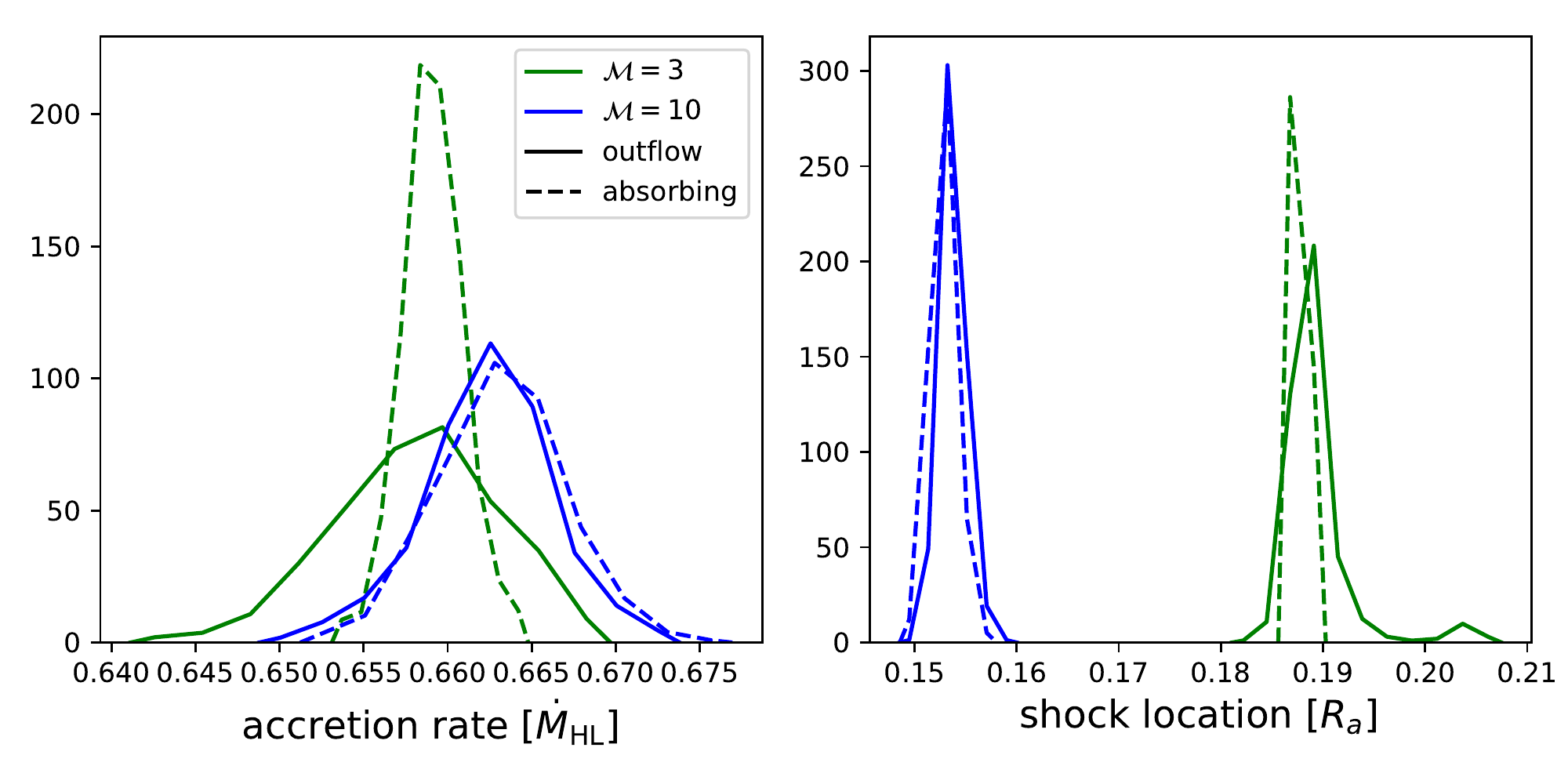}
\caption{The distribution of mass accretion rate (left panel) and shock standoff distance (right panel) for different inner boundary conditions. 
Results for absorbing and outflow boundary conditions are similar, except that absorbing boundary condition tends to suppress the formation of vortices.}
\label{fig:2D_acc_absorb}
\end{figure}

\subsubsection{Effect of different boundary condition}\label{subsec:compare_bc}
From Figure \ref{fig:2D_snapshot_3} and \ref{fig:2D_snapshot_10}, we observe that the sonic surface is in contact with the inner boundary, i.e. part of the inner boundary is subsonic. For $\gamma=5/3$, this has to be the case if the flow is in (laminar) steady state, as proved by \citet{FoglizzoRuffert1997}. Therefore, choosing a different inner boundary condition may affect the feedback at the inner boundary. To test whether this affects the stability of the flow, we redo two of our simulations ($r_{\rm in}=0.01,~4\times$ resolution, $\mathcal M=3$ and 10) with an absorbing inner boundary condition.
Figure \ref{fig:2D_acc_absorb} compares the distributions of accretion rate and shock standoff distance for different boundary conditions. For $\mathcal M=3$, absorbing boundary condition produces significantly less fluctuation because vortices are barely generated in this case. Meanwhile, for $\mathcal M=10$, the distributions of accretion rate and shock standoff distance, as well as the properties of the flow, are mostly independent of the boundary condition.
Therefore, changing the boundary condition does not significantly affect the properties of the flow, although an absorbing inner boundary condition tends to stabilize the flow more (by pulling the flow adjacent to the inner boundary more strongly).

\subsection{Comparison with previous studies}\label{subsec:axisym_comparison}
We find the axisymmetric accretion flow to be overall stable, with nearly constant accretion rate and shock standoff distance; this is in agreement with most previous studies (e.g. \citealt{Pogorelov2000}).

In some simulations, we observe intermittent vortex generation in front of the accretor, which is triggered by unphysical perturbation due to stair-stepping at the shock.
Most previous 2D axisymmetric simulations, which have relatively large $r_{\rm in}$ and relatively low resolution, do not produce such vortices;
this is consistent with our observation that vortex generation only happens at small $r_{\rm in}$ and high resolution (low numerical viscosity).
The only exception is \citet{Koide1991}, which observes a relatively large  (compared to ours) vortex for $r_{\rm in}=0.015R_a$ and $\mathcal M=10$ (see their Figure 12); it is likely that this vortex also originates from unphysical perturbation, similar to those in our simulations.

While all our simulations exhibit stable flow, observations of vortex generation (in response to unphysical perturbation at the shock) suggest that systems with larger $\mathcal M$ and smaller $r_{\rm in}$ are more sensitive to perturbation.
This $r_{\rm in}$ dependence qualitatively agrees with the 3D simulations of \citet{BR12}.
However, Blondin \& Raymer report a ``breathing-mode" which leads to a quasi-periodic variation of the accretion rate with amplitude $\sim 10\%$ at $\mathcal M=3,~~r_{\rm in}=0.01$; similar behavior is never observed in our simulations.
This breathing mode is likely a 3D effect: Although Blondin \& Raymer observes the flow to be highly axisymmetric, they also comment that this is not the case near the accretor for small accretor size, and significant asymmetry can be observed in the mass flux across the inner boundary shown on the lower panel of their Figure 5.
Since they also report a very low angular momentum accretion rate (the mean specific angular momentum of accreted material is always $<5\%$ of the Keplerian specific angular momentum at $r_{\rm in}$), it is possible that the breathing mode they observe originates from a reflection-symmetric but non-axisymmetric perturbation near the accretor,
which may be related to the Yin-Yang grid geometry they adopt.

\begin{table*}
\begin{tabular*}{\linewidth}{@{\extracolsep{\fill}}ccccccccc}
Name & $\mathcal M$ & $r_{\rm in}$ & gradient & root resolution & max refinement level & $r_{\rm in}/\delta$\\
 & & [$R_a$] & & $R_a/\delta$ & (root = 0) & \\

\\\hline\hline\\

AS1 & 10 & 0.04 & - & 8 & 5 & 10.24\\
AS2 & 10 & 0.02 & - & 8 & 6 & 10.24\\
AS3 & 10 & 0.01 & - & 8 & 7 & 10.24\\

\\\hline\\

B1 & 10 & 0.04 & $\epsilon_\rho=0.1$ & 8 & 5 & 10.24\\
B2 & 10 & 0.02 & $\epsilon_\rho=0.1$ & 8 & 6 & 10.24\\
B3 & 10 & 0.01 & $\epsilon_\rho=0.1$ & 8 & 7 & 10.24\\
B4 & 10 & 0.005 & $\epsilon_\rho=0.1$ & 8 & 8 & 10.24\\

\\\hline\\

C1 & 10 & 0.04 & $\epsilon_\rho=0.05$ & 8 & 5 & 10.24\\
D1 & 10 & 0.04 & $\epsilon_\rho=0.02$ & 8 & 5 & 10.24\\
D2 & 10 & 0.02 & $\epsilon_\rho=0.02$ & 8 & 6 & 10.24\\
D3 & 10 & 0.01 & $\epsilon_\rho=0.02$ & 8 & 7 & 10.24\\
D4 & 10 & 0.005 & $\epsilon_\rho=0.02$ & 8 & 8 & 10.24\\
E1 & 10 & 0.04 & $\epsilon_\rho=0.01$ & 8 & 5 & 10.24\\
E2 & 10 & 0.02 & $\epsilon_\rho=0.01$ & 8 & 6 & 10.24\\
E3 & 10 & 0.01 & $\epsilon_\rho=0.01$ & 8 & 7 & 10.24\\
E4 & 10 & 0.005 & $\epsilon_\rho=0.01$ & 8 & 8 & 10.24\\

\\\hline\\

V & 10 & 0.01 & $\epsilon_v=0.1$ & 8 & 7 & 10.24\\
M & 30 & 0.01 & $\epsilon_\rho=0.1$ & 8 & 7 & 10.24\\
R & 10 & 0.01 & $\epsilon_\rho=0.1$ & 8 & 8 & 20.48\\

\\\hline\\

VF1 & 10 & 0.01 & Vela X-1 (fast) & 8 & 7 & 10.24\\
VF2 & 10 & 0.005 & Vela X-1 (fast) & 8 & 8 & 10.24\\
VF3 & 10 & 0.0025 & Vela X-1 (fast)& 8 & 9 & 10.24\\
VF1R & 10 & 0.01 & Vela X-1 (fast) & 8 & 8 & 20.48\\
OAO & 10 & 0.01 & OAO 1657-415 & 16 & 6 & 10.24\\

\end{tabular*}
\caption{Parameters for 3D numerical simulations. We use a cartesian grid with static mesh refinement, choosing the refinement level at each location so that the local resolution $r/\delta$ is no smaller than $10$ (20 for R and VF1R) for $r\geq r_{\rm in}$. For the last five simulations, the initial and boundary conditions are implemented using our parametrized wind model (see Section \ref{subsec:wind_model}) with parameters drawn from corresponding systems (see Table \ref{tab:parameters1} and \ref{tab:parameters2}). All simulations (except the resolution studies R and VF1R) run for $\sim 40$ $t_a$.}
\label{tab:simulations}
\end{table*}

\begin{figure*}
\centering
\includegraphics[width=.8\textwidth]{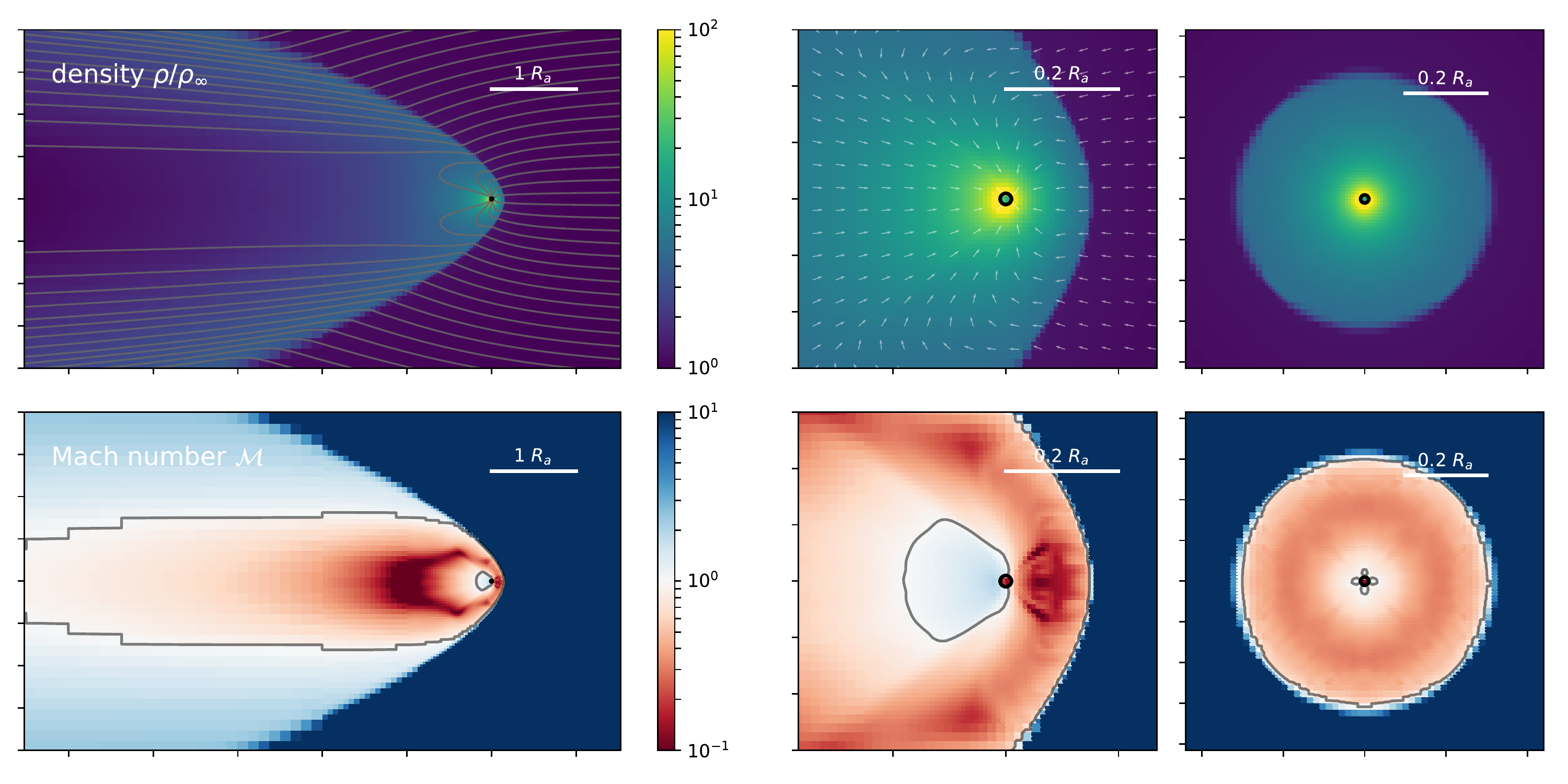}
\caption{A snapshot of the accretion flow for simulation AS3. The left and middle panels are similar to Figure \ref{fig:2D_snapshot_3}, showing cross sections at $z=0$. The right panels show cross sections at $x=0$. (Data have been downsampled when plotting. The actual resolution near the accretor is higher.)}
\label{fig:A3}
\end{figure*}

\begin{figure*}
\centering
\includegraphics[width=.8\textwidth]{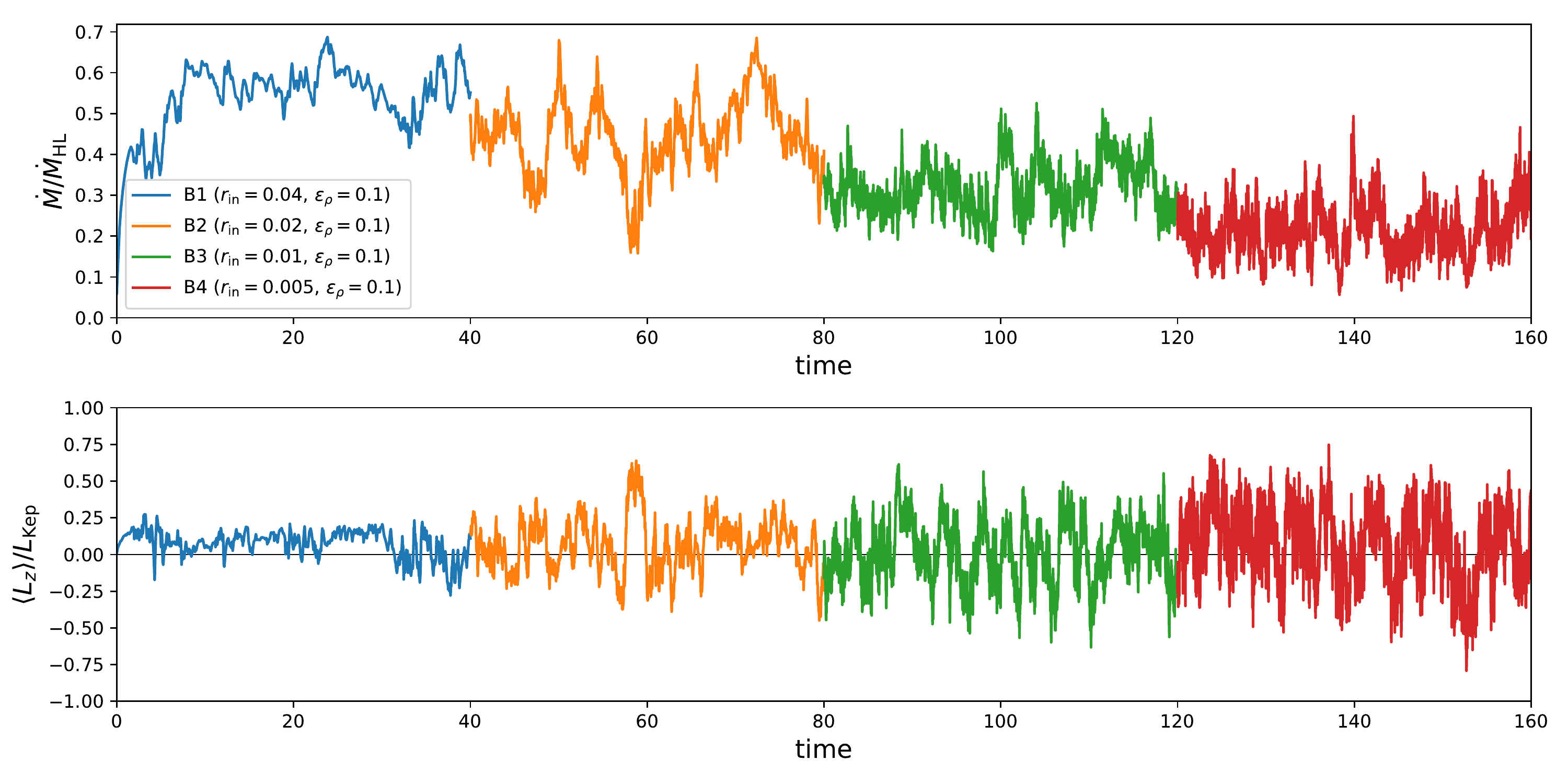}
\caption{Mass accretion rate $\dot M$ and mean specific angular momentum in $z$ direction of accreted material $\langle L_z\rangle$ (normalized by $L_{\rm Kep}\equiv\sqrt{GM_{\rm NS}r_{\rm in}}$, which is different for each simulation) for simulations B1 - B4 (same $\epsilon_\rho$, decreasing $r_{\rm in}$). As $r_{\rm in}$ decreases, the flow becomes more unstable, with accretion rate decreasing and the amplitude of  $\langle L_z\rangle/L_{\rm Kep}$ fluctuation increasing, which eventually becomes comparable to $L_{\rm Kep}$. $\dot M$ and $\langle L_z\rangle$ show no visible periodicity.}
\label{fig:B}
\end{figure*}
\begin{figure}
\centering
\includegraphics[width=.5\textwidth]{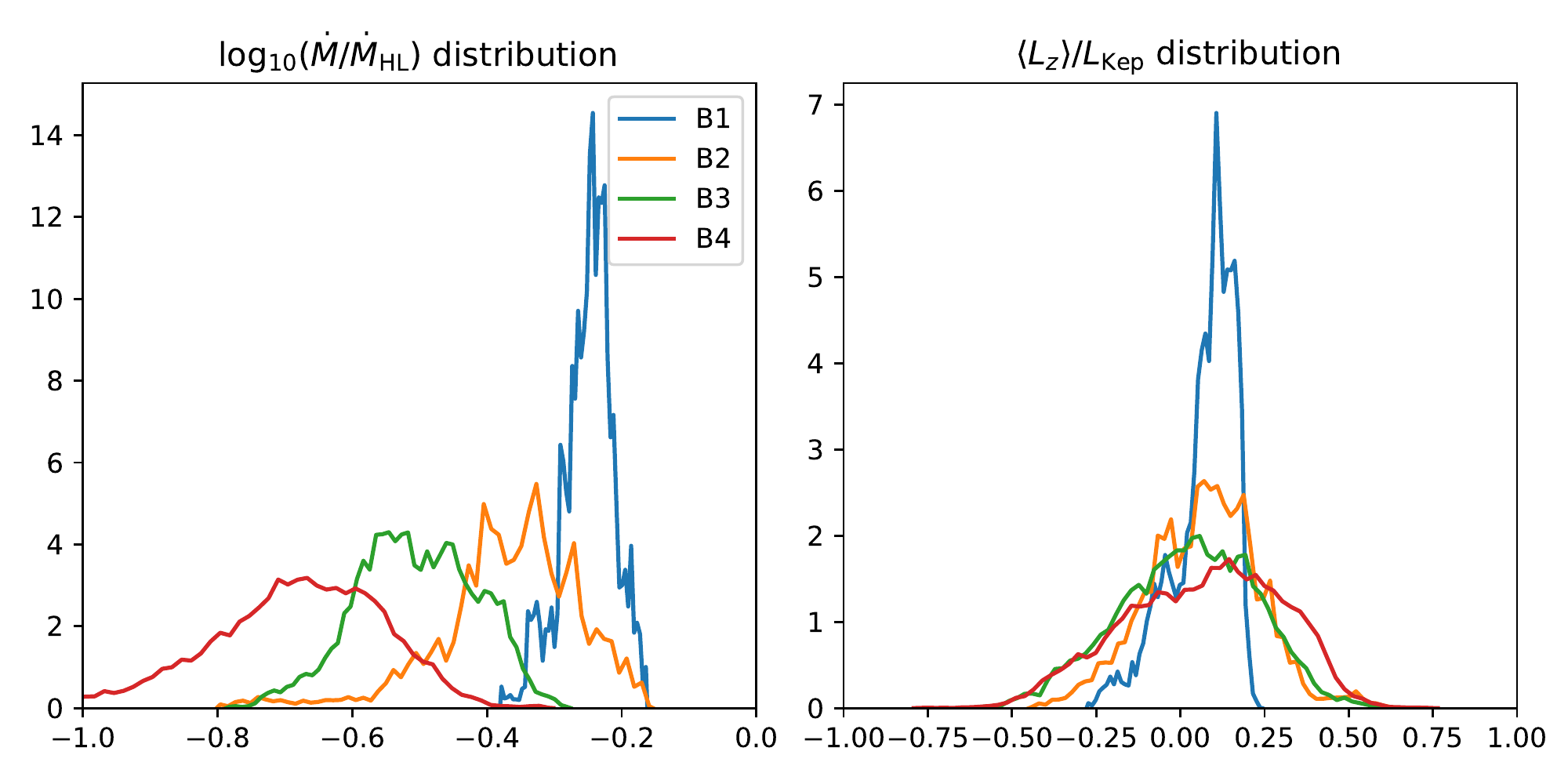}
\caption{Distribution of $\dot M$ and $\langle L_z\rangle/L_{\rm Kep}$ for simulations B1 - B4. As $r_{\rm in}$ decreases, $\dot M$ decreases and the distributions of $\dot M $ and $\langle L_z\rangle/L_{\rm Kep}$ widen. Note that the amplitude of $\langle L_z\rangle$ fluctuation does not necessarily increase, since $L_{\rm Kep}$ decreases for decreasing $r_{\rm in}$.}
\label{fig:B_hist}
\end{figure}

\begin{figure*}
\centering
\includegraphics[width=\textwidth]{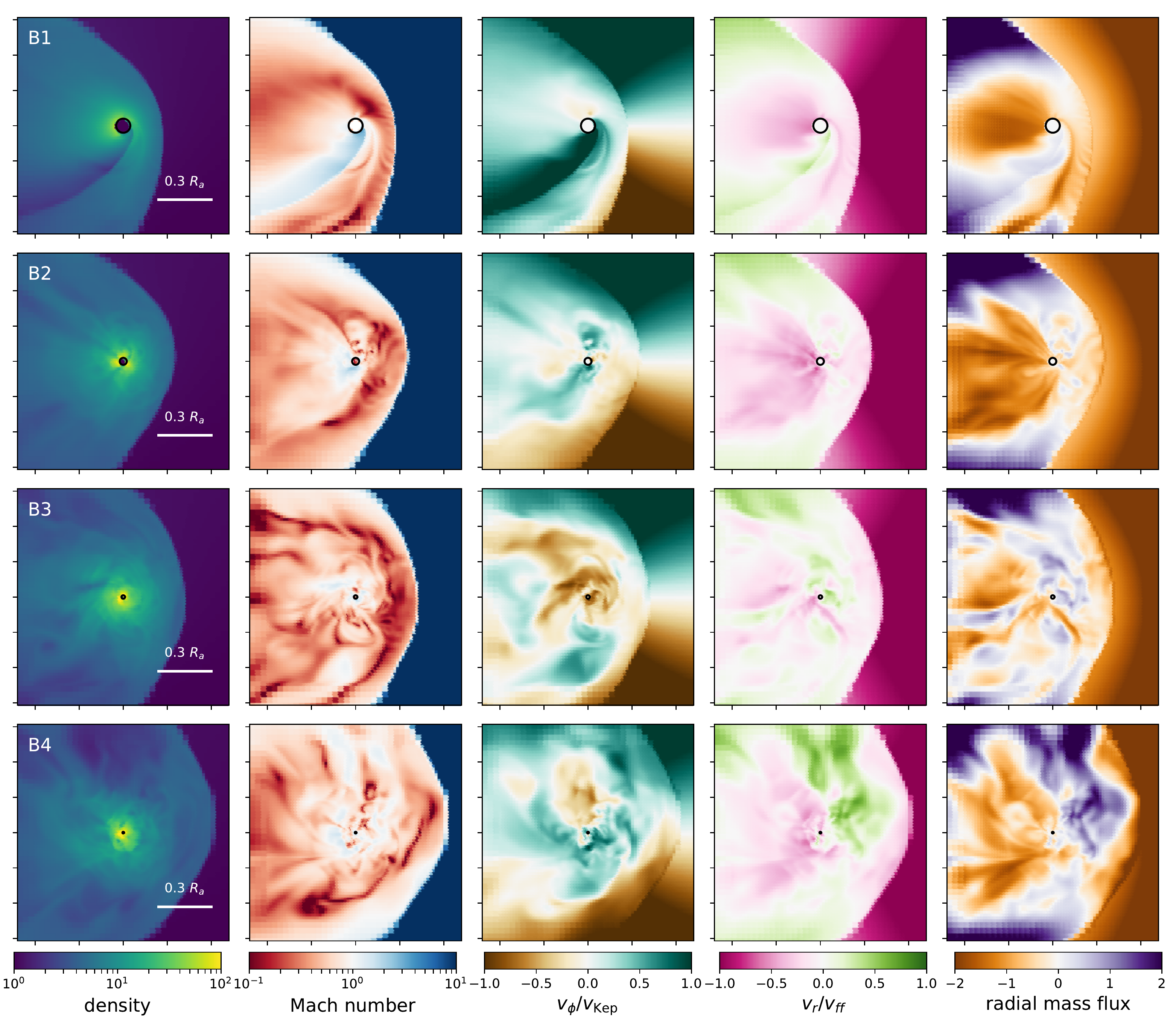}
\caption{Snapshots of accretion flow for B1 - B4. Each row gives a snapshot of one simulation, sliced at $z=0$. From left to right: Density ($\rho/\rho_\infty$); Mach number; azimuthal velocity $v_\phi/v_{\rm Kep}$ where $v_{\rm Kep} \equiv \sqrt{GM_{\rm NS}/r}$ and $r,\phi$ are defined for the spherical-polar coordinate with the north pole ($\theta=0$) in $+\hat{\boldsymbol{z}}$ direction; $v_r/v_{\rm ff}$ with the free-fall velocity $v_{\rm ff} \equiv \sqrt{2GM_{\rm NS}/r}$; radial mass flux per unit solid angle, defined as $d\dot M/d\Omega = r^2\rho v_r$, and normalized by $\dot M_{\rm HL}/4\pi$. As $r_{\rm in}$ decreases, the flow becomes more unstable and eventually turbulent.}
\label{fig:B_flow}
\end{figure*}

\section{3D BHL accretion with zero or small transverse upstream gradient}\label{sec:grad}
Simulations of BHL accretion in 3D with or without imposed transverse upstream gradients also have a long history (see a review in \citealt{Foglizzo2005}, which also includes simulations with $\gamma$ other than $5/3$).
The earliest works (e.g. \citealt{LSK86, MSS91}) used low resolution and did not resolve the accretor (i.e. diameter of accretor is only 1 - 2 cells).
\citet{SMA89} is perhaps the first to resolve the accretor (with $r_{\rm in}\approx 0.1R_a$) thanks to the use of a special radial grid with $\Delta r/r$ increasing near the accretor, and they find the flow to be stable when there are no upstream gradients and quasi-stable when there is a small velocity gradient.
A series of works by Ruffert \citep{RuffertArnett1994, R94, RA95, R97, R99} using a nested Cartesian grid manage to achieve a resolution of $r/\delta \sim 5$ (where $\delta$ is the cell size) for most simulations and $\sim 10$ for a few high-resolution simulations, allowing the accretor to be resolved. 
They find the flow to be always unstable, for $r_{\rm in}$ as large as $0.1R_a$, regardless of whether there is any upstream gradient. In all simulations, they randomly perturb the initial density of each cell by $3\%$ in order to break the symmetry of the grid and initial condition. We will show that such relatively large initial perturbation is the main reason for the instabilities they observe in the absence of upstream gradients (see the end of Section \ref{subsec:3D_sym}).

More recently, this problem has been revisited by \citet{BR12} and \citet{MR15} using higher resolution simulations, with a focus on how the stability of the flow depends on $r_{\rm in}$.
\citet{BR12} simulate BHL accretion with no upstream gradient, and report a stable flow for $r_{\rm in}=0.05R_a$ and an unstable flow with a near-axisymmetric breathing mode for $r_{\rm in}=0.01 R_a$; we discussed this breathing mode previously in Section \ref{subsec:axisym_comparison}.
\citet{MR15} study BHL accretion with finite upstream gradients in the context of accretion within a common envelope and mainly covers the regime of relatively low Mach number ($\mathcal M\lesssim 3$) and relatively large density gradient ($\epsilon_\rho = $0.1 - 5). They find the flow to be unstable for all simulations with finite $\epsilon_\rho$. Comparing results for $r_{\rm in}=0.05R_a$ and 0.01$R_a$, they also find that the flow becomes more unstable for smaller $r_{\rm in}$.

In this section, we consider high-$\mathcal M$ BHL accretion in full 3D, with small ($\epsilon_\rho \lesssim 0.1$) but finite transverse upstream gradient, which is relevant for many SgXB systems (Table \ref{tab:parameters2}).
Our simulations follow a formalism largely similar to that of \citet{MR15}, but we cover a different parameter space. We also use multiple $r_{\rm in}$ (between 0.005 and 0.04) to thoroughly investigate the $r_{\rm in}$ dependence.

\subsection{Setup}\label{subsec:3D_setup}
The upstream flow is now allowed to have some finite transverse gradient. We parametrize the transverse gradients such that wind at infinity has
\eq{
\rho = \rho_\infty\exp(\epsilon_\rho y),~~~v = v_\infty\exp(\epsilon_v y).\label{eq:parametrization}
}
Here $y$ is a direction normal to the direction of velocity (which we define as $-\hat{\boldsymbol{x}}$). The sound speed is set such that the flow is isentropic at infinity, with its value at $y=0$ specified by the Mach number $\mathcal M$.
We assume that $\epsilon_\rho,~\epsilon_v$ are both small, so the parameterization \eqref{eq:parametrization} is approximately a linear dependence on $y$.

\citet{MR15} have used the same setup to study BHL accretion with relatively large upstream density gradient ($\epsilon_\rho$ between 0.1 and 5).
However, in this section we only consider the case when $\epsilon_\rho,\epsilon_v$ are small ($\lesssim 0.1$) or zero, for two reasons: First, many observed systems show small transverse upstream gradients (\S \ref{subsec:parameters}), with $\epsilon_\rho,\epsilon_v\lesssim 0.1$, and it is important to study if the behavior at small gradients is different from that at larger gradient; especially, we want to see whether a realistic $r_{\rm in}$ (e.g. $r_{\rm in}\sim R_{\rm mag}$) can be sufficiently large to make the flow stable when $\epsilon_\rho,\epsilon_v$ are small.
Second, when transverse upstream gradients are large, a simple parametrization such as \eqref{eq:parametrization} may not be a good approximation of the actual wind profile; the effect of companion gravity and orbital motion are also often non-negligible, since systems with large upstream gradients usually also have large $R_a$. Later in this paper (Section \ref{subsec:OAO}), we will study systems with large transverse upstream gradients with an example that adopts a more realistic wind profile with parameters resembling OAO 1657-415.

A list of 3D simulations we perform (including simulations in Section \ref{sec:real}) is given in Table \ref{tab:simulations}.

\begin{figure}
\centering
\includegraphics[width=.5\textwidth]{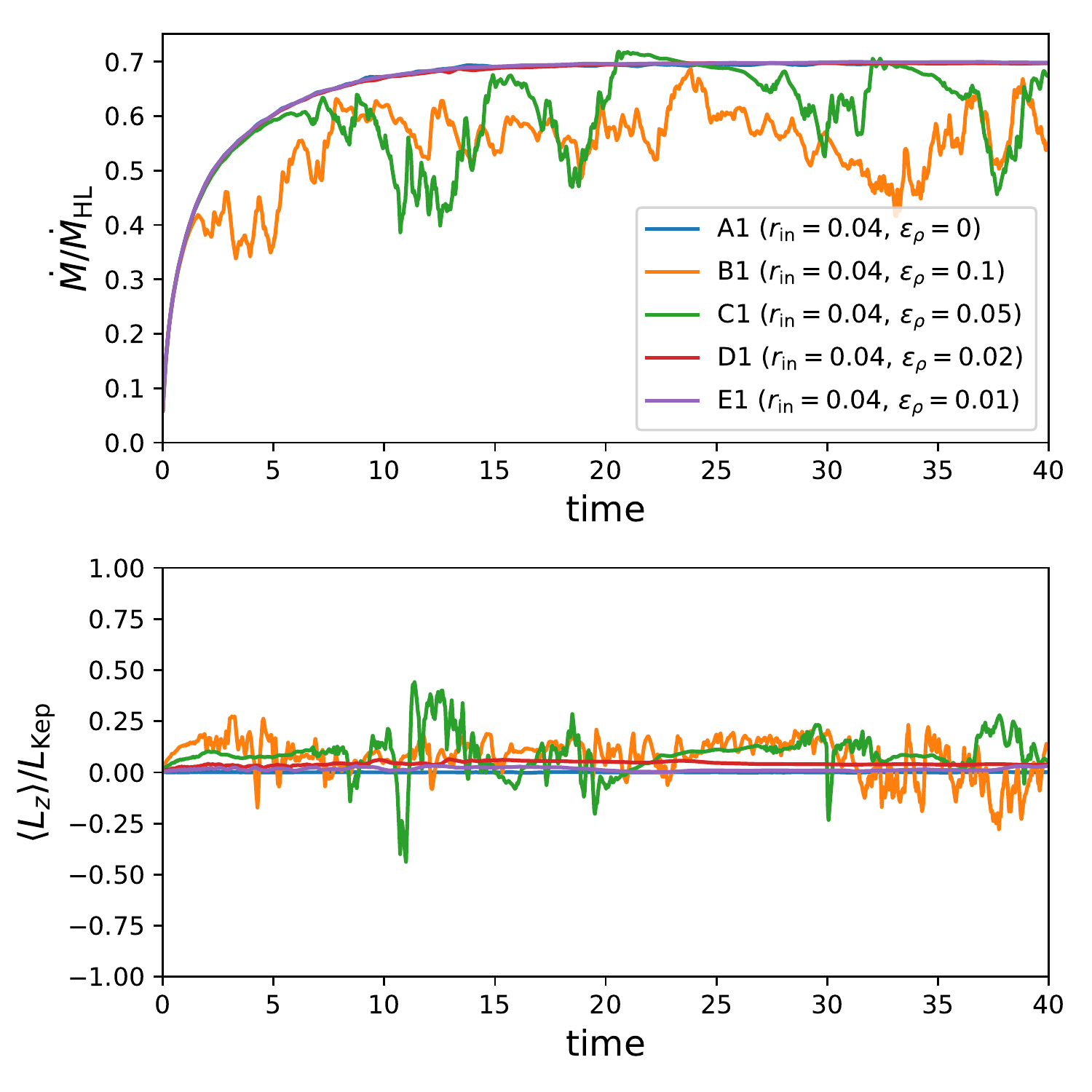}
\caption{Same as Figure \ref{fig:B}, but for simulations A1 - E1 (same $r_{\rm in}$, different $\epsilon_\rho$ values). The accretion rates for the stable simulations A1, D1, E1 are near identical.}
\label{fig:rho_grad}
\end{figure}

\subsubsection{Grid and resolution}
Since the flow is no longer axisymmetric, we need to switch to a 3D grid. A spherical-polar grid has the symmetry we desire, 
but to avoid problems near the pole, we adopt a nested cartesian grid.\footnote{A cartesian grid has the disadvantage that it does not conserve angular momentum exactly and cannot preserve axisymmetry; it is possible that the grid geometry affects the stability when there is no upstream gradient (see \S5.2). Meanwhile, the flow is turbulent for many of our 3D simulations, and failing to exactly conserve angular momentum is no longer problematic in this case.}
The domain has size $14R_a\times 8R_a\times 8R_a$, with range $[-10 R_a,4R_a]$ for $x$ and $[-4 R_a,4R_a]$ for $y,z$. Here the coordinate is defined such that the wind comes from $x\to \infty$, with initial velocity in $-\hat{\boldsymbol{x}}$ direction. This domain is large enough to ensure that no boundary can introduce unphysical feedback.

We also define a spherical-polar coordinate $(r,\theta,\phi)$ centered at the NS to facilitate our later discussions. This coordinate is oriented such that $\theta=0$ points to $+\hat{\boldsymbol{z}}$. Note that this orientation is different from that used in Section \ref{sec:axisym}.

Our nested grid has increased resolution at smaller $r$ so that the angular resolution remains roughly constant for $r\lesssim 1R_a$. The default root resolution is 8 cells per $R_a$, and we refine the mesh at smaller $r$ by multiple levels (each refinement level increases the resolution by a factor of two) such that $r/\delta$ (with $\delta$ being the cell size) is never below 10. This gives $10\lesssim r/\delta \lesssim 27$,
which is comparable to the lowest resolution for our 2D simulations ($r/\delta_r\approx 20,~1/\delta_\theta \approx 30$), and should be sufficient given the good convergence of our 2D simulations across all resolutions. The accretor is well resolved, with a diameter of $>20$ cells.
Our resolution is similar to that in \citet{MR15}, but lower than that of \citet{BR12}.

For a few simulations, we modify the domain size and resolution to test convergence, avoid grid effect or save computational cost, and such modifications will be individually introduced when discussing those simulations.

\subsubsection{Boundary conditions}
We use the $+x$ boundary as the upstream boundary, imposing a wind profile given by ballistic trajectories from infinity. This is also the default initial condition of the simulation.
The downstream boundary includes all other outer boundaries, and for them we use a free flow boundary condition.\footnote{Inflow may still occur on these ``downstream" boundaries, but the domain is large enough so that such inflow never affects the flow near the accretor.}
For the inner boundary, we use an absorbing boundary condition. Physically, this represents accretion with no feedback on the surrounding flow. We have also shown in our 2D simulations (see Section \ref{subsec:compare_bc}) that changing the boundary condition from absorbing to outflow barely affects the result.

To avoid physically unstable modes being suppressed by the symmetry of the grid, we introduce a random initial perturbation in each cell with amplitude $\delta\rho/\rho=10^{-4}$. We choose this value so that this random perturbation is always much smaller than the effect of upstream density gradient (with the smallest gradient we use being $\epsilon_\rho=0.01$). Here we do not attempt to use this random density perturbation to model any physical perturbation (e.g. clumps) in the wind; we discuss the effect of a perturbed clumpy wind in Section \ref{subsec:clumpy_wind}.

\begin{figure*}
\centering
\includegraphics[width=.8\textwidth]{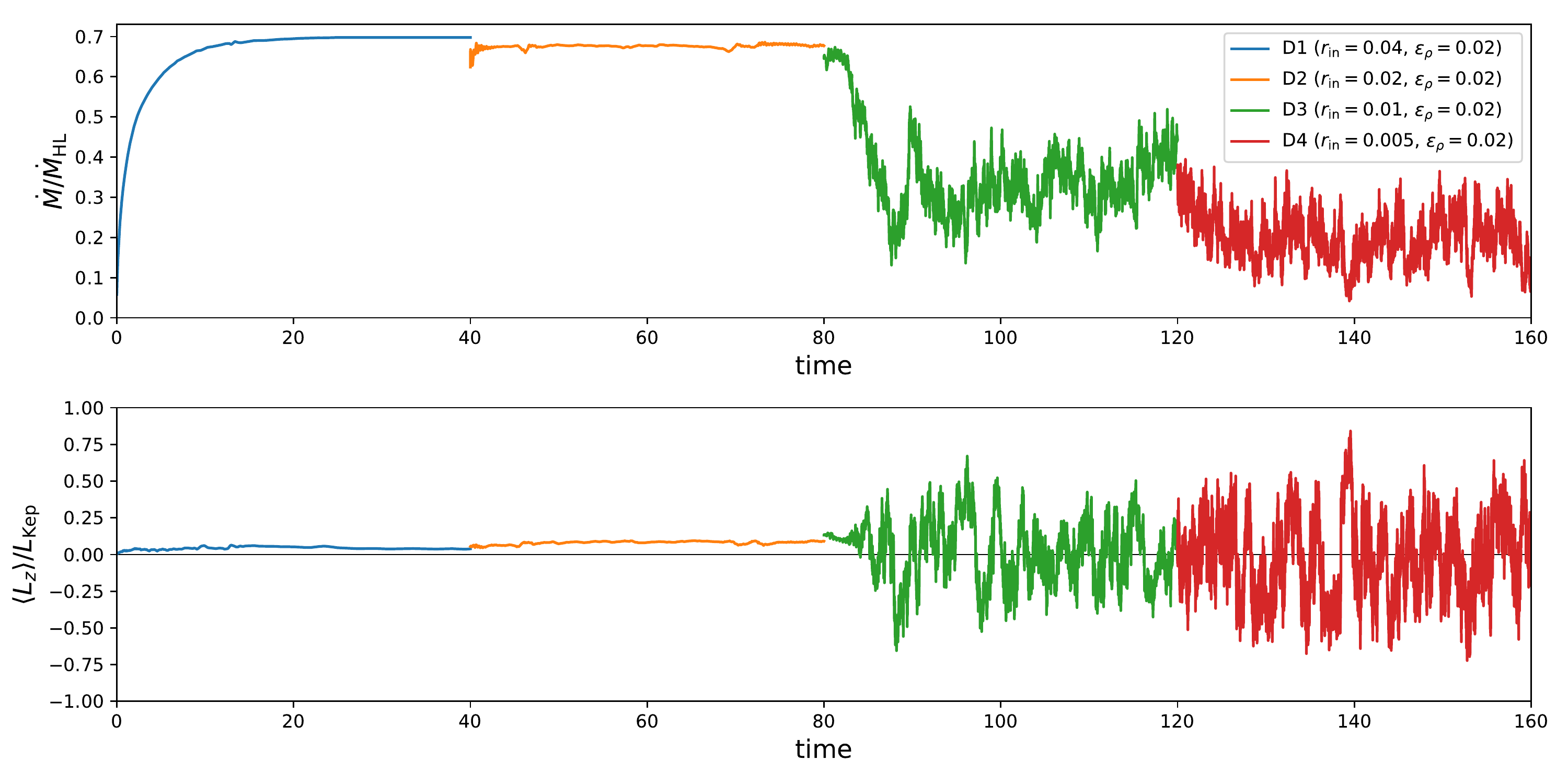}
\caption{Same as Figure \ref{fig:B}, but for simulations D1 - D4.}
\label{fig:D}
\end{figure*}

\begin{figure*}
\centering
\includegraphics[width=.8\textwidth]{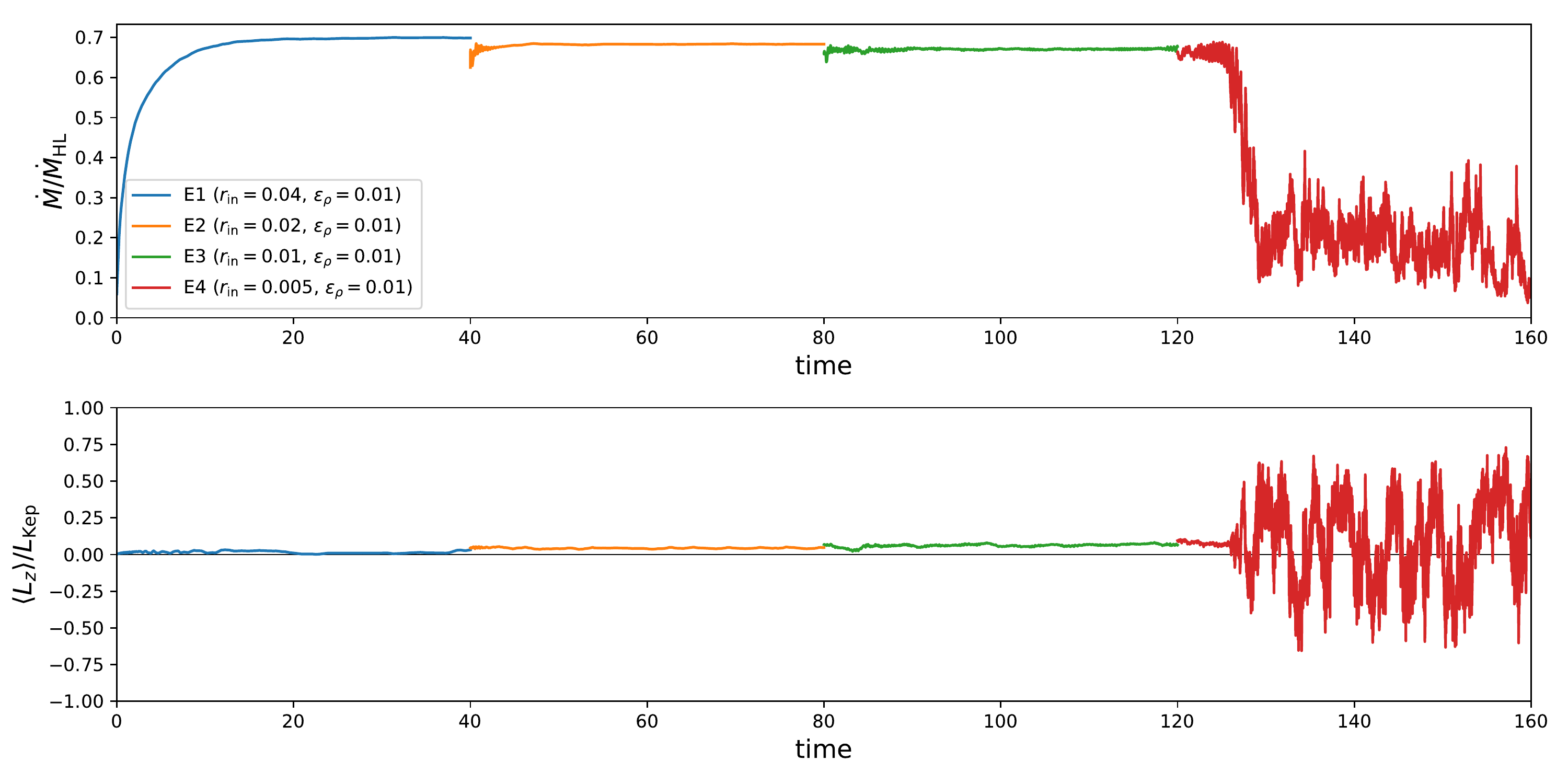}
\caption{Same as Figure \ref{fig:B}, but for simulations E1 - E4.}
\label{fig:E}
\end{figure*}

\subsection{Results: no transverse upstream gradient}\label{subsec:3D_sym}

For accretion with no transverse upstream gradients ($\epsilon_\rho,\epsilon_v=0$), the results of 3D simulations are largely similar to that of axisymmetric 2D simulations. A snapshot of the flow for simulation AS3 is shown in Figure \ref{fig:A3}. The flow is overall stable (with some minor perturbation due to finite grid spacing at the shock), and the accretion rate is approximately constant.
This is also the case for larger accretor size (AS1, AS2).
The flow does not show any vortex, which is reasonable given the relatively low resolution (compared to our 2D simulations).

As shown in the right panels in Figure \ref{fig:A3}, the flow appears largely axisymmetric, but there are visible artifacts due to the cartesian gird geometry, especially for the inner sonic surface (the black contour attached to the accretor in Figure \ref{fig:A3}).
To better understand how well our cartesian grid conserves angular momentum and preserves axisymmetry, we compute $v_\varphi/v_{\rm ff}$ where $v_\varphi$ is the non-axisymmetric component of the flow velocity and $v_{\rm ff}$ is the free-fall velocity, which is the typical velocity scale for flow near the accretor. We find $|v_\varphi|/v_{\rm ff}\lesssim 0.05$ for the flow behind the shock. Thanks to the approximately uniform angular resolution ensured by mesh refinement, this error does not further increase at smaller $r$.
One caveat is that the deviation from exact axisymmetry due to grid geometry, albeit small, may nontrivially affect the stability of a laminar, near-axisymmetric flow;
for example, the distorted inner sonic surface may affect the stability of acoustic modes in the subsonic region.
Nevertheless, when the result is less sensitive to small deviation from symmetry (e.g. when the flow is turbulent, as is the case for many simulations below), the cartesian grid should still produce reliable results.

Our result is in agreement with the $\epsilon_\rho=0$ simulations of \citet{MR15}, which is at a lower Mach number ($\mathcal M=2$).
It is also consistent with \citet{BR12} in that the flow remains axisymmetric, although we do not reproduce the breathing mode they observe for $r_{\rm in}=0.01$, probably due to our different grid geometry or lower resolution.
However, our result is very different from that of \citet{RuffertArnett1994}. Their simulations, despite having lower resolution and larger accretor size (both tend to make the flow more stable), show unstable flows.
This is likely because they use a much larger initial perturbation ($\delta\rho/\rho$ = 0.03) to break the symmetry.
We try to reproduce their results by running our simulation with their resolution and $r_{\rm in}$, and find that the flow is unstable for $\delta\rho/\rho = 0.03$ (and goes back to stable after a few flow crossing time, since such perturbation is only in the initial condition) and stable for $\delta\rho/\rho = 10^{-4}$.
Physically, this suggests that the accretion flow can become unstable when the upstream wind contains random perturbation at small length scale (e.g. small clumps in the wind) with sufficiently large amplitude.

\subsection{Results: small transverse upstream gradient}\label{subsec:small_grad}

For small transverse upstream gradient, we focus on the $r_{\rm in}$ dependence of stability, and how the strength of $\epsilon_\rho$ affects the $r_{\rm in}$ dependence. We also investigate whether increasing the Mach number and replacing the density gradient by a velocity gradient affect the result.
In this subsection, we focus on summarizing the simulation results; discussion of relevant physical mechanisms will be given in the next subsection.

\subsubsection{$r_{\rm in}$ dependence}

In simulation B1 - B4, we fix the upstream gradient at $\epsilon_\rho=0.1$ and vary $r_{\rm in}$ from 0.04 to 0.005 to investigate the $r_{\rm in}$ dependence of accretion.
The mass accretion rate $\dot M$ and mean (averaged spatially over the inner boundary but not temporally) specific angular momentum in $z$ direction of accreted material $\langle L_z\rangle$ are given in Figure \ref{fig:B}.
For all four simulations, the accretion flow is unstable and without visible periodicity.
(There is also no visible peak in the power spectra of $\dot M$ or $\langle L_z\rangle$.)
Figure \ref{fig:B_hist} shows the distribution of $\dot M$ and  $\langle L_z\rangle/L_{\rm Kep}$, where $L_{\rm Kep}\equiv\sqrt{GM_{\rm NS}r_{\rm in}}$ is the Keplerian specific angular momentum at $r_{\rm in}$.
As $r_{\rm in}$ decreases, $\dot M$ decreases and the distributions of $\dot M$ and $\langle L_z\rangle/L_{\rm Kep}$ both widen, suggesting an increase of instability. In addition, the centroid of $\langle L_z\rangle/L_{\rm Kep}$ becomes closer to zero, so the NS accretes less angular momentum.
These behaviors are in broad agreement with the observation of \citet{MR15} that the flow is more unstable for smaller $r_{\rm in}$.

Snapshots of the accretion flow, shown in Figure \ref{fig:B_flow}, confirm the trend of increasing instability for decreasing $r_{\rm in}$.
For B1 (largest accretor), the flow is only weakly unstable, and is often near-laminar.\footnote{Throughout this paper, ``unstable" refers to any flow that is not in a laminar steady state.
However, in some literature, weakly unstable flows like that in B1 would be considered as stable.}
For B2 - B4, the flow is turbulent everywhere near the accretor. The amount of turbulence kinetic energy increases as $r_{\rm in}$ decreases, as suggested by the overall increase of distance between the shock and the accretor. The turbulent nature of the flow explains the randomness of $\langle L_z\rangle$ and the lack of periodicity when $r_{\rm in}$ is small.

\subsubsection{$\epsilon_\rho$ dependence}

\begin{figure}
\centering
\includegraphics[width=.5\textwidth]{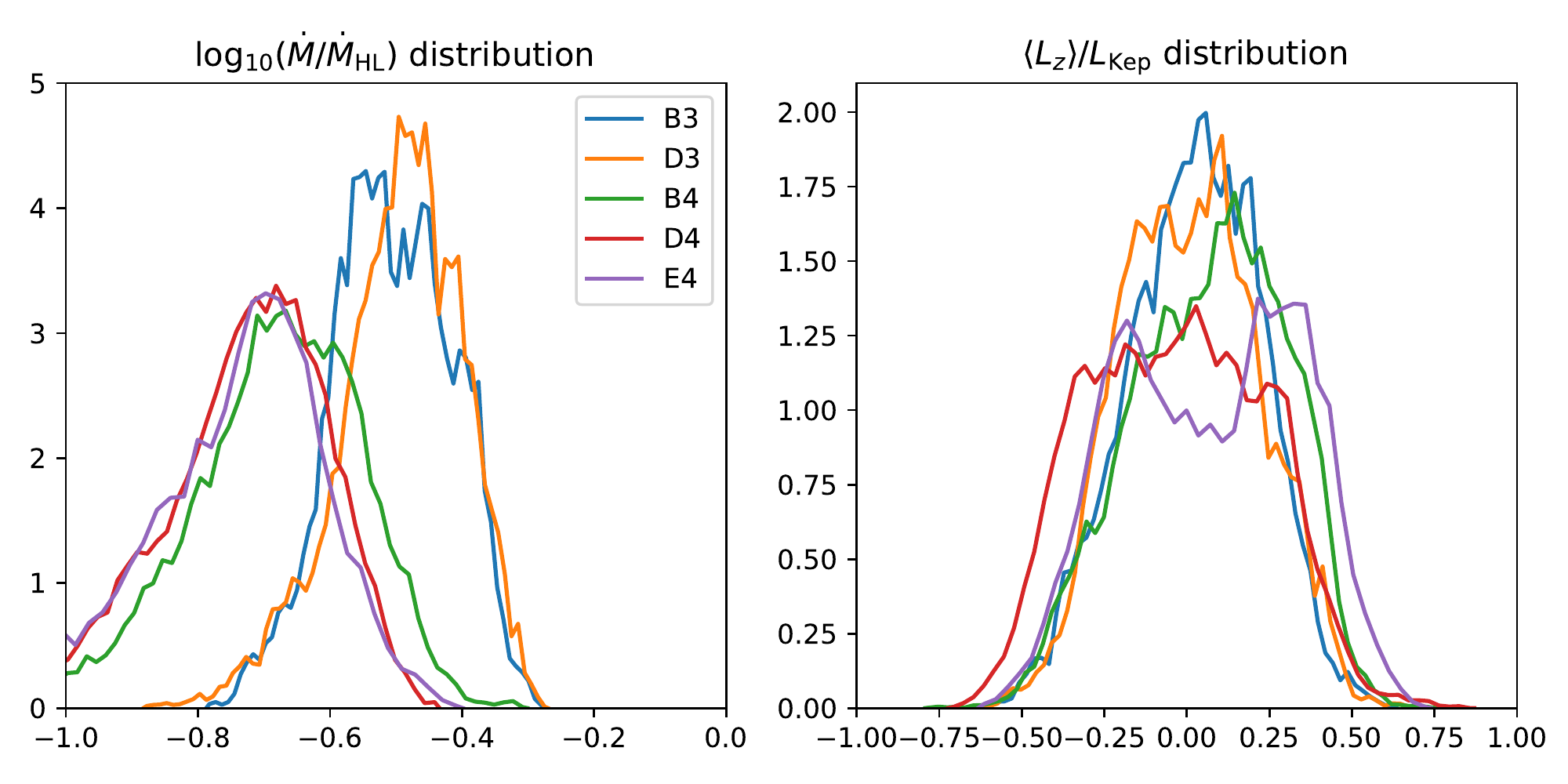}
\caption{Same as Figure \ref{fig:B_hist}, but for unstable simulations at $r_{\rm in} = 0.01$ (B3, D3) and $r_{\rm in}=0.005$ (B4, D4, E4). Distributions of $\dot M$ is sensitive to $\epsilon_\rho$, but show little dependence on $r_{\rm in}$ at given $\epsilon_\rho$.}
\label{fig:BDE_hist}
\end{figure}

In addition to simulations B1 - B4 in which we fix $\epsilon_\rho$ and vary $r_{\rm in}$, we also perform simulations at different $\epsilon_\rho$ (C1 - E4) to illustrate how the stability of the flow depends on $\epsilon_\rho$.

For all $\epsilon_\rho$ we used, the flow becomes unstable at sufficiently small $r_{\rm in}$, and the flow is in general more prone to instability at larger $\epsilon_\rho$.
In Figure \ref{fig:rho_grad}, we compare simulations B1 - E1, which have the same $r_{\rm in}=0.04$ and decreasing $\epsilon$. 
For B1 and C1 ($\epsilon_\rho = 0.1$ and 0.05), the flow is unstable, with C1 having $\dot M$ fluctuation with smaller amplitude; for D1 and E1 ($\epsilon_\rho=0.02$ and 0.01), the flow is stable, with accretion rate similar to the zero transverse gradient case (AS1).
This shows that at given $r_{\rm in}$, the flow is more stable for smaller $\epsilon_\rho$.
We also find that
the value of $r_{\rm in}$ below which the flow becomes unstable decreases as $\epsilon_\rho$ decreases.
For $\epsilon_\rho=0.1$ and 0.05, the flow is already unstable at $r_{\rm in}=0.04$; for $\epsilon_\rho=0.02$ and 0.01 (Figures \ref{fig:D} and \ref{fig:E}), the flow becomes unstable at $r_{\rm in}=$0.01 and 0.005 respectively.
We will further discuss the criterion of instability (which depends on both $\epsilon$ and $r_{\rm in}$) in Section \ref{subsec:instability_criterion}.

Another interesting feature is that for sufficiently small $r_{\rm in}$, the distribution of $\dot M$ and $\langle L_z\rangle$, although sensitive to $r_{\rm in}$, show little (if any) dependence on $\epsilon_\rho$, as shown in the comparison in Figure \ref{fig:BDE_hist}.
This result appears somewhat unexpected, but it is reasonable given that for small $r_{\rm in}$
the flow near the accretor is highly turbulent and should
not be sensitive to small gradients in the upstream flow.
Note that this may no longer be the case when $\epsilon_\rho$ is large; in that case the accretion flow contains a large angular momentum which can force the formation of disk-like structures even when the flow is turbulent. An example of this is simulation OAO, which we discuss in Section \ref{subsec:OAO}.

\subsubsection{Accretion rate scaling}
\begin{figure}
\centering
\includegraphics[width=.45\textwidth]{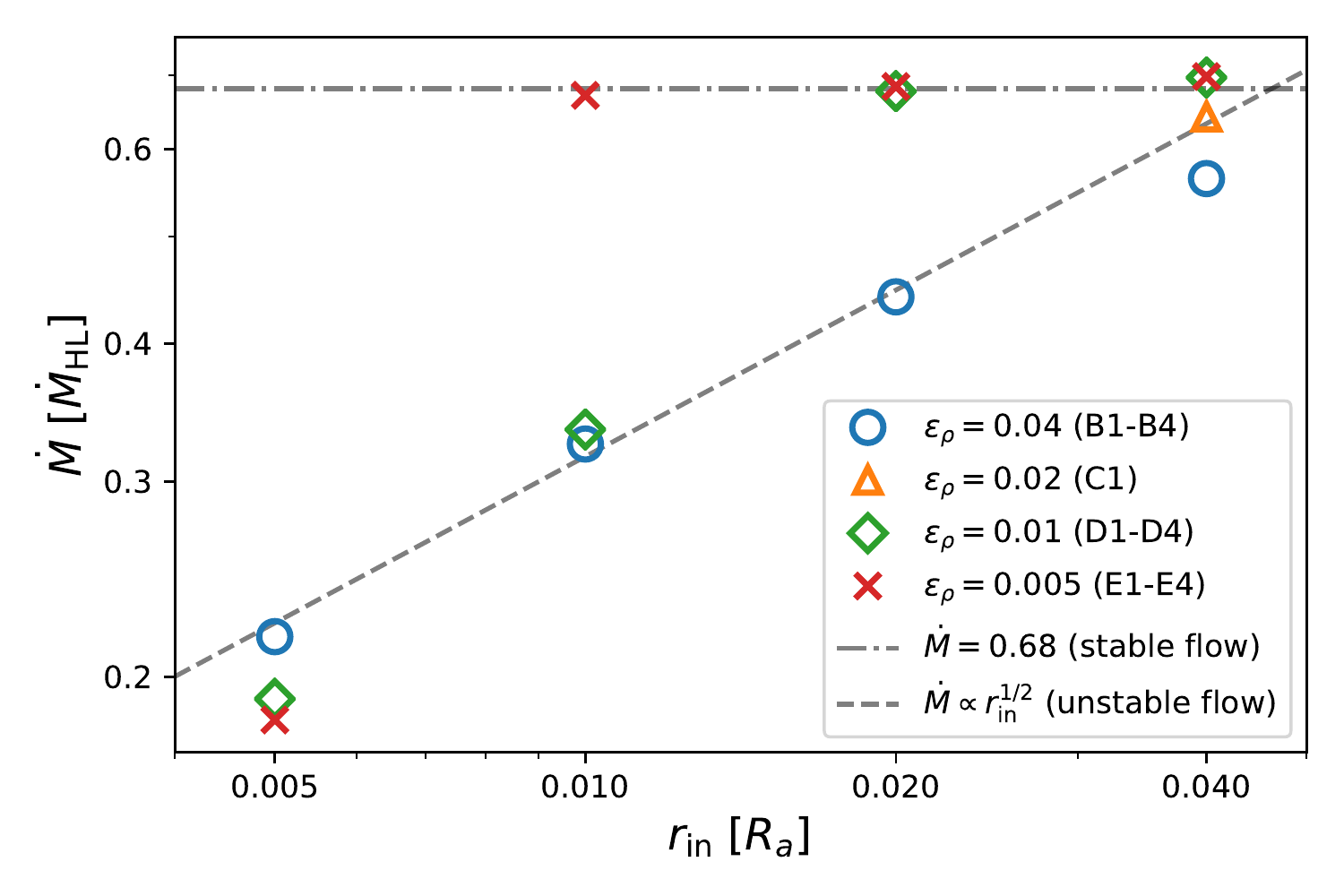}
\caption{Mean accretion rate of simulations with finite $\epsilon_\rho$. Stable simulations (D1-D2, E1-E3) all have $\dot M\approx 0.68\dot M_{\rm HL}$. The accretion rate of unstable simulations depends mainly on $r_{\rm in}$ and barely on $\epsilon_\rho$, with approximately $\dot M\propto r_{\rm in}^{1/2}$.}
\label{fig:Mdot_simulations}
\end{figure}
Figure \ref{fig:Mdot_simulations} shows the mean accretion rate of simulations with finite $\epsilon_\rho$. When the flow is stable (D1-D2, E1-E3), the accretion rate is independent of $r_{\rm in}$ or $\epsilon_\rho$, and $\dot M\approx 0.68\dot M_{\rm HL}$. When the flow is unstable, $\dot M$ decreases as $r_{\rm in}$ decreases, with little dependence on $\epsilon_\rho$. The data show an approximate power-law relation $\dot M\propto r_{\rm in}^{1/2}$.
However, the physical origin of this scaling remains unclear,\footnote{A similar $\dot M\propto r_{\rm in}^{1/2}$ scaling has been observed (and explained) in the simulations of wind accretion onto Sgr A* by \citet{Ressler2018}. However, their explanation does not directly apply to our problem.}
and we do not recommend extrapolating this relation to predict the accretion rates of real systems since the quality of the fit is not very good, especially at small $r_{\rm in}$.

\subsubsection{Effect of higher Mach number and velocity gradient}
\begin{figure}
\centering
\includegraphics[width=.5\textwidth]{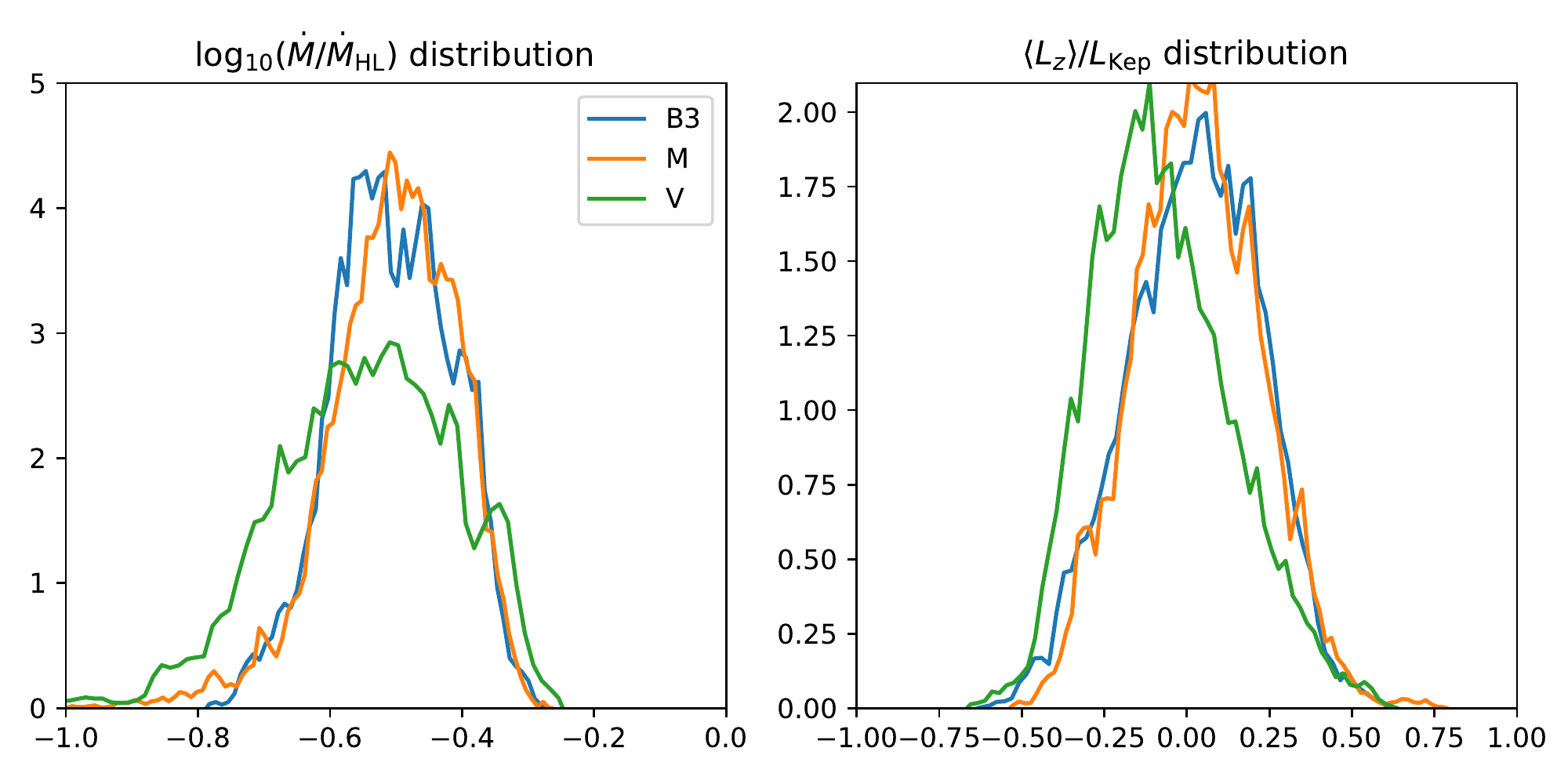}
\caption{Same as Figure \ref{fig:B_hist}, but for B3, M (higher Mach number $\mathcal M=30$) and V (finite velocity gradient $\epsilon_v=0.1$ but no density gradient). Increasing the Mach number barely affects the distributions of $\dot M$ and $\langle L_z\rangle$. At the same magnitude, velocity gradient makes the flow slightly more unstable than density gradient.}
\label{fig:MV_hist}
\end{figure}
So far, we only discussed the behavior at a fixed Mach number ($\mathcal M=10$) for finite upstream density gradient ($\epsilon_\rho\neq 0,~\epsilon_v=0$). Here we use two examples (simulations M and V) to briefly discuss the effect of higher Mach number and nonzero velocity gradient.

The distributions of $\dot M$ and $\langle L_z\rangle$ for B3, M and V are shown in Figure \ref{fig:MV_hist}. For simulation M, parameters are identical to B3 except $\mathcal M$ is increased to 30. Increasing the Mach number barely affects the behavior of the flow near the accretor and the distributions of $\dot M$ and $\langle L_z\rangle$.
Therefore, our simulations at $\mathcal M=10$ should be applicable to real SgXB systems, although most of them have $\mathcal M>10$ (see Table \ref{tab:parameters2}).

For simulation V, parameters are identical to B3 except density gradient $\epsilon_\rho=0.1$ is changed to velocity gradient $\epsilon_v=0.1$. The result for V is qualitatively similar to B3, but the distribution of $\dot M$ is slightly wider, suggesting more instability. 
A perhaps more interesting result is that the distribution of $L_z$ is now clearly centered at a negative value, unlike the case for finite density gradient.
This is because the flow from $y>0$, with more velocity, requires a smaller impact parameter to be accreted (note that $R_a\propto v_\infty^{-2}$). Although the accreted material from $y>0$ contains more specific angular momentum, more mass is accreted from $y<0$. When the latter effect overpowers the former, $\langle L_z\rangle$ tends to be negative.\footnote{This analysis assumes a laminar accretion flow, and it is not obvious {\em a priori} whether the same argument holds when the accretion flow is highly turbulent.}
Our observation of a preferentially negative $\langle L_z\rangle$ is consistent with previous works with finite $\epsilon_v$, such as \citet{RA95} and \citet{R97}.

\begin{figure*}
\centering
\includegraphics[width=\textwidth]{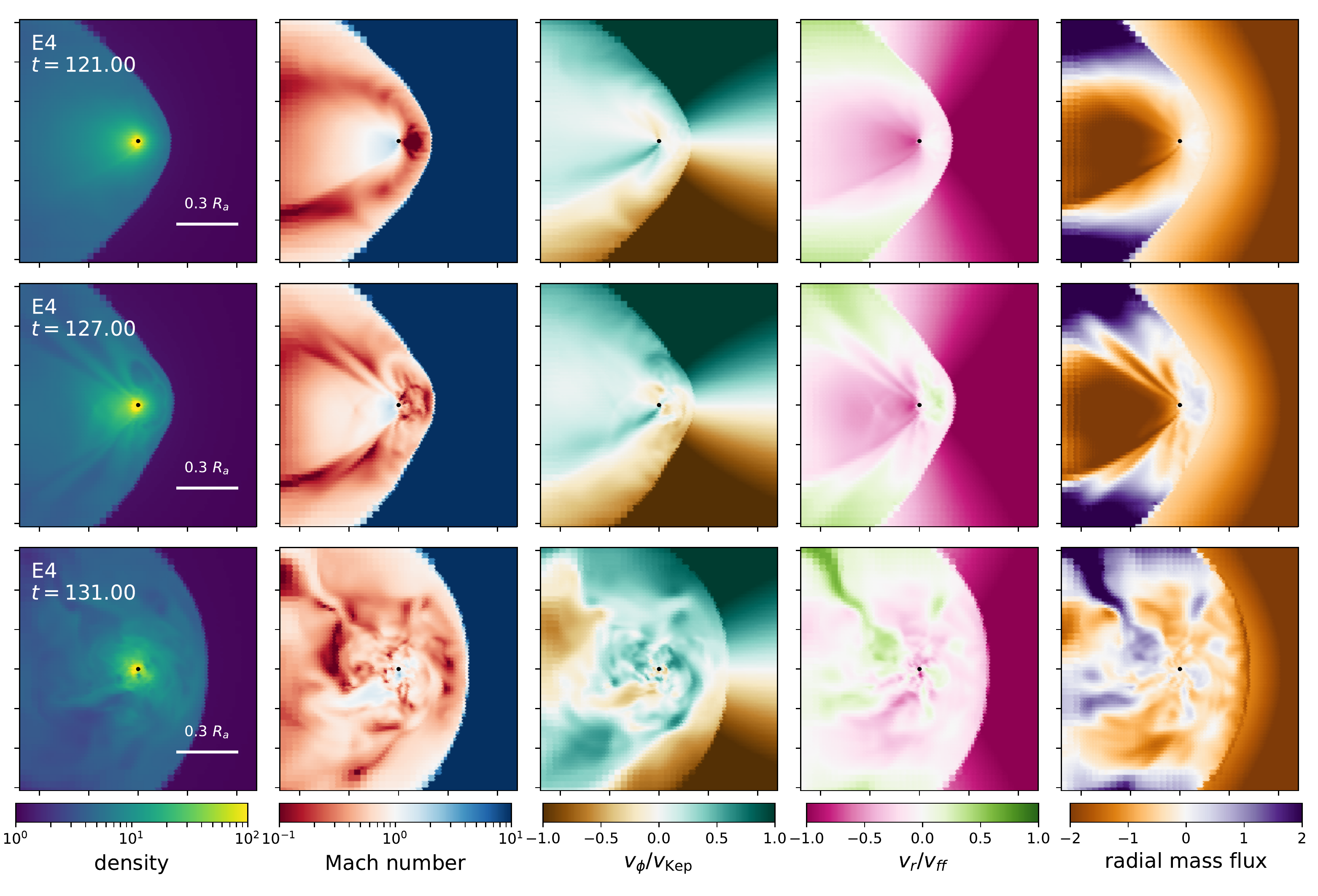}
\caption{Simular to Figure \ref{fig:B_flow}, but for simulation E4 at different epochs, showing development of instability. The simulation starts at $t=120t_a$. At $t=121t_a$ (first row), flow is still overall stable; at $t=127t_a$ (second row), the flow in front of the accretor is already turbulent due to the flow behind the accretor (with $L_z\gtrsim L_{\rm Kep}$) overshooting the accretor; at $t=131t_a$ (third row), the turbulent region expands and the flow is turbulent everywhere near the accretor.}
\label{fig:E4_flow}
\end{figure*}

\subsection{Physical explanation of flow morphology}\label{subsec:morphology}
\subsubsection{Flow morphology}
To study the physical origin of the instability, we first discuss the morphology of the flow for our simulations. Here we consider B1 - E4, which have identical parameters and setups except $r_{\rm in}$ and $\epsilon_\rho$.
Overall, the flow pattern observed in these simulations can be classified into three types: (a) Stable flow (D1, D2, E1 - E3), with negligible variation of $\dot M$ and $\langle L_z\rangle$. For stable flow, $\langle L_z\rangle$ always aligns with $\epsilon_\rho$ (i.e. they have the same sign). (b) Weakly unstable flow (B1, C1), characterized by relatively small variation of $\dot M$ and $\langle L_z\rangle$ (typically, $\dot M$ fluctuation amplitude is $\lesssim 20\%$), and $\langle L_z\rangle$ has a strong preference to align with $\epsilon_\rho$. In this case, the flow is in general unstable, but not always turbulent; especially, the flow behind the accretor is mostly laminar (e.g. see top panels in Figure \ref{fig:B_flow}). (c) Highly unstable and turbulent flow (B2 - B4, D3, D4, E4), characterized by an always turbulent flow near the accretor; variation of $\dot M$ and $\langle L_z\rangle$ are both large, and $\langle L_z\rangle$ no longer show a significant preference to align with $\epsilon_\rho$.

\begin{figure}
\centering
\includegraphics[width=.45\textwidth]{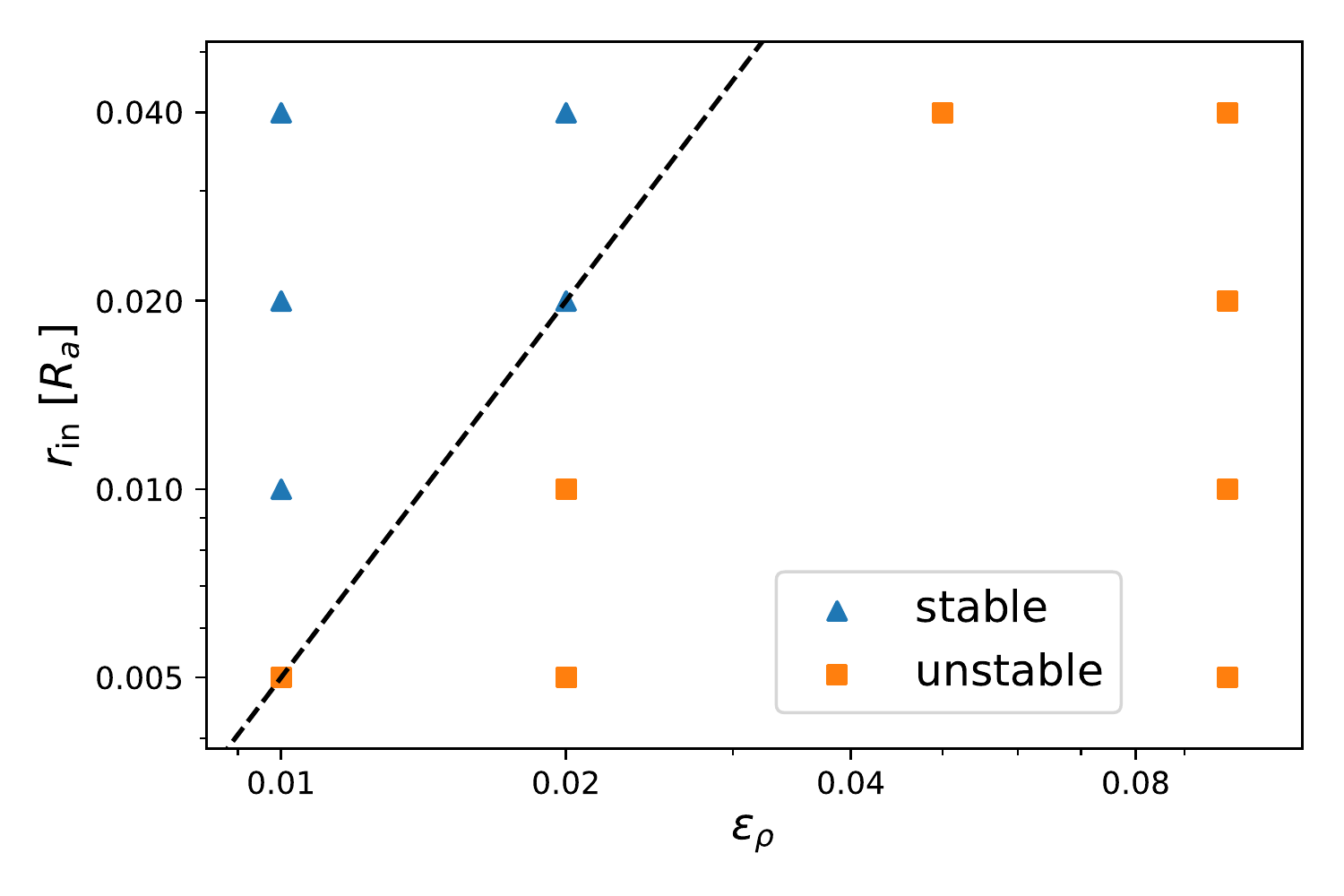}
\caption{Stability of simulations with finite $\epsilon_\rho$ (B1 - E4). The black dashed line marks the instability threshold $r_{\rm in}=50\epsilon_\rho^2$ [see Eq. \eqref{eq:rin_scaling_rho}], which agrees with all simulations. Two simulations (D2 and E4) lie on this threshold with different stability, tightly constraining the prefactor if we assume that the critical $r_{\rm in}$ is approximately $\propto \epsilon_\rho^2$.}
\label{fig:threshold}
\end{figure}

\begin{figure}
\centering
\includegraphics[width=.5\textwidth]{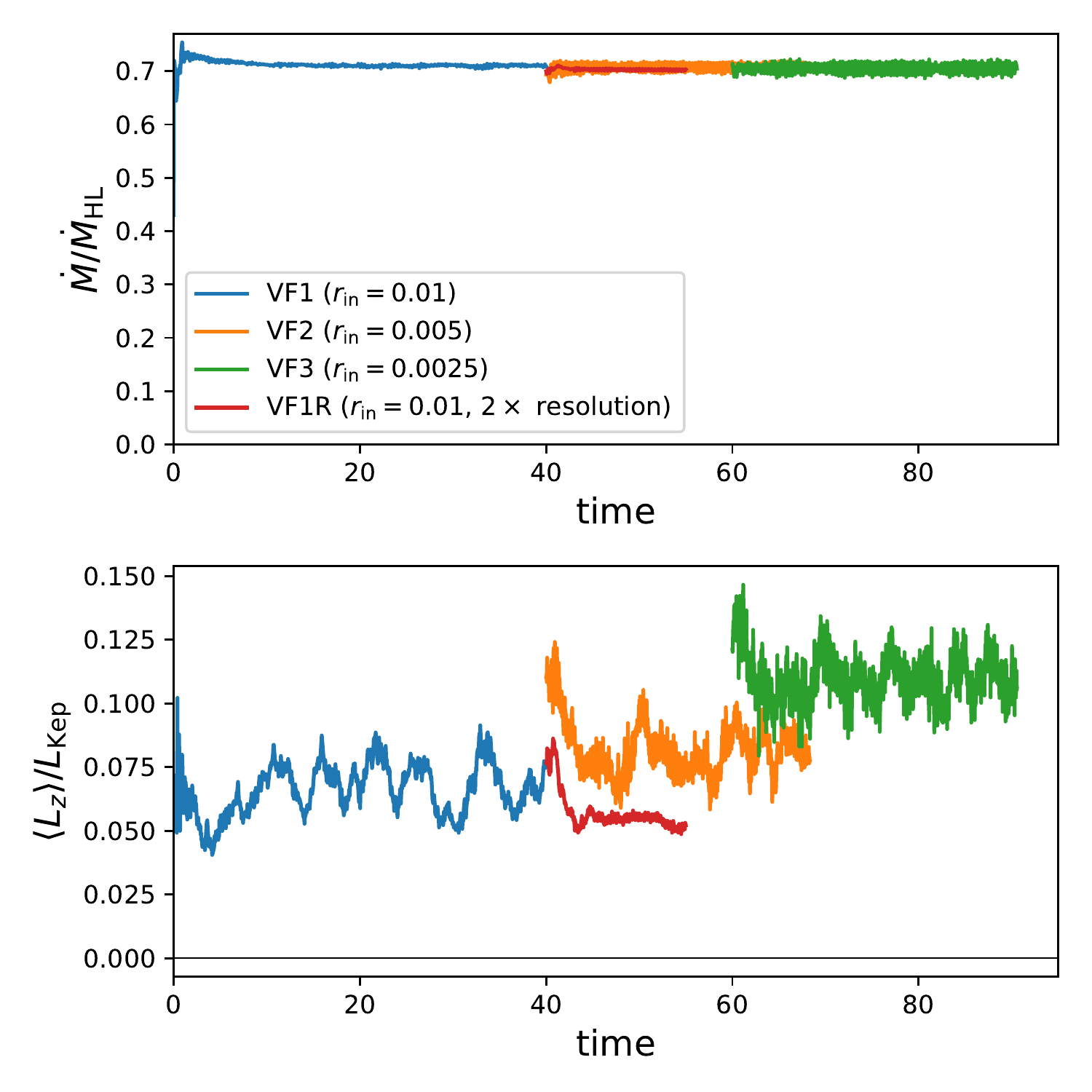}
\caption{Same as Figure \ref{fig:B}, but for simulations VELA1 - VELA3. VELA3 uses the flow of VELA2 at $t=60$ as the initial condition. The flow is always stable. $\langle L_z\rangle$ for the three simulations are nearly identical.}
\label{fig:VELA}
\end{figure}

\subsubsection{Mechanism of instability}\label{subsec:instability_mechanism}
The distinction between the three regimes can be analyzed by considering the angular momentum of the accretion flow.
Consider, for instance, $\epsilon_\rho>0,~\epsilon_v=0$.
Since the flow initially coming from $y>0$ has larger density, when the flow from $y>0$ and $y<0$ meet behind the accretor the resulting flow should have a positive $L_z$.
Of course, $L_z$ is not conserved as the flow goes towards the accretor, but it is reasonable to assume that there is no order-of-magnitude change in $L_z$ if the flow is laminar.\footnote{A weak spiral shock (hardly spiral due to the large pitch angle) attached to the accretor can appear (see top panels in Figure \ref{fig:B_flow}), leading to a reduction of $L_z$ as the flow crosses the shock. Still, this reduction does not affect the order-of-magnitude result.}
When $L_z\lesssim L_{\rm Kep}$ everywhere in the inflow near the accretor, all material can be directly accreted through a near-radial accretion flow, leading to a stable flow and a constant $\langle L_z\rangle$, which is $\propto \epsilon_\rho$ for flow with small density gradient.
This produces the case of stable accretion flow.

However, when the accretor size becomes smaller, $L_{\rm Kep}$ decreases and part of the flow cannot be directly accreted; instead, it tends to overshoot the accretor, with $v_\phi\gtrsim v_{\rm Kep}$. This can destabilize the flow by forming a strong velocity shear, since the $v_\phi$ of the incoming flow near the accretor is in general small. This is qualitatively similar to the case for 2D axisymmetric simulation where accretion flow that overshoots the accretor (due to unphysically produced perturbation in the flow instead of upstream gradient) creates vortices. However, the 3D geometry allows the flow to be more unstable: the flow in front of the accretor eventually becomes turbulent.

The turbulent region in front of the accretor tends to expand to the downstream side, and produces two opposing effects. First, the accretion flow that would miss the accretor due to its angular momentum excess can now dispose its angular momentum by colliding with the turbulent flow, which reduces overshooting and suppress instability. Second, when the turbulence is strong enough, it can significantly perturb the flow behind the accretor and increase its transverse velocity, which promotes overshooting and increases instability.

When the first effect is more prominent, the flow is only weakly unstable because
the instability tends to shut itself once it becomes strong enough, and the flow switch between turbulent and near-laminar, as is observed in B1 and C1 (the snapshot for B1 in Figure \ref{fig:B_flow} is in the near-laminar phase). Since the flow from behind the accretor remains less perturbed, the variation of $\dot M$ is not very significant. $\langle L_z\rangle$ preferentially aligns with $\epsilon_\rho$ because it does so when the flow is near-laminar.
Note that this weakly unstable behavior may be unique to cases when the accretor size is comparable to the distance between the accretor and the shock front, since we only observe this behavior when $r_{\rm in}=0.04$.

When the second effect is more prominent, the flow become highly unstable, since the instability tends to self-amplify until the flow becomes fully turbulent everywhere near the accretor. The turbulent nature of the flow near the accretor makes it barely sensitive to the upstream gradient, and $\langle L_z\rangle$ no longer show a clear preference to align with $\epsilon_\rho$. This also explains why the distribution of $\dot M$ and $\langle L_z\rangle$ in this regime shows dependence only on $r_{\rm in}$ but not on $\epsilon_\rho$.

To confirm our interpretation, we study the development of instability in D3 and E4. These are examples where the simulation starts with a near-laminar flow and eventually becomes unstable. The stable steady-state flow of another simulation with larger $r_{\rm in}$ (D2 and E3) is used as the initial condition, so the flow is laminar for the first few $t_a$ except very closer to the accretor. Consider the snapshots of E4 at different times shown in Figure \ref{fig:E4_flow}: At $t=121$ (the simulation starts at $t=120$), the flow is still overall stable, similar to the case when $r_{\rm in}$ is larger. At $t=127$, the instability has grown so that the flow in front of the accretor is turbulent (most visible in $\mathcal M$ and $v_\phi$ panels), but the flow behind the accretor is less affected (see mass flux panel). At $t=131$, as the instability continues to grow, the flow eventually becomes turbulent everywhere near the accretor. The case for D3 is largely similar.

\subsubsection{Threshold of instability}\label{subsec:instability_criterion}
Our physical interpretation of the mechanism of instability allows an estimate of the threshold of instability.
The flow should turn unstable when $L_z\gtrsim L_{\rm Kep}$. When flow from $y\sim \pm R_a$ meet behind the accretor, the specific angular momentum of the flow should be (in code unit, assuming only density gradient and $\epsilon_\rho\ll 1$)
\eq{
L_z \sim \epsilon_\rho.
}
Meanwhile, $L_{\rm Kep} = r_{\rm in}^{1/2}$. Therefore, the condition of instability should be
\eq{
r_{\rm in} \lesssim \epsilon_\rho^2.\label{eq:rin_scaling_rough}
}
We can calibrate this scaling using our simulations.
The stability of simulations with finite $\epsilon_\rho$ (B1 - E4) are summarized in Figure \ref{fig:threshold}. If we assume that the scaling of the threshold follows \eqref{eq:rin_scaling_rough}, D2 and E4 which have opposite stability but the same $r_{\rm in}/\epsilon_\rho^2$ must lie on the instability threshold, and the condition of instability needs to be
\eq{
r_{\rm in}\lesssim 0.005\left(\frac{\epsilon_\rho}{0.01}\right)^2 = 50~ \epsilon_\rho^2.\label{eq:rin_scaling_rho}
}
This result is consistent with all our simulations in Figure \ref{fig:threshold}.
Note that this result should only be applicable for high Mach number and small accretor, i.e. when $r_{\rm in}$ is much smaller than the distance between the shock and the accretor.

In general, for a flow with both density and velocity gradient, we expect
\eq{
r_{\rm in}\lesssim 50({\epsilon_\rho+\alpha\epsilon_v})^2,\label{eq:rin_scaling_rhov}
}
with $\alpha$ being some $\mathcal O(1)$ constant, and is likely negative since positive $\epsilon_v$ produce negative $\langle L_z\rangle$. $|\alpha|$ is likely $>1$, given the more unstable flow shown in simulation V. The instability threshold may also depend on upstream entropy gradient (which is taken to be 0 in our simulations), but the scaling should remain overall similar.

The factor of 50 in Eq. \eqref{eq:rin_scaling_rho} seems rather surprising, since our rough analytic argument and previous analytic estimates (see a review in \citealt{Ho1988}; note that these estimates are in general not rigorous enough, as Ho pointed out) would both predict this factor to be $\mathcal O(1)$ (i.e. unstable when $\langle L_z\rangle$ becomes comparable to $L_{\rm Kep}$). In other words, the flow is significantly more prone to instability as one might naively expect.
This is because instability only requires that the maximum $L_z$ (instead of the mean, $\langle L_z\rangle$) of the material going towards the accretor is greater than $L_{\rm Kep}$.
The maximum $L_z$ can be much greater than its mean value, and the latter is indeed $\propto \epsilon_\rho$ with a $\mathcal O(1)$ prefactor as we show in Section \ref{subsec:compare_others}.
One caveat here is that our analysis implicitly assumes the ratio between maximum and mean $L_z$ to be independent of $\epsilon_\rho$, and wether this assumption is appropriate is not obvious.

\begin{figure*}
\centering
\includegraphics[width=\textwidth]{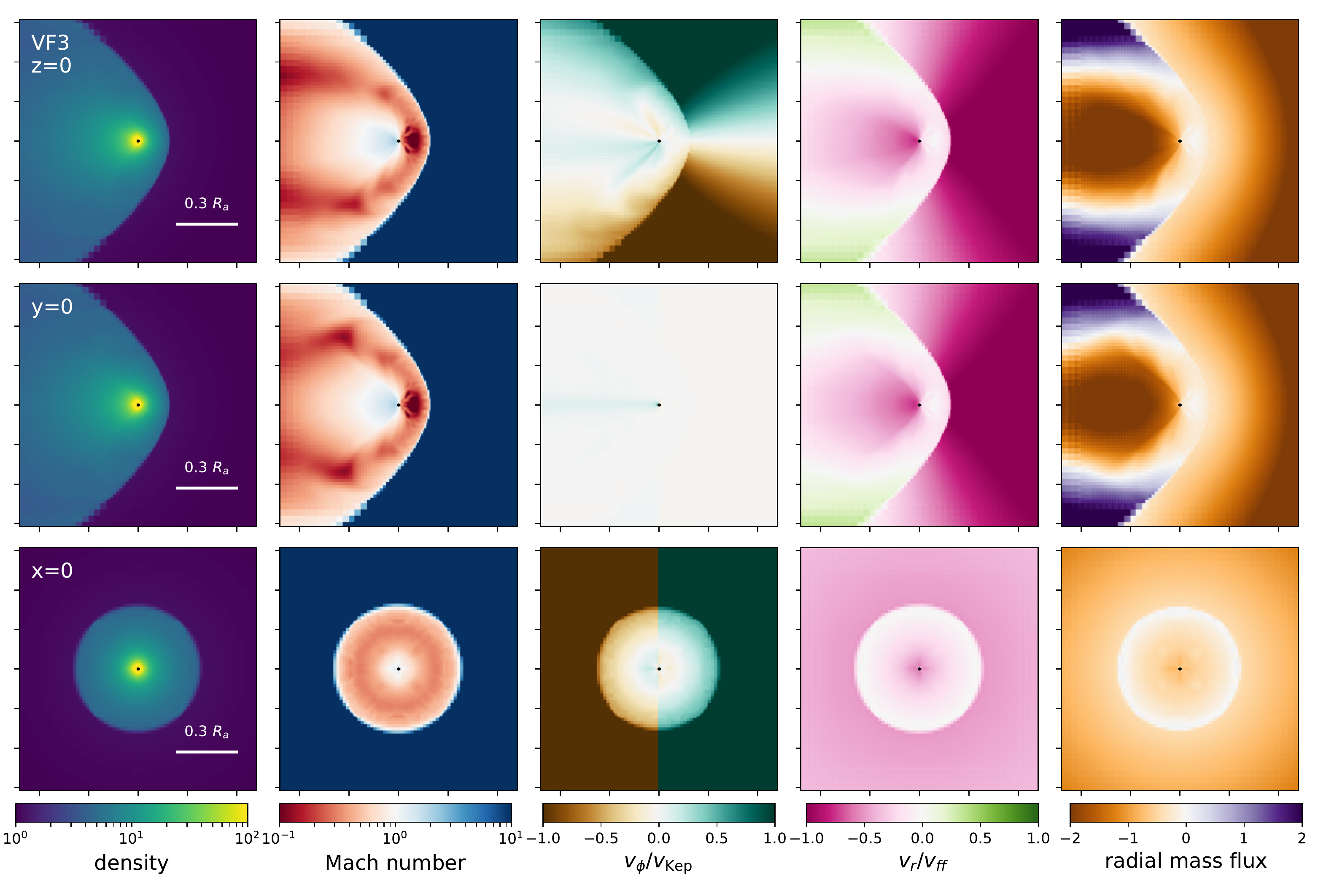}
\caption{A snapshot of accretion flow for VELA3, similar to Figure \ref{fig:B_flow}. The three rows show slices at $z=0,y=0$ and $x=0$ respectively. The flow is almost axisymmetric.}
\label{fig:VELA_flow}
\end{figure*}

\subsection{Difference between 2D and 3D BHL accretion: flip-flop instability and disk formation}\label{subsec:flip_flop}
The simulations in this section illustrate the key difference between 2D planar and 3D BHL accretion with zero or small upstream gradient.
 the flow exhibits a flip-flop instability, which leads to the overstable oscillation (or wobbling) of the shock cone behind the accretor even when there is no upstream gradient.
However, similar to previous studies \citep{BR12,MR15}, we never observe flip-flop instability in 3D:
The instability we observe at sufficiently large $\epsilon_\rho$ and sufficiently small $r_{\rm in}$ is due to a completely different mechanism (see \S\ref{subsec:instability_mechanism}), and the downstream flow beyond the stagnation point shows little transverse perturbation even when the flow is highly unstable.

As a consequence of the different instability mechanisms, 2D planar and 3D simulations also give different results regarding disk formation.
For 2D planar simulations, at sufficiently small $r_{\rm in}$ flip-flop instability can produce enough angular momentum to cause the spontaneous formation of a disk around the accretor, and such disk is stable for $\gamma=5/3$ \citep{Blondin2013}.
In 3D, however, this is no longer possible since the instability no longer involves transverse motion of the shock and cannot significantly increase the angular momentum of the accretion flow.
Instead, turbulence produced by instability can effectively reduce the flow's mean specific angular momentum, inhibiting disk formation.
Moreover, the high temperature of the flow near the accretor (partly due to turbulent heating) prevents the formation of a thin disk and makes any disk-like structure less stable.
Consistent with this argument, no persistent, rotationally-supported accretion disk has been observed in B1 - E4.
In some simulations, we do observe transient, turbulent disk-like structures around the accretor with $|v_\phi|\sim v_{\rm Kep}$ (e.g. in the snapshot for B3 in Figure \ref{fig:B_flow}). However, such disk-like structures are unstable and have a very short lifespan ($\lesssim$ a few $t_a$, which is $\lesssim$ several hours for real systems). They should not be described or studied using models for stable, rotationally-supported disks, and are probably unimportant for accretion.

It is worth noting that the angular momentum in the accretion flow is relatively small for all simulations in this section.
At sufficiently large $\epsilon_\rho$, the accretion flow can contain enough angular momentum to circulate before reaching the accretor; this may produce a persistent disk (or disk-like structure).
We will use one numerical example to explore this regime in Section \ref{subsec:OAO}, and discuss the criterion of disk formation based on analytic arguments in Section \ref{subsec:compare_others}.

\section{Application to real systems}\label{sec:real}
In this section, we use a more realistic wind model to investigate accretion in SgXB systems. We perform simulations with parameters resembling two systems in Table \ref{tab:parameters2} that can serve as limiting cases: Vela X-1 with fast wind and OAO 1657-415.

Vela X-1 with fast wind has been studied in several previous simulations (e.g. \citealt{Blondin1990,ManousakisWalter2015}), in accordance to early observations.
It has small upstream gradients and weak orbital effect ($R_a/R_H$ and $\Omega_bt_a$ are both small), and is representative of SgXBs with high wind speed (e.g. 4U 1907+097).
Although more recent observations favor a lower wind speed for Vela X-1, we adopt this potentially unrealistic fast wind to allow direct comparison with previous studies.

OAO 1657-415 has the largest upstream gradients among the systems in Table \ref{tab:parameters2}, and the orbital effect is non-negligible. In previous sections, we only considered the case of small or zero upstream gradient; the large upstream gradients ($\epsilon_\rho\sim \mathcal O(1)$) of OAO 1657-415 may produce qualitatively different behaviors. We will also compare our results with observations of the system, and discuss whether an accretion disk can be formed.

\subsection{Setup}
The setup is similar to simulations in the previous section, with the following differences.
We no longer ignore the orbital motion of the binary, and the simulations are performed in a frame rotating at the binary orbital frequency. We assume a circular binary orbit, so the location of the NS and the companion remain fixed in the rotating frame. We include the gravity of the companion and centrifugal and Coriolis force, as discussed in Section \ref{subsec:eqs}.

For the initial condition and the upstream boundary condition, we now use the single-star wind model described in Section \ref{subsec:wind_model}. To make our initial condition self-consistent, we introduce additional acceleration and heating (which depend only on location and remain constant in time) so that our single-star wind profile will be in steady state if the NS gravity is turned off.
Physically, such acceleration and heating correspond to line-driven acceleration of the wind.
We assume them to be constant for simplicity, and this should serve as a reasonable approximation in the upstream flow since the upstream flow shows no temporal variation and remains similar to the single-star wind.
Behind the shock, the effect of these additional acceleration and heating should be negligible (compared to, for instance, the effect of NS gravity).

The parameters of the simulations are chosen to agree with those of Vela X-1 with fast wind (simulations VF1 - VF3) and OAO 1657-415 (simulation OAO), except that we increase the temperature of the wind so that $\mathcal M=10$ at the NS. (This reduces computational cost by allowing larger time steps.)

The coordinate is still centered at the NS, and is oriented such that the single-star wind velocity is in $-\hat{\boldsymbol{x}}$ at the NS and the binary orbit lies on the $xy$ plane.
Simulations VF1 - VF3 have small gradient, and we use the same domain and resolution as in the previous section.
For simulation OAO, since the streamlines are significantly curved in the whole domain (due to orbital motion and companion spin), we double the root resolution to avoid grid effects in the upstream wind. The resolution near the accretor remains the same. We also change the domain to $[-6,2]\times[-4,4]\times[-2,2]$. Under this choice, the companion is partially inside the domain. We impose the single-star wind profile for $1.02R_c<R<1.25R_c$ to represent the surface of the companion.

\begin{figure}
\centering
\includegraphics[width=.5\textwidth]{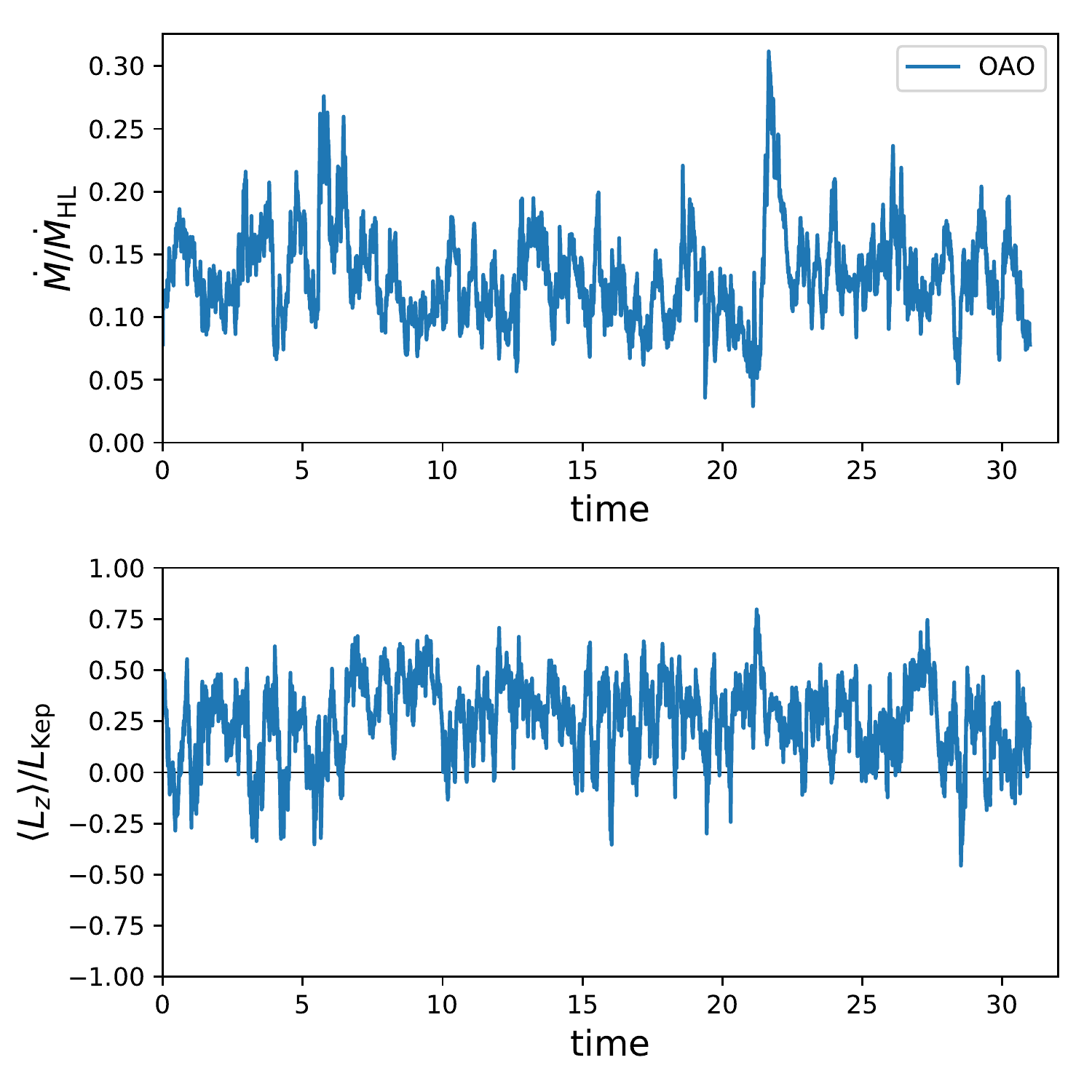}
\caption{Same as Figure \ref{fig:B}, but for simulation OAO.}
\label{fig:OAO}
\end{figure}

\begin{figure}
\centering
\includegraphics[width=.5\textwidth]{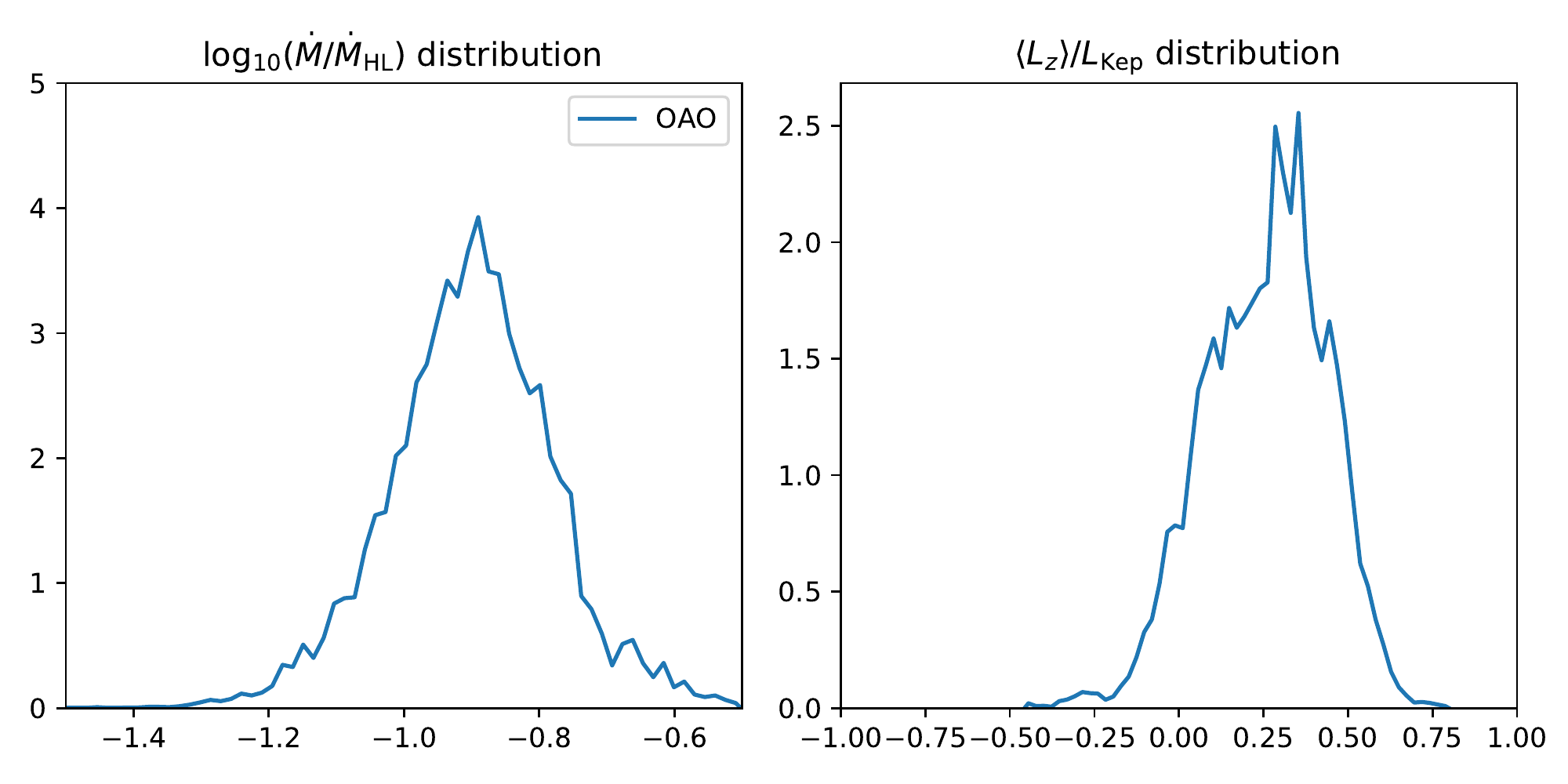}
\caption{Same as Figure \ref{fig:B_hist}, but for simulation OAO.}
\label{fig:OAO_hist}
\end{figure}

\begin{figure*}
\centering
\includegraphics[width=\textwidth]{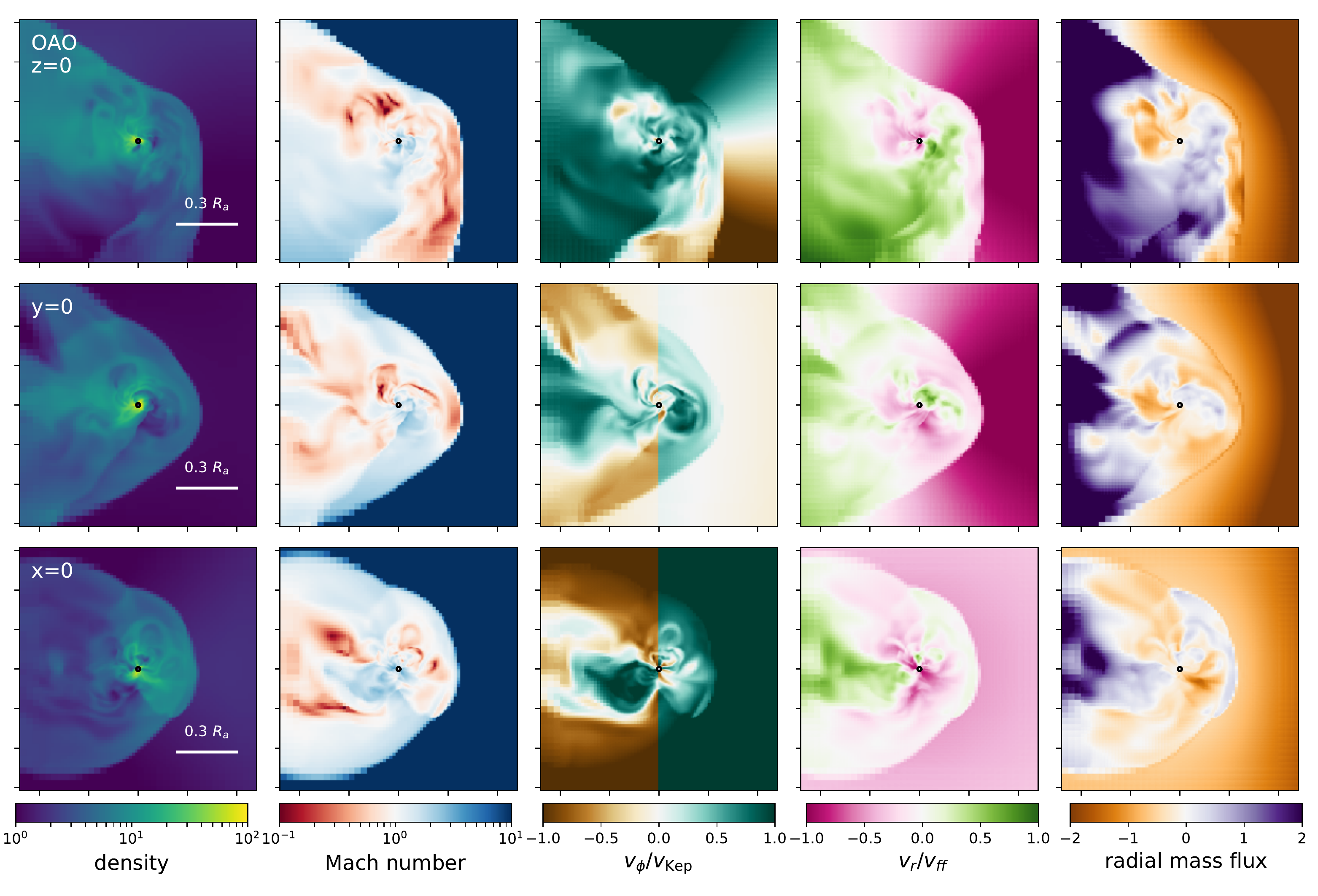}
\caption{Same as Figure \ref{fig:VELA_flow}, but for simulation OAO. The flow is highly turbulent, and shows a persistent disk-like structure due to the large upstream gradient.}
\label{fig:OAO_flow}
\end{figure*}

\subsection{Results: Vela X-1 with fast wind}\label{subsec:vela}

VF1 - VF3, which represent Vela X-1 with fast wind at $r_{\rm in}=0.01,0.005$ and 0.0025, all show stable accretion flow, as shown in Figure \ref{fig:VELA}.
The accretion rate remains approximately constant, and $\langle L_z\rangle$ undergoes some fluctuation.
To verify the origin of the $\langle L_z\rangle$ fluctuation and confirm that the scale of $\langle L_z\rangle$ is not affected by our relatively low resolution (the angular size of the cell, $\delta/r$, is larger than $\epsilon_\rho$), we continue VF1 with double resolution (simulation VF1R). In VF1R, fluctuation of $\langle L_z\rangle$ is significantly reduced, and the scale of $\langle L_z\rangle$ remains the same, suggesting that the fluctuation is only due to grid effect related to finite resolution.
A snapshot of VF3 is shown in Figure \ref{fig:VELA_flow}; the flow is nearly axisymmetric.

Comparing this result with simulations including only $\epsilon_\rho$ shows that that using a realistic wind profile increases stability. In E4, the flow becomes unstable at $r_{\rm in}=0.005$ for $\epsilon_\rho=0.01$, while here for similar $\epsilon_\rho$ the flow is still stable at $r_{\rm in}=0.0025$.
The reason of this increased stability is still unclear.
It should not be due to the inclusion of velocity gradient $\epsilon_v$, since 
a negative $\epsilon_v$ should further increase the angular momentum in the accretion flow, making it more unstable.
One possibility is that
stability is increased by the inclusion of orbital dynamical effects, especially the Coriolis force: The relative strength of Coriolis force (compared to NS gravity) at $r\sim R_a$ is $\sim \Omega_b t_a$, which is comparable to $\epsilon_\rho$. [Meanwhile, the relative strength of centrifugal force is $\sim(\Omega_b t_a)^2$, and companion gravity $\sim (R_a/R_H)^3$; both are much smaller.]

For Vela X-1 with fast wind, assuming $R_{\rm NS}=11$ km, $B_0\sim 10^{12}$ G and $L_x = 4\times 10^{36}{\rm ~erg/s}$ \citep{Walter2015} gives $R_{\rm mag}/R_a \sim 0.0065 > 0.0025$.
This result implies that the accretion flow for systems with very high wind speed is likely stable, if angular momentum transport is efficient within the magnetosphere and there is no other upstream perturbation (e.g. clumps in wind; see \ref{subsec:other_factors}).

Our result for Vela X-1 with fast wind is very different from the 2D planar simulations by \citet{Blondin1990} and \citet{ManousakisWalter2015}, both of which use similar stellar and wind parameters but find the flow (and the accretion rate) to be variable. 
This difference is mainly because these simulations include both the radiative acceleration of the wind and the suppression of it by NS X-ray photoionization.
\citet{ManousakisWalter2015} produce the observed quasi-periodicity and off-states of Vela X-1; the fact that this behavior is not produced in our 3D hydrodynamic simulations suggests that it is likely driven by radiative acceleration and X-ray photoionization feedback.
Additionally, the 2D planar geometry adopted in these simulations can produce strong flip-flop instability, which is not present in 3D (see \S \ref{subsec:flip_flop}).

\subsection{Results: OAO 1657-415}\label{subsec:OAO}

Simulation OAO lies in a very different regime of the parameter space: the upstream gradients are large ($\epsilon_\rho=0.44$), and companion gravity and orbital effects are very important ($R_a/R_H\sim 1$). It is worth noting that our results for OAO 1657-415 may not be directly generalized to other systems with large upstream gradient, since qualitative behavior of the flow may depend on the relative strength between different effects.

The $\dot M$ and $\langle L_z\rangle$ evolution, their distribution, and a snapshot of the flow are shown in Figure \ref{fig:OAO}, \ref{fig:OAO_hist}, \ref{fig:OAO_flow} respectively. The flow is highly asymmetric due to large curvature of the upstream wind and small separation between the NS and the companion ($R_a/D=0.41$).
It is also highly turbulent, with lower and more variable $\dot M$ (Figure \ref{fig:OAO_hist}) and stronger shocks around the accretor (see density panels in Figure \ref{fig:OAO_flow})
compared to simulations with smaller $\epsilon_\rho$, agreeing with the trend we observe in the previous section.

\begin{figure}
\centering
\includegraphics[width=.5\textwidth]{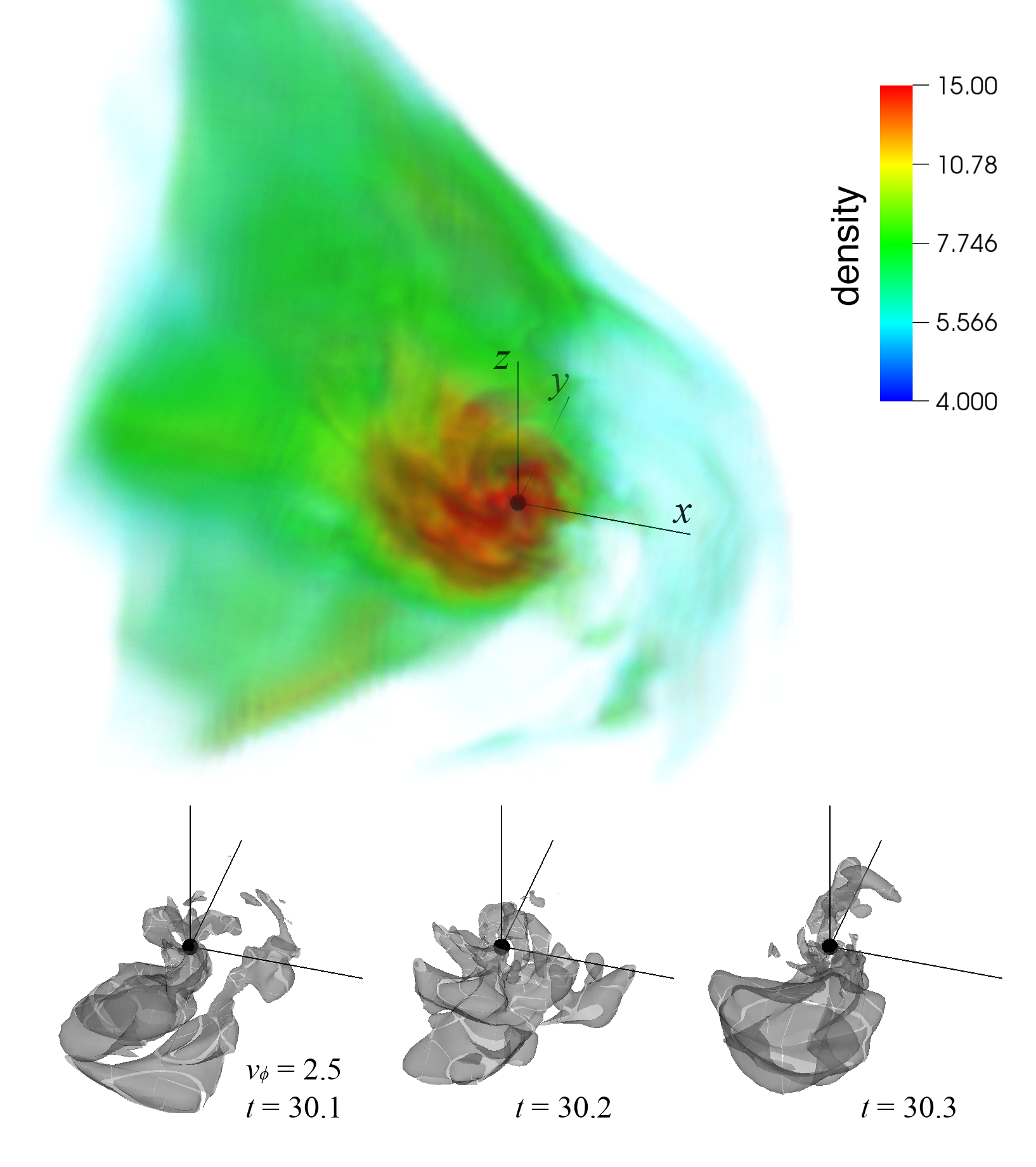}
\caption{\textbf{Top panel:} Volumetric density snapshot for simulation OAO, taken at $t=30.2t_a$. Overdensity corresponding to an inclined, turbulent disk-like structure is visible.
\textbf{Bottom panel:} Contours of $v_\phi$, taken at $t=30.1, 30.2$ and $30.3t_a$. The scale is the same as the top panel. The flow is structured but highly variable.}
\label{fig:OAO_volume}
\end{figure}

One distinctive feature of simulation OAO is that it exhibits a persistent disk-like structure around the accretor: the flow on the x-y plane near the accretor mostly have $v_\phi \sim 0.2$ - $1.5v_{\rm Kep}$ in Figure \ref{fig:OAO_flow}, and disk-like overdensity is visible in Figure \ref{fig:OAO_volume}. 
The NS is thus allowed to accrete angular momentum efficiently (Figure \ref{fig:OAO} and \ref{fig:OAO_hist}), perhaps explaining why OAO 1657-415 has significantly larger spin rate than other systems in Table \ref{tab:parameters1}.
This disk-like structure, however, is not rotationally supported due to the large variation of $v_\phi$. Instead,
it is highly turbulent, thick and variable (see bottom panel of Figure \ref{fig:OAO_volume}). Moreover, accretion does not happen mainly through this disk-like structure; a significant amount of accretion happens near the poles.

Our result is consistent with the observations that OAO 1657-415 undergoes periods of steady spin-up \citep{Jenke2012}\footnote{\citet{Jenke2012} also observe a mode where the NS spins down at a rate uncorrelated with the flux. This may correspond to occasional disruption of the disk-like structure (possibly due to physical effects that we do not include, such as radiative feedback from NS), which is not observed in our simulation.} and that its accretion rate is inconsistent with BHL-like (i.e. $\dot M\sim \dot M_{\rm HL}$, corresponding to a stable or weakly unstable flow) wind-fed accretion \citep{Taani2018}.
Although these observations are usually used to suggest the existence of an accretion disk, our result show that a turbulent disk-like structure (which is neither thin nor rotationally supported) is also consistent with the observations. Note that we do not rule out the possibility of disk formation; in principle, disk formation is still possible provided sufficient cooling (see \S\ref{subsec:compare_others}).

\begin{figure}
\centering
\includegraphics[width=.5\textwidth]{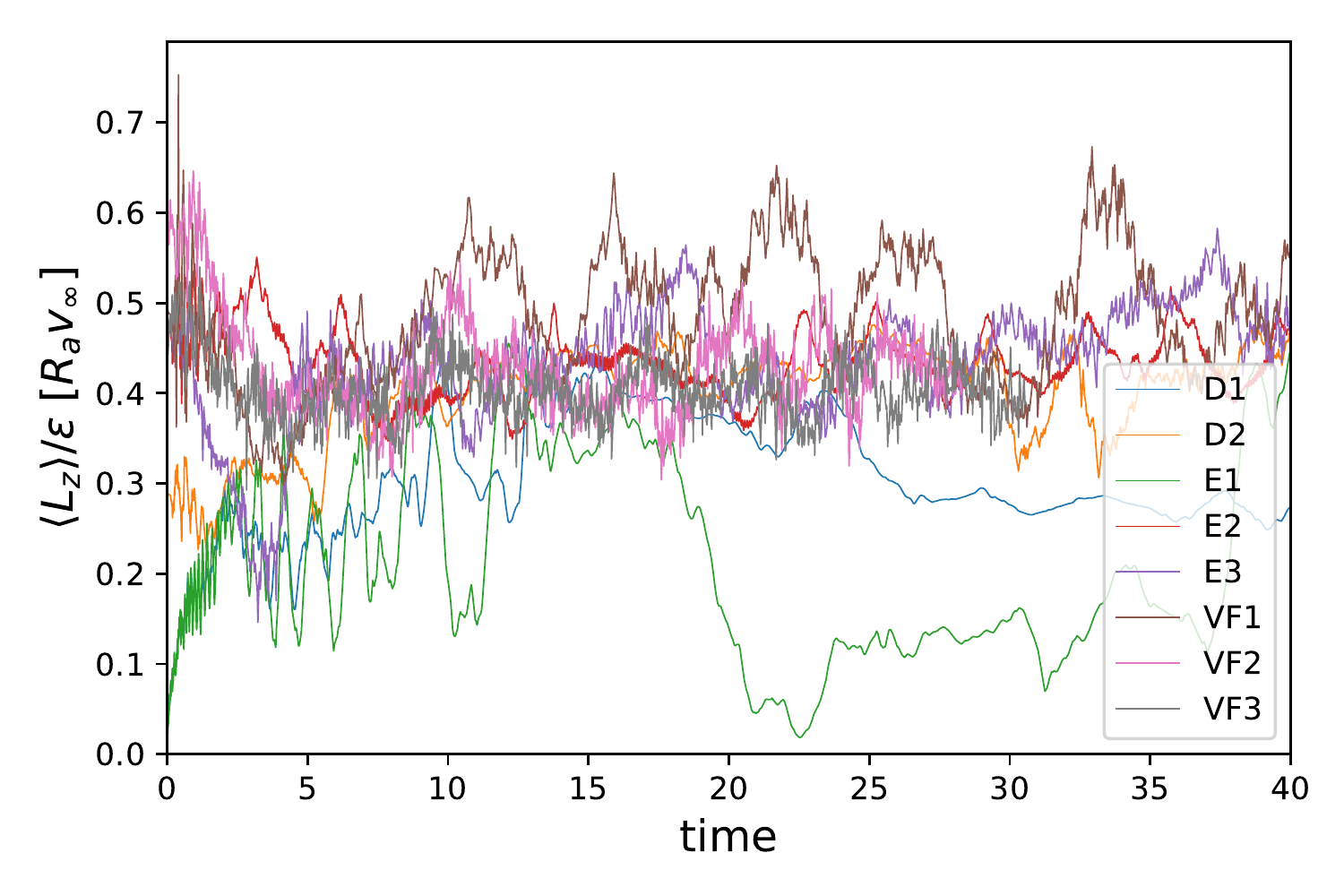}
\caption{$\langle L_z\rangle / \epsilon_\rho$ for simulations with a stable accretion flow. When $r_{\rm in}$ is small (i.e. all simulations in this figure except D1 and E1), $\langle L_z\rangle \approx 0.4 \epsilon_\rho R_a v_\infty$ and is independent of $r_{\rm in}$. (For VF1 - VF3, the angular momentum of the accretion flow is also affected by upstream velocity gradient and orbital effects, and their similarity to simulations with only density gradient is coincidental.)}
\label{fig:L_dot}
\end{figure}

\section{Discussion}
\subsection{Disk formation and regimes of parameter space}\label{subsec:compare_others}
Based on the flow stability and the possibility of forming a disk-like structure,
we can divide the parameter space into regimes of different behaviors. Despite the large number of relevant parameters, we will mainly focus on the $(\epsilon_\rho,r_{\rm in}/R_a)$ parameter space and assume that the behavior is less sensitive to other parameters (e.g. $\mathcal M$, $\epsilon_v$ and orbital effects).

The $(\epsilon_\rho,r_{\rm in}/R_a)$ parameter space can be divided into the following three regimes\footnote{Here we only consider $\epsilon_\rho\lesssim 1$. For $\epsilon_\rho>1$, the behavior may depend on the shape of $\rho_\infty(y)$, since approximating $\rho_\infty(y)$ by linear or exponential will no longer be appropriate; this regime is less relevant to SgXBs and is beyond the scope of this paper.}:
\begin{enumerate}

\item \textbf{$\epsilon_\rho^2\lesssim r_{\rm in}/50R_a$: stable flow, no disk formation.} This and the next regime have been discussed in \S5. The accretion flow is in a laminar steady-state, with $\dot M \sim \dot M_{\rm HL}$. The angular momentum in the accretion flow (which originally comes from the upstream gradient) is approximately conserved, and the mean specific angular momentum of accreted material is $\langle L_z\rangle \sim 0.4\epsilon_\rho R_a v_\infty$ (Figure \ref{fig:L_dot}). Note that we always have $\langle L_z\rangle \ll L_{\rm Kep}$ in this regime.

\item \textbf{$r_{\rm in}/50R_a \lesssim \epsilon_\rho^2\lesssim R_{\rm shock}/R_a$: turbulent flow, no disk formation.} The flow near the accretor is highly turbulent (\S\ref{subsec:morphology}),
and $\dot M$ and $\langle L_z\rangle$ undergo large random variation.
The variation of $\dot M$ and $\langle L_z\rangle$ increase and their mean values decrease as $r_{\rm in}$ decreases, but they barely depend on $\epsilon_\rho$.
This suggests that the accretion flow's memory of the (weak) upstream gradient is lost as it is disrupted by the turbulence.
The mean $\langle L_z\rangle / L_{\rm Kep}$ decreases as $r_{\rm in}$ decreases. As a result, the accretion flow cannot circulate before reaching the accretor, and disks cannot form.

\item \textbf{$R_{\rm shock}/R_a\lesssim \epsilon_\rho^2\lesssim 1 $: formation of turbulent disk-like structures.} Our simulation OAO lies in this regime (\S6.3). The flow circulates at a distance comparable to (or larger than) the shock standoff distance $R_{\rm shock}$.  The assumption of small upstream gradient used in the analysis of \S5 is no longer valid, and the flow (as well as the shock) should be highly asymmetric.
In this case, the incoming flow contains too much angular momentum to be fully disrupted by turbulence, and forms a turbulent disk-like structure.
This gives highly variable $\dot M$ (due to the turbulence) and variable but mostly positive $\langle L_z\rangle$ whose mean value is of order $L_{\rm Kep}$. The NS can spin up due to efficient angular momentum accretion.
\end{enumerate}

\begin{figure}
\centering
\includegraphics[width=.5\textwidth]{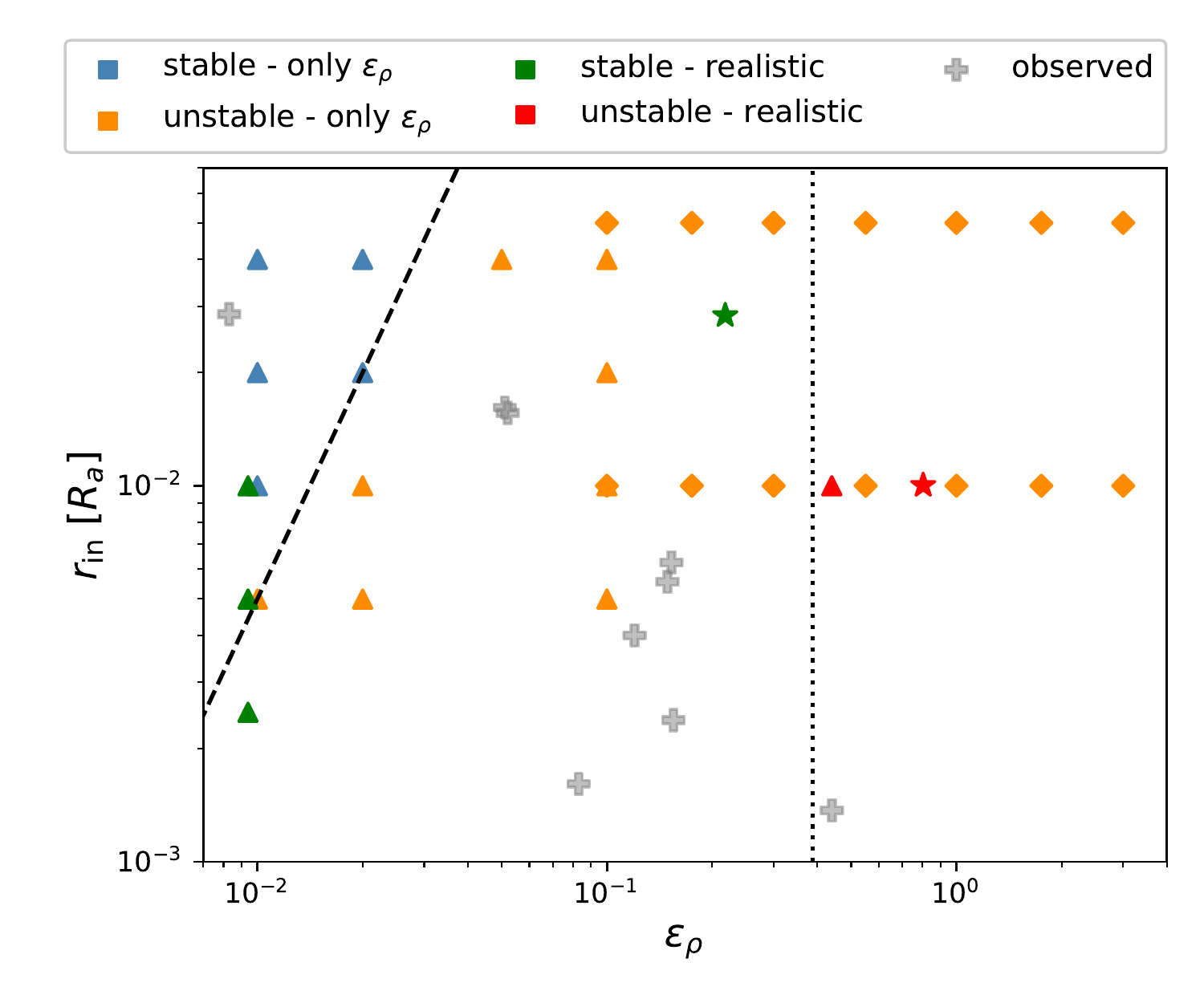}
\caption{Transverse upstream gradient $\epsilon_\rho$ and inner boundary size $r_{\rm in}$ for simulations in this work and other recent studies. Color marks the stability of the flow and the type of the simulation (with only density gradient, or using more realistic setups that also include velocity gradient and orbital effects such as Coriolis force). Marker shape signifies the source of the simulations: Triangles are from this work (B1 - E4, VF1 - 3, OAO), diamonds from \citet{MR15}, and stars from \citet{ElMellah2018b}. (We caution that our estimate of $\epsilon_\rho$ may not be accurate for the last work, which adopts a different simulation setup.) Estimated $\epsilon_\rho$ and $R_{\rm mag}/R_a$ (assuming $R_{\rm mag}\sim 10^9$ cm) for some observed systems are plotted in light grey crosses; we include all systems in Table \ref{tab:parameters1} and \ref{tab:parameters2} except ``Vela X-1 (fast)", which is based on early observations but is disfavored by more recent ones.
The dashed and dotted lines mark $r_{\rm in}=50\epsilon_\rho^2$ and $R_{\rm shock}=\epsilon_\rho^2$ (with $R_{\rm shock}\approx 0.15$; all lengths are in $R_a$) respectively; these are approximately the boundaries between different parameter regimes (see text).}
\label{fig:parameter_all}
\end{figure}

Approximate boundaries between these regimes are shown in Figure \ref{fig:parameter_all}, together with estimated parameters of observed systems and recent simulations.
Simulations of BHL accretion with only upstream density gradient ($\epsilon_\rho$) agree well with this picture.
However, when velocity gradient and orbital effects are also included (in attempt to model realistic systems), the stability of the flow seem to be increased (see \S\ref{subsec:vela}, and green points in Figure \ref{fig:parameter_all}).
Especially, one of the simulations in \citet{ElMellah2018b} (the green star in Figure \ref{fig:parameter_all}) shows stable flow but lies deep inside a regime where a turbulent flow is expected.
If this is not due to our inaccurate estimation of $\epsilon_\rho$ in their simulation (which is possible, since our method of estimating $\epsilon_\rho$ based on single-star wind profile may not be appropriate in their numerical setup where radiative wind acceleration inside the NS Roche lobe, which extends to a few $R_a$, is ignored), it will suggest that the stability of the flow is in fact sensitive to velocity gradient and orbital effects even though they appear weaker than the density gradient.
In that case, directly using our above results to predict the stability of the accretion flow in real systems may not be appropriate.

In our simulations, we observe disk-like structures when upstream gradients are large, but rotationally supported disks are never formed.
This is mainly due to our neglect of cooling.
For a $\gamma=5/3$ adiabatic flow, the pressure becomes high near the accretor due to compression and turbulent heating, and the strong (and asymmetric) pressure gradient prevents the formation of a rotationally supported disk.
When cooling is included \citep{ElMellah2018b} or a more compressible equation of state ($\gamma<5/3$) is adopted \citep{MacLeod2017,Berthier2017}, forming a rotationally supported thin disk becomes possible.

\begin{figure}
\centering
\includegraphics[width=.5\textwidth]{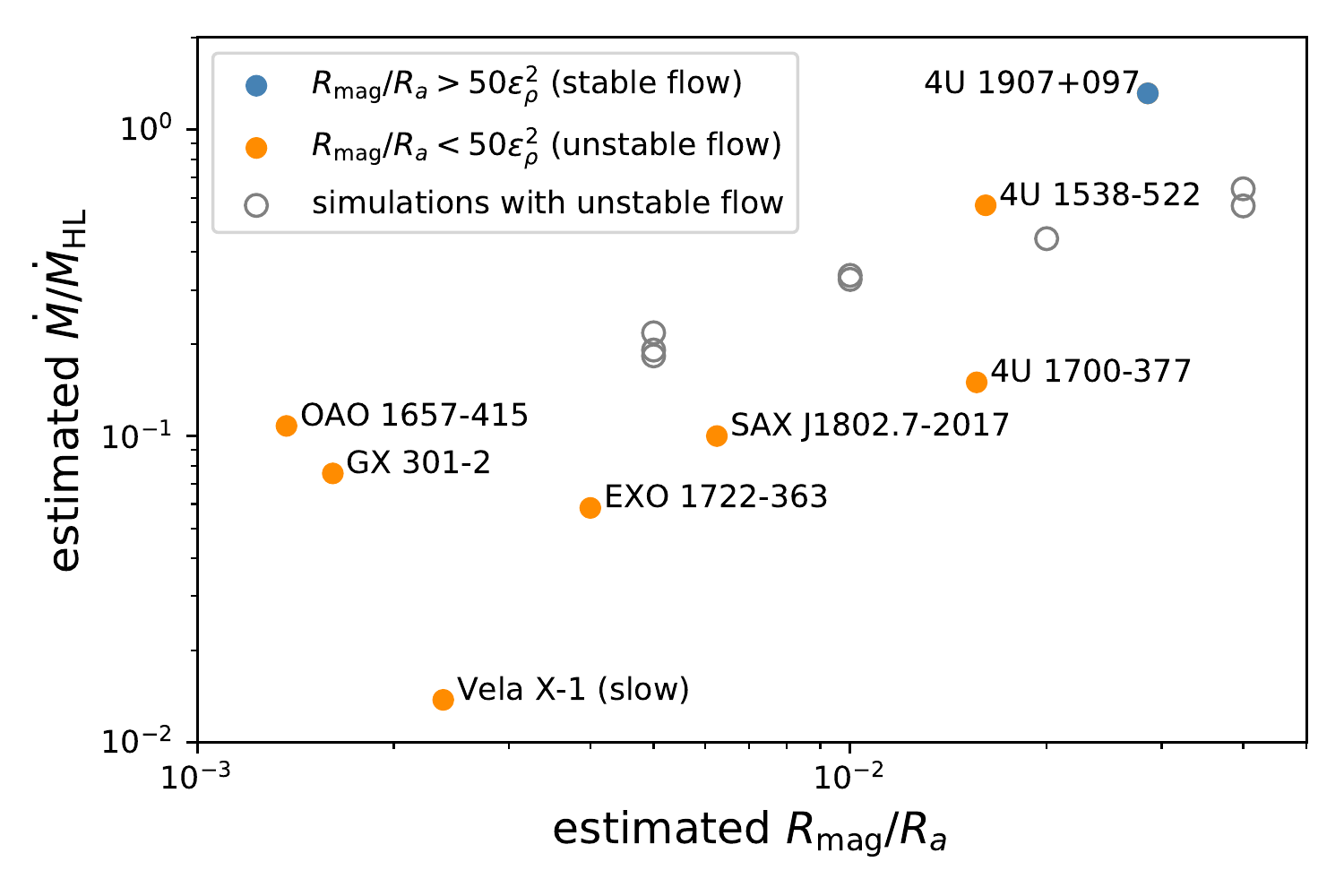}
\caption{Estimated $\dot M/\dot M_{\rm HL}$ and $R_{\rm mag}/R_a$ for SgXBs in Table \ref{tab:parameters1}.
Color marks the expected stability of the accretion flow.
$\dot M/\dot M_{\rm HL}$ and $r_{\rm in}/R_a$ of simulations with unstable flow are plotted (grey circles) for comparison.
The estimated data show trends that are consistent with our simulations.}
\label{fig:observed_M_dot}
\end{figure}

\subsection{Consistency with observation}
In order to further compare our results with observations, we estimate $\dot M/\dot M_{\rm HL}$ for SgXBs in Table \ref{tab:parameters1}.
$\dot M$ is estimated using the 17-40 keV flux (and distance) from \citet{Walter2015}, assuming that $R_{\rm NS}=10^6$ cm (10 km) and $10\%$ of the accretion power ($GM_{\rm NS}\dot M/R_{\rm NS}$) goes into isotropic radiation in the observed band.
$\dot M_{\rm HL}$ is estimated using the mass loss rate $\dot M_{\rm wind}$, assuming the single-star wind profile from \S\ref{subsec:wind_model}. We also estimate $R_{\rm mag}/R_a$ and $R_{\rm mag}/R_a\epsilon_\rho^2$, assuming $R_{\rm mag} = 10^3 R_{\rm NS} =10^9$ cm.

The estimated SgXB accretion rates show trends that are consistent with our simulation results,
despite the large uncertainties introduced by our crude assumptions.
The estimated $\dot M/\dot M_{\rm HL}$ and $R_{\rm mag}/R_a$ are summarized in Figure \ref{fig:observed_M_dot}.
There is only one system (4U 1907+097) with $R_{\rm mag}/R_a>50\epsilon_\rho^2$, which implies a stable accretion flow.
This is also the system with the largest $\dot M/\dot M_{\rm HL}$, consistent with our simulations which show reduced accretion rate when the flow is unstable.
For unstable systems ($R_{\rm mag}/R_a<50\epsilon_\rho^2$), the estimated data show the same trend as our simulations, with $\dot M/\dot M_{\rm HL}$ decreasing as $R_{\rm mag}/R_a$ decreases.

\subsection{Other factors that may affect accretion dynamics}\label{subsec:other_factors}
\subsubsection{Perturbations in the line-driven wind}\label{subsec:clumpy_wind}
In reality, the stellar wind from the companion is not necessarily smooth and unperturbed. Variability at the windbase \citep{Poe1990} together with the line-deshadowing instability \citep{LucyWhite1980} can make the wind ``clumpy". Observational results are also consistent with a clumpy wind \citep{Surlan2013}. Numerical simulation by \citet{Sundqvist2018} shows that for typical O star parameters, the clumps in the wind can be over one order of magnitude denser, with size $\sim 0.05 R_c$ at $\sim 2R_c$ (this is also the typical location of the NS in a SgXB undergoing wind accretion).
For systems with fast wind (see Table \ref{tab:parameters2}) this clump size is $\sim R_a$; this should make the accretion flow highly variable and reduce the accretion rate, as is shown in the simulation of \citet{ElMellah2018}.

Still, this does not mean that investigating accretion from a smooth (or slightly perturbed) upstream flow is irrelevant for SgXB systems, since the perturbation in the wind may not be as strong for different stellar parameters, and other physical mechanisms may suppress the initial perturbation that triggers the instability, making the variability small (if not zero) at the NS.

\subsubsection{Radiation from NS}
Radiation from the NS ionizes the wind, reducing the wind speed near the accretor by suppressing radiative acceleration. This is likely related to the observed quasi-periodicity and off-states of Vela X-1 (see discussion in \S\ref{subsec:vela}).

\subsubsection{Dynamics near and inside the magnetosphere}
All our simulations ignores the effect of NS magnetic field, whose effect can be important for $r\lesssim R_{\rm mag}$. Therefore, our simulations are only meaningful for flow outside the magnetosphere (and we do achieve $r_{\rm in}\sim R_{\rm mag}$ in some cases). What happens inside the magnetosphere directly affects the accretion rate and its variability, and feedback (e.g. outflows and jets) from inside the magnetosphere may modify the dynamics of larger scale flows.

Accretion near and inside the magnetosphere has been discussed, for example, in \citet{Shakura2013}, which considers quasi-spherical accretion (i.e. the outer boundary condition is a radial, laminar inflow).
However, our simulations show that the magnetosphere is likely embedded in a highly turbulent environment, and whether this affects the accretion dynamics inside the magnetosphere requires further investigation.

\section{Conclusion}
In this paper, we investigate BHL accretion with and without transverse upstream gradients using 2D axisymmetric and 3D hydrodynamic simulations. We use a $\gamma=5/3$ adiabatic equation of state, and focus on the regime of high (upstream) Mach number, weak upstream gradients and small accretor size; this regime is relevant to most observed wind-fed SgXBs but has not been systematically explored before.

When there are no upstream gradients, 2D axisymmetric (\S4) and 3D (\S5.2) simulations at different Mach number and accretor size ($r_{\rm in}$) all show stable, axisymmetric accretion flow. The ``flip-flop" instability commonly observed in 2D planar BHL accretion does not occur in 3D.

When the upstream gradients are small but nonzero (\S5.3), however, the flow is significantly more prone to instability than previously expected. When there is only upstream density gradient ($\epsilon_\rho$), the flow becomes unstable at $\epsilon_\rho^2\gtrsim r_{\rm in}/50R_a$ (\S5.4).
When unstable, the flow near the accretor eventually becomes highly turbulent.
The accretion rate ($\dot M$) and mean (averaged over the accretor but not over time) specific angular momentum of accreted material ($\langle L_z\rangle$) depend on $r_{\rm in}$ but barely on $\epsilon_\rho$.
As $r_{\rm in}$ decreases, the time-averaged $\dot M$ and $\langle L_z\rangle$ decrease (with approximately $\dot M\propto r_{\rm in}^{1/2}$) and their variation increase.

The turbulent flow near the accretor reduces the mean specific angular momentum of the flow near the accretor and can suppress disk formation (\S5.5).
A persistent disk-like structure cannot form until $\epsilon_\rho^2\gtrsim R_{\rm shock}/R_a$, with $R_{\rm shock}$ being the shock standoff distance (\S7.1). 
This requires significantly larger upstream gradients compared to simple analytic estimates, which often suggest that disk should form when $\epsilon_\rho^2\gtrsim r_{\rm in}/R_a$.

To apply our results to real SgXBs, we develop a model which adopts a more realistic wind profile and includes orbital effects (\S2.2, \S6.1). We perform two sets of simulations with this model, one resembling Vela X-1 with (likely unrealistic) fast wind, the other resembling OAO 1657-415.
These two systems serve as examples of small and large upstream gradients respectively.
For Vela X-1 with fast wind (\S6.2), we observe a stable accretion flow down to $r_{\rm in}=0.0025R_a$; this is more stable compared to simulations with only upstream density gradient. This trend of increased stability when adopting a realistic wind model and including orbital effects is also observed in previous studies, but its reason remains unclear (\S\ref{subsec:compare_others}).
For OAO 1657-415 (\S6.3), the large upstream gradients lead to the formation of a persistent but turbulent and geometrically thick disk-like structure, agreeing with our analytic estimate.

Regimes of different behaviors in $(\epsilon_\rho, r_{\rm in}/R_a)$ parameter space are summarized in \S\ref{subsec:compare_others}.
The regime of turbulent and disk-less flow occupies a large parameter space\footnote{Meanwhile, simple analytic estimates often suggest that the flow can be unstable without disk-like structure only when $\epsilon_\rho^2\sim r_{\rm in}/R_a$ (which is a very small region in the parameter space) since (naively) smaller $\epsilon_\rho$ should give stable flow and larger $\epsilon_\rho$ should allow disk formation.} since the accretion flow is prone to instability, and most observed SgXBs lie inside this regime.
Overall, without cooling accretion disks are harder to form than expected. Future work to explore the effect of radiative cooling is required.

\section*{Acknowledgements}
We thank Morgan MacLeod for insightful discussions and comments. JMS was supported by NSF grant AST-1715277.

\bibliographystyle{mnras}
\bibliography{ms}

\end{document}